\pdfoutput=1
\newcommand*{\ATLASLATEXPATH}{latex/}
\documentclass[UKenglish,texlive=2016,txfonts,cernpreprint]{\ATLASLATEXPATH atlasdoc}

\usepackage[csquotes=false,default,subfigure,backend=biber]{\ATLASLATEXPATH atlaspackage}
\usepackage{\ATLASLATEXPATH atlasbiblatex}

\usepackage{\ATLASLATEXPATH atlascontribute}

\usepackage{\ATLASLATEXPATH atlasphysics}

\graphicspath{{logos/}{figures/}}

\usepackage{TPPaper-defs}

\AtlasTitle{Electron reconstruction and identification in the ATLAS experiment using the 2015 and 2016 LHC proton--proton collision data at $\sqrt{s} = 13$~\tev}

\author{The ATLAS Collaboration}

\AtlasRefCode{PERF-2017-01}
\AtlasNote{PERF-2017-01}

\PreprintIdNumber{CERN-EP-2018-273}

\AtlasJournal{EPJC}
\AtlasJournalRef{\EPJC 79 (2019) 639}
\AtlasDOI{DOI:10.1140/epjc/s10052-019-7140-6}

\AtlasAbstract{%
Algorithms used for the reconstruction and identification of electrons in the central region of the ATLAS detector at the Large Hadron Collider (LHC) are presented in this paper; these algorithms are used in ATLAS physics analyses that involve electrons in the final state and which are based on the 2015 and 2016 proton--proton collision data produced by the LHC at $\sqrt{s}$ = 13~\TeV. The performance of the electron reconstruction, identification, isolation, and charge identification algorithms is evaluated in data and in simulated samples using electrons from \Zee\ and \Jee\ decays. Typical examples of combinations of electron reconstruction, identification, and isolation operating points used in ATLAS physics analyses are shown.}

\hypersetup{pdftitle={ATLAS document},pdfauthor={The ATLAS Collaboration}}

\begin{document}

\maketitle

\tableofcontents

\clearpage

\newpage
\section{Introduction}
\label{sec:intro}

Stable particles that interact primarily via the electromagnetic interaction, such as electrons, muons, and photons, are found in many final states of proton--proton ($pp$) collisions at the Large Hadron Collider (LHC) located at the CERN Laboratory. These particles are essential ingredients of the ATLAS experiment's Standard Model and Higgs-boson physics programme as well as in searches for physics beyond the Standard Model. Hence, the ability to effectively reconstruct electrons\footnote{Throughout this paper, the term ``electron'' usually indicates both electrons and positrons.} originating from the prompt decay of particles such as the \Zboson boson, to identify them as such with high efficiency, and to isolate them  from misidentified hadrons, electrons from photon conversions, and non-isolated electrons originating from heavy-flavour decays are all essential steps to a successful scientific programme.

The ATLAS Collaboration has presented electron-performance results in several publications since the start of the high-energy data-taking in 2010~\cite{PERF-2010-04,PERF-2013-03,PERF-2016-01}. The gradual increase in peak luminosity and the number of overlapping collisions (pile-up) in ATLAS has necessitated an evolution of the electron reconstruction and identification techniques.
In addition, the LHC shutdown period of 2013--2014 brought a new charged-particle detection layer to the centre of ATLAS and a restructuring of the trigger system, both of which impact physics analyses with electrons in the final state. These changes require a new benchmarking of electron-performance parameters.   
The electron efficiency measurements presented in this paper are from the  data recorded during the 2015--2016 LHC $pp$ collision run at centre-of-mass energy $\sqs = 13$~\TeV.  During the period relevant to this paper, the LHC circulated \SI{6.5}{\TeV} proton beams with a \SI{25}{ns} bunch spacing.
The peak delivered instantaneous luminosity was $\mathcal{L}=1.37\times 10^{34}$~cm$^{-2}$s$^{-1}$ and the mean number of $pp$ interactions per bunch crossing (hard scattering and pile-up events) was $\langle \mu \rangle = 23.5$.  The total integrated luminosity~\cite{DAPR-2013-01} used for most of the measurements presented in this paper is \lumi~\ifb. 
Another important goal of this paper is to document the methods used by the ATLAS experiment at the start of Run~2 of the LHC (2015 and beyond) to reconstruct, identify, and isolate prompt-electron candidates  with high efficiency, as well as to suppress electron-charge misidentification. The methods presented here would be of value to other experiments with similar experimental conditions of fine granularity detection devices but also substantial inactive material in front of the active detector, or with significant activity from pile-up events.


The structure of the paper is described in the following, highlighting additions and new developments with respect to Ref.~\cite{PERF-2016-01}. 
Section~\ref{sec:detector} provides a brief summary of the main components of the detector germane to this paper, with special emphasis on the changes since the 2010-2012 data-taking period.
Section~\ref{sec:samples} itemises the datasets and simulated-event samples used in this paper. 
Given that the method for calculating efficiencies is common to all measurements, it is described in Section~\ref{sec:efficiencies}, before the individual measurements are presented.  
The algorithms and resulting measurements for electron reconstruction efficiencies are described in Section~\ref{sec:reconstruction}, including a detailed discussion of the Gaussian Sum Filter algorithm.
Electron identification and the corresponding measurement of efficiencies are described in Section~\ref{sec:identification}. New developments here include the optimisation based on simulated events and the treatment of electrons with high transverse momentum.
The algorithms used to identify isolated electron candidates and the resulting measured benchmark efficiencies are published for the first time; these are presented in Section~\ref{sec:isolation}. 
This paper also presents detailed discussion of studies of the probability to mismeasure the charge of an electron; these are presented in Section~\ref{sec:chargemisid}. This section also includes a discussion of the sources of charge misidentification and a new Boosted Decision Tree algorithm that reduces the rate of charge-misidentified electrons significantly. 
A few examples of combined reconstruction, identification, and isolation efficiencies for typical working points used in ATLAS physics analyses but illustrated with a common \Zee\ sample are shown in Section~\ref{sec:examples}. 
The summary of the work is given in Section~\ref{sec:conclusion}.

\section{The ATLAS detector}
\label{sec:detector}
\newcommand{\AtlasCoordFootnote}{%
ATLAS uses a right-handed coordinate system with its origin at the nominal interaction point (IP) in the centre of the detector and the $z$-axis along the beam pipe. The $x$-axis points from the IP to the centre of the LHC ring,
and the $y$-axis points upwards. Cylindrical coordinates $(r,\phi)$ are used in the transverse plane, $\phi$ being the azimuthal angle around the $z$-axis.
The pseudorapidity is defined in terms of the polar angle $\theta$ as $\eta = -\ln \tan(\theta/2)$.
Angular distance is measured in units of $\Delta R \equiv \sqrt{(\Delta\eta)^{2} + (\Delta\phi)^{2}}$.}

The ATLAS detector~\cite{PERF-2007-01} is designed to observe particles produced in the high-energy $pp$ and heavy-ion LHC collisions. It is composed of an inner detector, used for charged-particle tracking, immersed in a 2~T axial magnetic field
produced by a thin superconducting solenoid; electromagnetic (EM) and
hadronic calorimeters outside the solenoid; and a muon spectrometer. A two-level triggering system reduces the total
data-taking rate 
to approximately 1~kHz. 
The second level, the high-level trigger (HLT), employs selection algorithms using full-granularity detector information; likelihood-based electron identification and its HLT variant are described in Section~\ref{sec:identification}.

The inner  detector provides precise reconstruction of tracks
within a pseudorapidity range\footnote{\AtlasCoordFootnote}
$|\eta| \lesssim 2.5$. The innermost part of the inner detector consists of a high-granularity silicon pixel detector and includes the insertable B-layer~\cite{Abbott:2018ikt,ATLAS-TDR-19}, a new tracking layer closest to the beamline designed to improve impact parameter resolution, which is important primarily for heavy-flavour identification.
The silicon pixel detector provides typically four measurement points for charged particles originating in the beam-interaction region. A
semiconductor tracker (SCT) consisting of modules with two layers of silicon microstrip sensors
surrounds the pixel detector and provides typically eight
hits per track at intermediate radii. The outermost region of the inner detector
is covered by a transition radiation tracker (TRT) consisting of straw
drift tubes filled with a xenon-based gas mixture, interleaved with polypropylene/polyethylene radiators.  The TRT offers
electron identification capability via the detection of transition-radiation photons generated by the radiators for highly relativistic particles.
Some of the TRT modules instead contain an argon-based gas mixture, as mitigation for gas leaks that cannot be repaired without an invasive opening of the inner detector. The presence of this gas mixture is taken into account in the simulation. 
ATLAS has developed a TRT particle-identification algorithm that partially mitigates the loss in identification power caused by the use of this argon-based gas mixture.
For charged particles with transverse momentum
$\pT>0.5$~\GeV\ within its pseudorapidity coverage ($|\eta| \lesssim
2$), the TRT provides typically 35 hits per track. 

The ATLAS calorimeter system has both electromagnetic and hadronic
components and covers the pseudorapidity range $|\eta|<4.9$, with
finer granularity over the region matching the inner detector. The
central EM calorimeters are of an accordion-geometry design made from
lead/liquid-argon (LAr) detectors, providing a full $\phi$ coverage.
These detectors are divided into two half-barrels ($-1.475<\eta<0$ and
$0<\eta<1.475$) and two endcap components ($1.375<|\eta|<3.2$),
with a transition region between the barrel and the endcap
($1.37<\abseta<1.52$) which contains a relatively large amount of
inactive material. Over the region devoted to precision measurements
($|\eta|<2.47$, excluding the transition regions), the EM calorimeter
is segmented into longitudinal (depth) compartments called the 
\emph{first} (also known as \emph{strips}), \emph{second}, and
\emph{third} layers. The first layer consists of strips finely segmented in 
$\eta$, offering excellent discrimination between photons
and $\pi^{0}\to\gamma\gamma$ decays. At electron or photon energies relevant to this paper, most of the energy is
collected in the second layer, which has a lateral granularity of
$0.025 \times 0.025$ in $(\eta,\phi)$ space, while the third layer
provides measurements of energy deposited in the tails of the shower.
The central EM calorimeter is complemented by
two presampler detectors in the region $|\eta|<1.52$ (barrel) and
$1.5<|\eta|<1.8$ (endcaps), made of a thin LAr layer, providing a
sampling for particles that start showering in front of the EM
calorimeters.
Hadronic calorimetry is provided by the steel/scintillating-tile calorimeter,
segmented into three barrel structures within \(|\eta| < 1.7\), and two copper/LAr hadronic endcap calorimeters.
They surround the EM calorimeters and provide
additional discrimination through further energy measurements of
possible EM shower tails as well as rejection of events with activity of hadronic origin. 

\section{Datasets and simulated-event samples}
\label{sec:samples}
All data collected by the ATLAS detector undergo careful scrutiny to ensure the quality of the recorded information; data used for the efficiency measurements are filtered by requiring that all detector subsystems needed in the analysis (calorimeters and tracking detectors) are operating nominally. After all data-quality requirements (94\% efficient), \lumi~\ifb\ of $pp$ collision data from the 2015--2016 dataset are available for analysis. Some results in this paper are based on the 2016 dataset only, and contain approximately 10\% less data.

Samples of simulated \Zee\ and \Jee\ decays as well as single-electron samples are used to benchmark the expected electron efficiencies and to define the electron-identification criteria.  
The \Zee\  Monte Carlo (MC) samples were generated with the \POWHEG-Box~v2 MC program~\cite{Nason:2004rx,Frixione:2007vw,Alioli:2010xd,Alioli:2008gx,Frixione:2007nw} interfaced to the \PYTHIA~v.8.186~\cite{Sjostrand:2007gs} parton shower model. The CT10 parton distribution function (PDF) set~\cite{Lai:2010vv} was used in the event generation with the matrix element, and the AZNLO~\cite{STDM-2012-23} set of generator-parameter values (tune) with the CTEQ6L1~\cite{Pumplin:2002vw} PDF set were used for the modelling of non-perturbative effects.
The \Jee\  samples were generated with \PYTHIA~v.8.186; the A14 set of tuned parameters~\cite{ATL-PHYS-PUB-2014-021} was used together with the CTEQ6L1 PDF set for event generation and the parton shower. The simulated single-electron samples were produced with a flat distribution in $\eta$ as well as in \pT\ in the region $3.5~\GeV$ to $100~\GeV$, followed by  a linear ramp down to 300~\GeV, and then a flat distribution again to 3~\TeV. 
For studies of electrons in simulated event samples, the reconstructed-electron track is required to have hits in the
inner detector which originate from the true electron during simulation.

Backgrounds that may mimic the signature of prompt electrons were simulated
with two-to-two processes in the \PYTHIA~v.8.186 event generator using the A14 set of tuned parameters and the NNPDF2.3LO PDF set~\cite{Ball:2012cx}.
These processes include multijet production, $qg\rightarrow q\gamma$,
$q\bar{q}\rightarrow g\gamma$, \Wboson- and \Zboson-boson production (as well as other
electroweak processes), and top-quark production.
A filter was applied to the simulation to enrich the final sample in electron
backgrounds.
This filter retains events in which particles produced in the hard scatter
(excluding muons and neutrinos) have a summed energy that exceeds 17~\GeV\
in an area of $\Delta \eta \times \Delta \phi = 0.1 \times 0.1$,
which mimics the highly localised energy deposits that are characteristic
of electrons. 
When using this background sample, prompt electrons from \Wboson- and \Zboson-boson decays
are excluded using generator-level simulation information.

Multiple overlaid $pp$ collisions were simulated with the soft QCD processes of \PYTHIA~v.8.186 using the {\textsc{MSTW2008LO}} PDF~\cite{Martin:2009iq}. The Monte Carlo events were reweighted so that the $\langle \mu \rangle$ distribution matches the one observed in the data. All samples were processed with the {\textsc{Geant}4}-based simulation~\cite{Agostinelli:2002hh,SOFT-2010-01} of the ATLAS detector.

\section{Electron-efficiency measurements}
\label{sec:efficiencies}
Electrons isolated from other particles are important ingredients in Standard Model
measurements and in searches for physics beyond the Standard Model.
However, the experimentally determined electron spectra must be
corrected for
the selection efficiencies, such as those related to the
trigger, as well as particle isolation, identification, and reconstruction, before absolute
measurements can be made. These efficiencies may be
estimated directly from data using tag-and-probe
methods. These methods select, from known
resonances such as \Zee\ or \Jee, unbiased samples of electrons (\emph{probes})
by using strict selection requirements on the second object (\emph{tags}) produced
from the particle's decay. The events are selected on the basis of the
electron--positron invariant mass.
The efficiency of a given requirement can then be determined by applying it to the
probe sample after accounting for residual background contamination.

The total efficiency \epstot\ 
may be factorised as a product of different efficiency terms:
\begin{equation} \label{eq:EffDef}
  \epstot = \epsemclus \times \epsreco \times \epsid \times \epsiso \times \epstrig =
  \left( \frac{ \Nclus }{ \Nall }\right) \times \left(\frac{ \Nreco }{  \Nclus }\right) \times \left( \frac{ \Nid }{ \Nreco } \right) \times \left( \frac{ \Niso }{ \Nid } \right) \times \left( \frac{ \Ntrig }{ \Niso } \right) .
\end{equation}
The efficiency to reconstruct in the electromagnetic calorimeter EM-cluster candidates (localised energy deposits) associated with all produced electrons, \epsemclus, is given by the number of reconstructed EM calorimeter clusters \Nclus divided by the number of produced electrons \Nall. This efficiency is evaluated entirely from simulation, where the reconstructed cluster is associated to a genuine electron produced at generator level.   The reconstruction efficiency, \epsreco, is given by the number of reconstructed  electron candidates \Nreco\ divided by the number of EM-cluster candidates \Nclus. This reconstruction efficiency, as well as the efficiency to reconstruct electromagnetic clusters, is described in Section~\ref{sec:reconstruction}. The identification efficiency, \epsid, is given by the number of identified and reconstructed electron candidates \Nid\ divided by \Nreco, and is described in Section~\ref{sec:identification}. The isolation efficiency is calculated as the number of identified electron candidates satisfying the isolation, identification, and reconstruction requirements \Niso\ divided by \Nid, and is explained in Section~\ref{sec:isolation}. Finally, the trigger efficiency is calculated as the number of triggered (and isolated, identified, reconstructed) electron candidates \Ntrig\ divided by \Niso (see for example Ref.~\cite{TRIG-2016-01}; trigger efficiency is not discussed further in this paper). 
 
Isolated electrons selected for physics analyses are subject to large backgrounds from
misidentified hadrons, electrons from photon conversions, and
non-isolated electrons originating from heavy-flavour decays. 
The biggest challenge in the efficiency measurements presented in this paper is the estimation of probes that originate from background rather than signal processes.  This background is largest for the sample of cluster probes, but the fraction of such events is reduced with each efficiency step, from left to right, as given in Eq.~(\ref{eq:EffDef}).

The accuracy with which the detector simulation models the observed electron efficiency plays an important role when using simulation to predict physics processes, for example the signal or background of a measurement. In order to achieve reliable results, the simulated events need to be corrected to reproduce as closely as possible the efficiencies measured in data. This is achieved by applying a multiplicative correction factor to the event weight in simulation. This correction factor is defined as the ratio of the efficiency measured in data to that determined from Monte Carlo events. These correction weights are normally close to unity; deviations from unity usually arise from mismodelling in the simulation of tracking properties or shower shapes in the calorimeters. 

Systematic uncertainties in the correction factors are evaluated by varying the requirements on the selection of both the tag and the probe electron candidates as well as varying the details of the background-subtraction method. The central value of the measurement is extracted by averaging the measurement results over all variations.  The statistical uncertainty in a single variation of the measurement is calculated following the approach in Ref.~\cite{cdf7168_eff_uncertainties}, i.e.\ assuming a binomial distribution. If the evaluation of the number of events (before or after the selection under investigation) is the result of a background subtraction, the corresponding statistical uncertainties are also included in the overall statistical uncertainty, rather than in the systematic uncertainty. 
The systematic uncertainty in the averaged result is obtained from the root-mean-square (RMS)  of the individual results, and in the case of non-Gaussian behaviour, it is inflated to cover 68\% of the variations.  

The tag-and-probe measurements are based on samples of \Zee\ and \Jee\ events. Whereas the \Zee sample is used to extract all terms in Eq.~(\ref{eq:EffDef}), the \Jee\ sample is only used to extract the identification efficiency \epsid since the significant background as well as the difficulties in designing a trigger for this process prevent its use in determining the reconstruction efficiency. The combination of the two samples
allows identification efficiency measurements over a significant transverse energy \et~range of 4.5~\GeV\ to 20~\GeV\ for the \Jee\ sample, and above 15~\GeV\ (4.5~\GeV\ for the isolation efficiency measurement) for the \Zee\ sample,  while still providing overlapping measurements between the samples in the \et\ range 15--20~\GeV\ where the correction factors of the two results are combined using a $\chi^2$ minimisation~\cite{Aaron:2009bp,PERF-2013-03}.  Combining the correction factors instead of the individual measured and simulated efficiencies reduces the dependence on kinematic differences of the physics processes as they cancel out in the ratio. 

Due to the number of events available in the sample, the \Zee\ tag-and-probe measurements provide limited information about electron efficiencies beyond approximately electron $\et=150$~\GeV. The following procedure is used to assign correction factors for candidate electrons with high \et:
\begin{itemize}
\item \textit{reconstruction:} the same $\eta$-dependent correction factors are used for all $\et>80$~\GeV,
\item \textit{identification:} correction factors determined up to $\et=250$~\GeV\ are applicable beyond,
\item \textit{isolation:} correction factors of unity  are used for $\et>150$~\GeV.
\end{itemize}

The following subsections give a brief overview of the methods used to extract efficiencies in the data. Efficiency extraction using simulated events is performed in a very similar fashion, except that no background subtraction is performed. More detailed descriptions may be found in Ref.~\cite{PERF-2016-01}.

\subsection{Measurements using \Zee\ events}
\label{sec:ZeeEffMeasurements}

\Zee~events with two electron candidates in the central region of the detector, $|\eta| < 2.47$, were collected using two triggers designed to identify at least one electron in the event. One trigger has a minimum \Et\ threshold of 24 GeV (which was changed to 26 GeV during 2016 data-taking), and requires \Tight trigger identification (see Section~\ref{sec:identification}) and track isolation (see Section~\ref{sec:isolation}), while the other trigger has a minimum \Et\ threshold of 60 GeV and \Medium trigger identification. The tag electron is required to have $\et>27$~\GeV\ and to lie outside of the calorimeter transition region, $1.37< |\eta| <1.52$. It must be associated with the object that fired the trigger, and must also pass \tight-identification (see Section~\ref{sec:identification}) and isolation requirements. If both electrons pass the tag requirements, the event will provide two probes. The invariant-mass distribution constructed from the tag electron and the cluster probe is used to discriminate prompt electrons from background. The signal efficiency is extracted in a window of $\pm 15$~\gev\ around the \Zboson-boson mass peak at 91.2~\gev. Approximately 35 million electron-candidate probes from \Zee\ data events are available for analysis.

The probe electrons in the denominator of the reconstruction-efficiency measurement (see Eq.~(\ref{eq:EffDef})) are electromagnetic clusters both with and without associated tracks, while those in the numerator consist of clusters with matched tracks, i.e.\ reconstructed electrons (see Section~\ref{sec:reconstruction}). These tracks are required to have at least seven hits in the silicon detectors (i.e.\ both pixel and SCT) and at least one hit in the pixel detector. 
The background for electron candidates without a matched track is estimated by fitting a polynomial to the sideband regions of the invariant-mass distribution of the candidate electron pairs, after subtracting the remaining signal contamination using simulation. The background for electron candidates with a matched track is estimated by constructing a background template by inverting identification or isolation criteria for the probe electron candidate and normalising it to the invariant-mass sideband regions, after subtraction of the signal events in both the template and the sidebands.  

The probe electrons used in the denominator of the identification efficiency measurement are the same as those used in the numerator of the reconstruction efficiency measurement, with an additional opposite-charge requirement on the tag--probe pair; this method assumes that the charge of the candidate is correctly identified. The numerator of the identification measurement consists of probes satisfying the identification criteria under evaluation. 
Two methods are used in the identification measurements to estimate the non-prompt background~\cite{PERF-2013-03,PERF-2016-01}; they are treated as variations of the same measurement: the \textit{Zmass} method uses the invariant mass of the tag--probe pair while the \textit{Ziso} method uses the isolation distribution of probes in the signal mass window around the \Zboson-boson peak. In both cases, and as discussed for the  reconstruction-efficiency measurement, background templates are formed and normalised to the sideband regions, after subtraction of the signal events. The contamination from charge-misidentified candidates is negligible in this sample. In the \textit{Zmass} method, the numerator of the identification efficiency uses same-charge events to obtain a normalisation factor for the template in opposite-charge events, in order to reduce the contamination from signal events.

The isolation-efficiency measurements are performed using the \textit{Zmass} method, as described above. The denominator in the efficiency ratio is the number of identified electron candidates, while the numerator consists of candidates that also satisfy the isolation criteria under evaluation.

In all cases, systematic uncertainties in the data-MC correction factors are evaluated from the background-subtraction method as well as variations of the quality of the probed electrons via changes in the window around the \Zboson-boson mass peak. They are also evaluated by varying the identification and isolation requirements on the tag, the sideband regions used in the fits, and the template definitions. 

\subsection{Measurements using $\Jee$ events}

\Jee~events with at least two electron candidates with \et\ > 4.5~\gev\ and $|\eta|<2.47$ were collected with dedicated dielectron triggers with electron \et\ thresholds ranging from 4~to 14~\GeV.
Each of these triggers requires \Tight trigger identification and \Et\ above a certain threshold for one trigger object, while only demanding the electromagnetic cluster \Et\ to be higher than some other (lower) threshold for the second object.
The \Jee\ selection consists  of one electron candidate passing a \tight-identification selection (see Section~\ref{sec:identification}) and one reconstructed-electron candidate (see Section~\ref{sec:reconstruction}). The tag electron is required to be outside the calorimeter transition region $1.37 < |\eta| < 1.52$ and  to be associated with the \tight\ trigger object. The probe electron must be matched to the second trigger object. Due to the nature of the sample (a mixture of prompt and non-prompt decays) as well as significant background, isolation requirements are applied on both the tag and the probe electrons, although for the latter the requirement is very loose so as to not bias the identification-efficiency measurement. Furthermore, the tag and the probe electron candidates must be separated from each other in $\eta$--$\phi$ space by $\Delta R > 0.15$. If both electrons pass the tag requirements, the event will provide two probes. Approximately 80 thousand electron-candidate probes from  \Jee\ data events are available for analysis.

The invariant-mass distribution of the two electron candidates in the range 1.8--4.6~\GeV\ 
is fit with functions to extract three contributions: \jpsi\ events, \psip\ events, and the background from hadronic jets, heavy flavour, and electrons from conversions. 
The \jpsi\ and \psip\ contributions are each modelled with a Crystal Ball function convolved with a Gaussian function, and the background is estimated using same-charge events and fit with a second-order Chebyshev polynomial.

\Jee\ events come from a mixture of prompt and non-prompt \jpsi\ production, with relative fractions depending both on the triggers used to collect the data and on the \et\ of the probe electrons. 
Prompt \jpsi\ mesons are produced directly in $pp$ collisions and in radiative decays of directly produced heavier charmonium states. Non-prompt \jpsi\ production occurs when the \jpsi\ is produced in the decay of a $b$-hadron. Only the prompt production yields isolated electrons, which are expected to have efficiencies similar to those of electrons from physics processes of interest such as $H \rightarrow ZZ^* \rightarrow 4\ell$.
Given the difficulties associated with the fact that electrons from non-prompt decays are often surrounded by hadronic activity, two methods have been developed to measure the efficiency for isolated electrons at low \et, both
exploiting the pseudo-proper time variable\footnote{The pseudo-proper time is defined  as $t_{0}=L_{xy}\cdot m^{\jpsi}_\mathrm{PDG}/\pT^{\jpsi}$, where $L_{xy}$ is the  displacement of the \jpsi\ vertex from the primary vertex  projected onto the flight direction of the \jpsi\ in the transverse  plane, $m^{\jpsi}_\mathrm{PDG}$ is the nominal \jpsi~mass~\cite{Patrignani:2016xqp} and $\pT^{\jpsi}$ is the  \jpsi-reconstructed transverse momentum.}  $t_{0}$.
In the \textit{cut} method, a requirement is imposed on the pseudo-proper time, so that the prompt component is enhanced, thereby limiting the non-prompt contribution. The residual non-prompt fraction is estimated using simulated samples and ATLAS measurements of \Jmm~\cite{BPHY-2010-01}. In the \textit{fit} method, a fit to the pseudo-proper time distribution is used to extract the prompt fraction, after subtracting the background using the pseudo-proper time distribution in sideband regions around the \jpsi\ peak.

The systematic uncertainties in the data-to-simulation correction factors of both methods are estimated by varying the isolation criteria for the tag and the probe electron candidates, the fit models for the signal and background, the signal invariant-mass range, the pseudo-proper time requirement in the \textit{cut} method, and the fit range in the \textit{fit} method.

\section{Electron reconstruction}
\label{sec:reconstruction}
An electron can lose a significant amount of its energy due to bremsstrahlung when interacting with the material it traverses. The radiated photon may convert into an electron--positron pair which itself can interact with the detector material. These positrons, electrons, and photons  are usually emitted in a very collimated fashion and are normally reconstructed as part of the same electromagnetic cluster. These interactions can occur inside the inner-detector volume or even in the beam pipe, generating multiple tracks in the inner detector, or can instead occur downstream of the inner detector, only impacting the shower in the calorimeter. As a result, it is possible to produce and match multiple tracks to the same electromagnetic cluster, all originating from the same primary electron.
 
The reconstruction of electron candidates within the kinematic region encompassed by the high-granularity electromagnetic calorimeter and the inner detector is based on three fundamental components characterising the signature of electrons: localised clusters of energy deposits found within the electromagnetic calorimeter, charged-particle tracks identified in the inner detector, and close matching in $\eta \times \phi$ space of the tracks to the clusters to form the final electron candidates.  Therefore, electron reconstruction in the precision region of the ATLAS detector ($|\eta| < 2.47$) proceeds along those steps, described below in this order. Figure~\ref{fig:schematic_id_inputs} provides a schematic illustration of the elements that enter into the reconstruction and identification (see Section~\ref{sec:identification}) of an electron.

\begin{figure}[t]
\begin{center}
\includegraphics[width=0.98\textwidth]{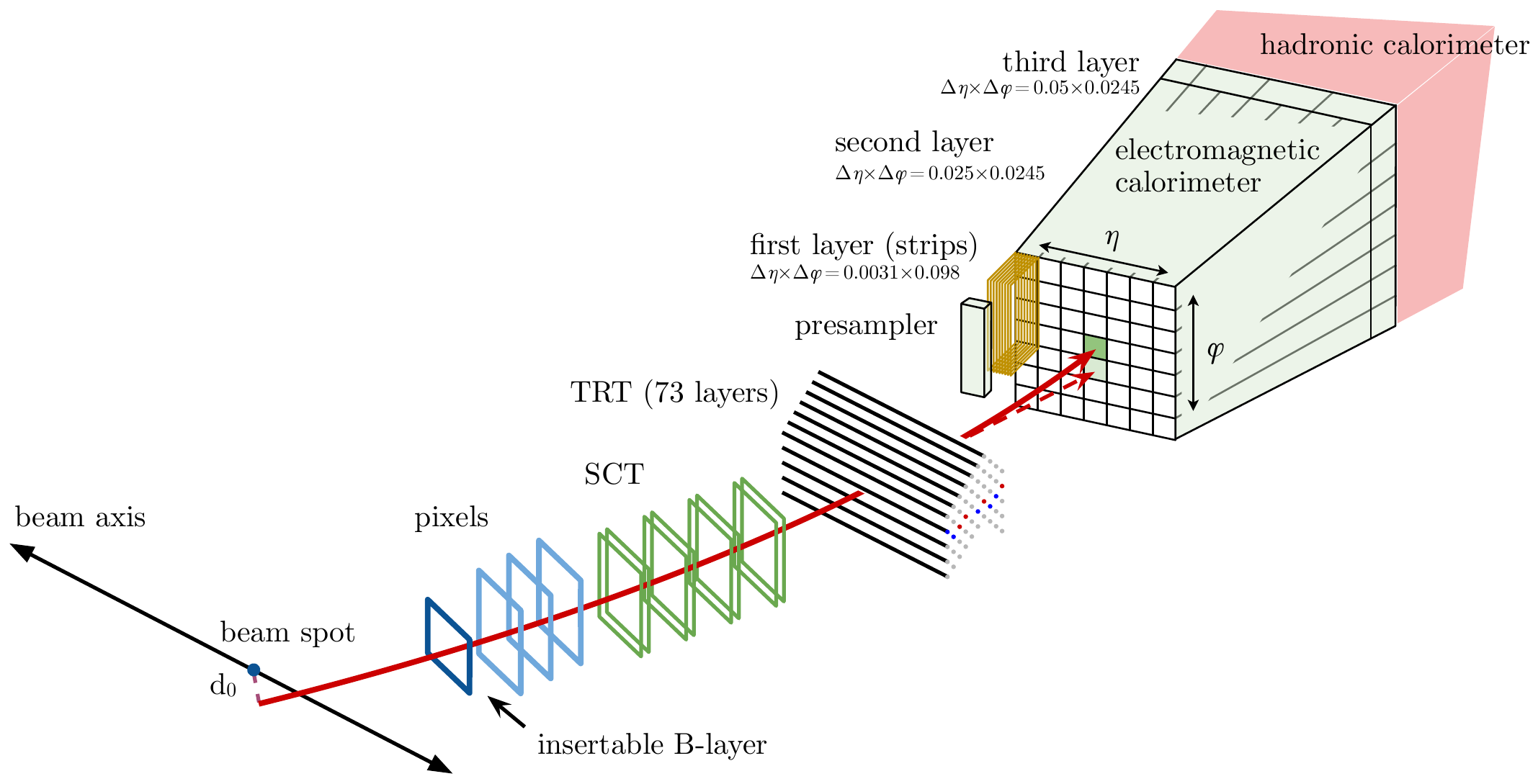}
\caption{
  A schematic illustration of the path of an electron through the detector.
The red trajectory shows the hypothetical path of an electron, which first traverses
the tracking system (pixel detectors, then silicon-strip detectors and lastly the TRT)
and then enters the electromagnetic calorimeter.
The dashed red trajectory indicates the path of a photon produced by the interaction of
the electron with the material in the tracking system. 
}
\label{fig:schematic_id_inputs}
\end{center}
\end{figure}

\subsection{Seed-cluster reconstruction} 
\label{sec:recoseed}

The $\eta \times \phi$ space of the EM calorimeter is divided into a grid of $200 \times 256$ elements (towers) of size $\Delta\eta \times \Delta\phi = 0.025 \times 0.025$, corresponding to the granularity of the second layer of the EM calorimeter. For each element, the energy (approximately calibrated at the EM scale),  collected in the first, second, and third calorimeter layers as well as in the presampler (only for $|\eta| <1.8$, the region where the presampler is located) is summed to form the energy of the tower. Electromagnetic-energy cluster candidates are then seeded from localised energy deposits using a sliding-window algorithm~\cite{ATL-LARG-PUB-2008-002} of size $3\times5$ towers in $\eta \times \phi$, whose summed transverse energy exceeds 2.5~\GeV. The centre of the $3\times5$ seed cluster moves in steps of $0.025$ in either the $\eta$ or $\phi$ direction, searching for localised energy deposits; the seed-cluster reconstruction process is repeated until this has been performed for every element in the calorimeter. If two seed-cluster candidates are found in close proximity (if their towers overlap within an area of $\Delta\eta \times \Delta\phi = 5 \times 9$ units of $0.025 \times 0.025$), the candidate with the higher transverse energy is retained, if its \ET\ is at least 10\% higher than the other candidate. If their \ET\ values are within 10\% of each other, the candidate containing the highest-\ET\ central tower is kept. The duplicate cluster is thereby removed.
The reconstruction efficiency of this seed-cluster algorithm (effectively \epsemclus in Eq.~(\ref{eq:EffDef})) depends on $|\eta|$ and \Et. As a function of \Et, it ranges from 65\% at $\et = 4.5$~\GeV, to 96\% at $\et = 7$~\GeV, to more than 99\% above $\et = 15$~\GeV, as can be seen in Figure~\ref{fig:reco:singleElectronRecoEff}. This efficiency is determined entirely from simulation. Efficiency losses due to seed-cluster reconstruction for $\eT>15$~\GeV\ are negligible compared with the uncertainties attributed to the next two steps of the reconstruction (track reconstruction and track--cluster matching).

\begin{figure}[t]
\centering
\includegraphics[width=0.65\textwidth]{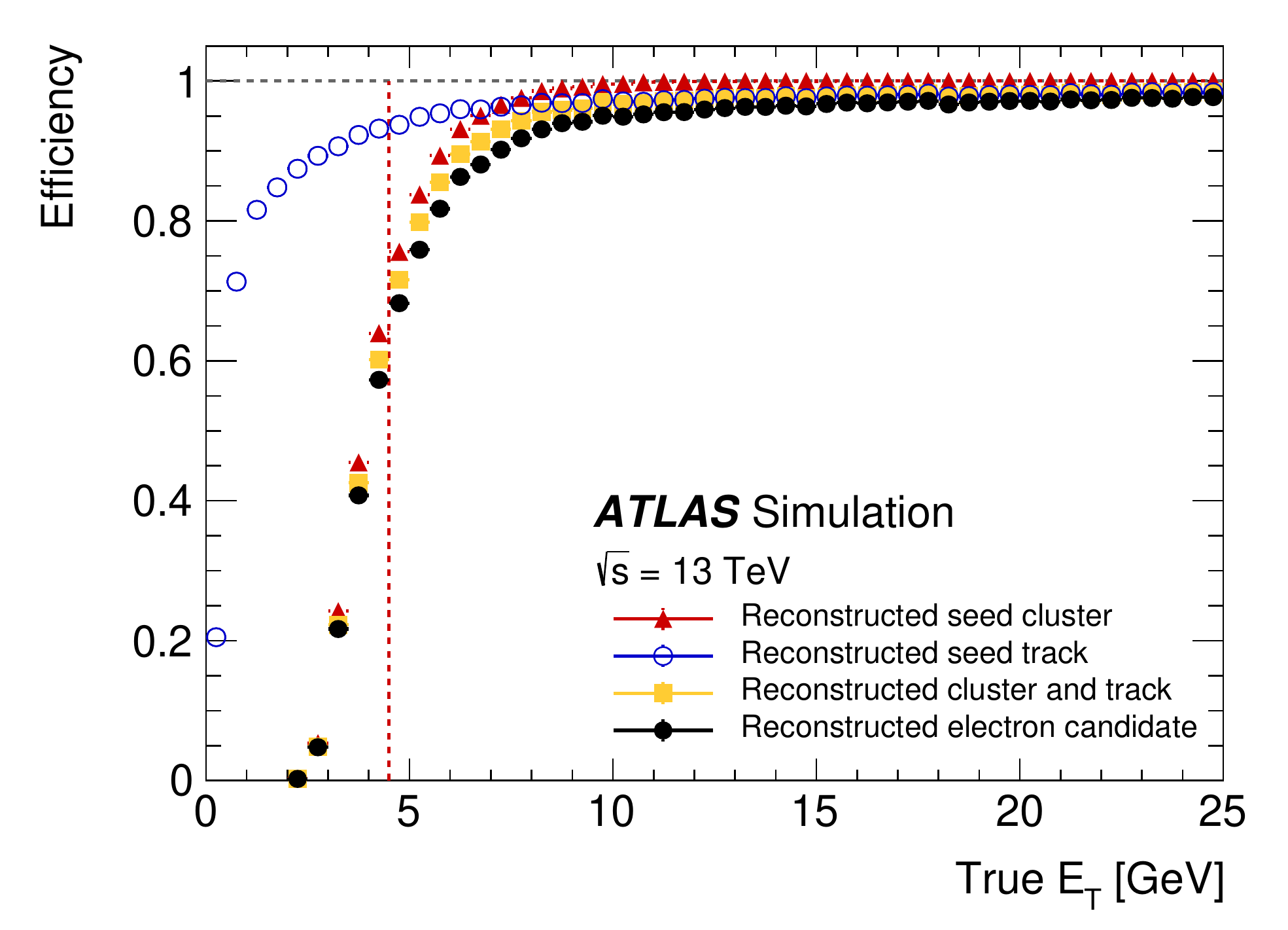}
\includegraphics[width=0.65\textwidth]{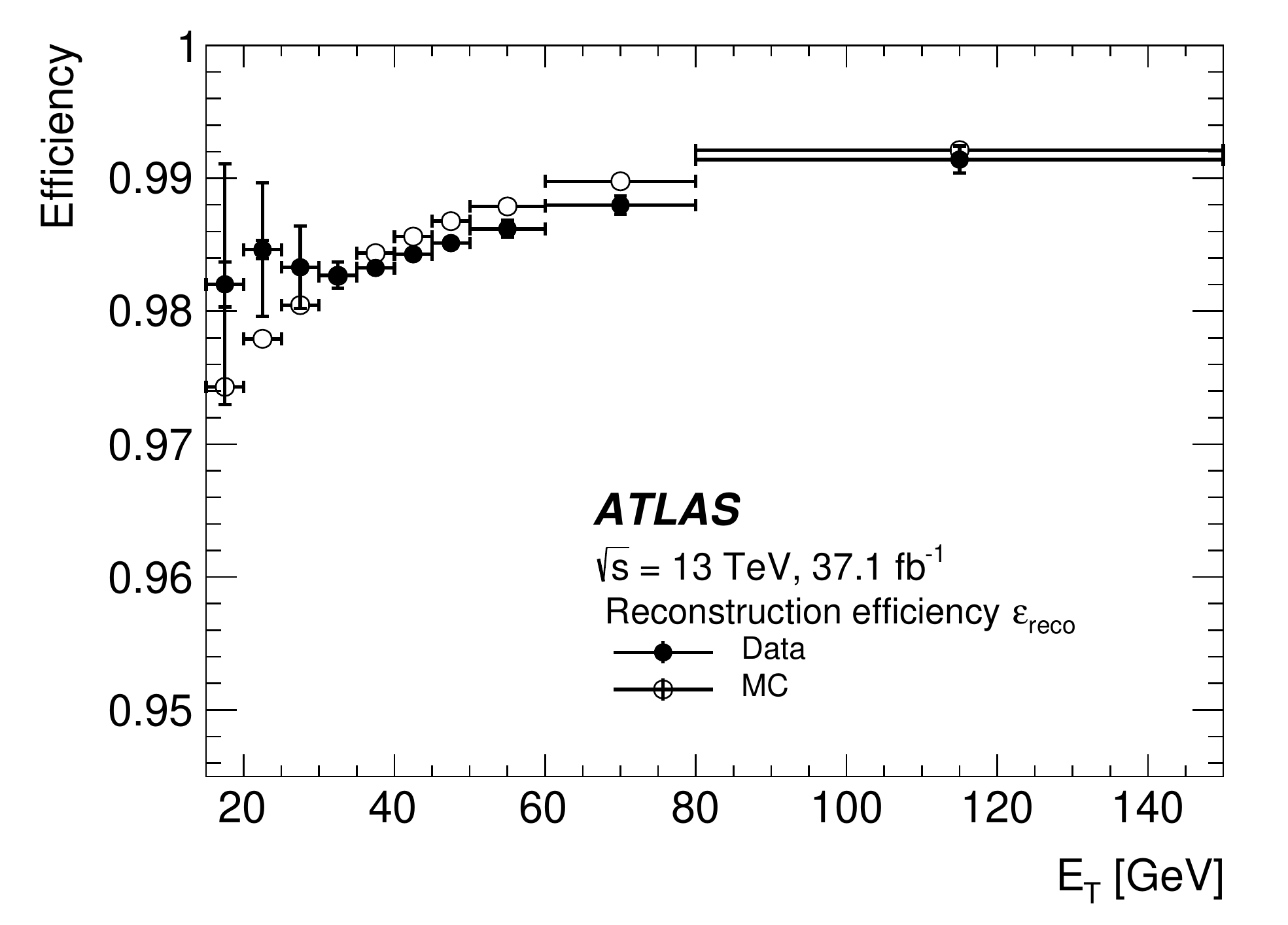}
\caption{
  Top: the total reconstruction efficiency for simulated electrons in a single-electron sample is shown as a function of the true (generator) transverse energy \eT\ for each step of the electron-candidate formation: $\Delta\eta \times \Delta\phi = 3\times5$ (in units of $0.025 \times 0.025$) seed-cluster reconstruction (red triangles), seed-track reconstruction using the Global $\chi^2$ Track Fitter (blue open circles), both of these steps together but instead using GSF tracking (yellow squares), and the final reconstructed electron candidate, which includes the track-to-cluster matching (black closed circles). 
As the cluster reconstruction requires uncalibrated cluster seeds  with $\et > 2.5~\GeV$, the total reconstruction efficiency is less than 60\% below $4.5~\GeV$ (dashed line). Bottom: the reconstruction efficiency relative to reconstructed clusters, \epsreco, as a function of electron transverse energy \ET\ for \Zee events, comparing data (closed circles) with simulation (open circles). The inner uncertainties are statistical while the total uncertainties include both the statistical and systematic components. 
}
\label{fig:reco:singleElectronRecoEff}
\end{figure}
\clearpage

\subsection{Track reconstruction} 
\label{sec:trackreco}

The basic building block for track reconstruction is a `hit' in one of the inner-detector tracking layers. Charged-particle reconstruction in the pixel and SCT detectors begins by assembling clusters from these hits~\cite{PERF-2015-08}.  From these clusters, three-dimensional measurements referred to as space-points are created. In the pixel detector, each cluster equates to one space-point, while in the SCT, clusters from both stereo views of a strip layer must be combined to obtain a three-dimensional measurement. Track seeds are formed from sets of three space-points in the silicon-detector layers.  The track reconstruction then proceeds in three steps: pattern recognition, ambiguity resolution,  and TRT extension (for more details of the TRT extension, see Ref.~\cite{Cornelissen:2008zzc}).  The pattern-recognition algorithm uses the pion hypothesis for the model of energy loss from interactions of the particle with the detector material. However, if a track seed with $\pT> 1~\GeV$ cannot be successfully extended to a full track of at least seven silicon hits per candidate track and
the EM cluster satisfies requirements on the shower width and depth,
a second attempt with modified pattern recognition, one which allows up to 30\% energy loss for bremsstrahlung at each intersection of the track with the detector material, is made. 
Track candidates with $\pt > 400~\MeV$ are fit, according to the hypothesis used in the pattern recognition, using the ATLAS Global $\chi^2$ Track Fitter~\cite{ATLASTrackFitter}. Any ambiguity resulting from track candidates sharing hits is resolved at the same stage. In order to avoid inefficiencies for electron tracks with significant bremsstrahlung, if the fit fails under the pion hypothesis and
its polar and azimuthal separation to the EM cluster is below a value,
a second fit is attempted under an electron hypothesis (an extra degree of freedom, in the form of an additional Gaussian term, is added to the $\chi^2$ to compensate for the additional bremsstrahlung losses coming from electrons;  such an energy-loss term is neglected in the pion-hypothesis fit). 
Figure~\ref{fig:reco:singleElectronRecoEff}~(top) shows that the reconstruction efficiency of the track-fitting step ranges from 80\% at $\eT = 1$~\GeV\ to more than 98\% above $\eT = 10$~\GeV.

A subsequent fitting procedure, using an optimised Gaussian-sum filter (GSF)~\cite{ATLAS-CONF-2012-047} designed to better account for energy loss of charged particles in material, is applied to the clusters of raw measurements.
This procedure is used for tracks which have at least four silicon hits and that are loosely matched to EM clusters. The separation of the cluster-barycentre position 
and the position of the track extrapolated from the perigee to the second layer of the calorimeter must satisfy $|\eta_\text{cluster}~-~\eta_\text{track}|~<~0.05$ and one of two alternative requirements on the azimuthal separation between the cluster position and the track: $-0.20~<\Delta\phi<~0.05$ or $-0.10~< \deltaphires <~0.05$, where $q$ is the sign of the electric charge of the particle, and $\Delta\phi$ and \deltaphires are calculated as $-q \times(\phi_\text{cluster}~-~\phi_\text{track})$ with the momentum of the track rescaled to the energy of the cluster for $\deltaphires$. 
The asymmetric condition for the matching in $\phi$ mitigates the effects of energy loss due to bremsstrahlung where tracks with negative (positive) electric charge bend due to the magnetic field in the positive (negative) $\phi$ direction. 

The GSF method~\cite{FRUHWIRTH2003131} is based on a generalisation of the Kalman filter~\cite{Fruhwirth:1987fm} and takes into account the non-linear effects related to bremsstrahlung. Within the GSF, experimental noise is modelled by a sum of Gaussian functions. The GSF therefore consists of a number of Kalman filters running in parallel, the result of which is that each track parameter is approximated by a weighted sum of Gaussian functions. 
Six Gaussian functions are used to describe the material-induced energy losses and up to twelve to describe the track parameters. In the final step, the mode of the energy distribution is used to represent the energy loss.

Radiative losses of energy lead to a decrease in momentum, resulting in increased curvature of the electron's trajectory in the magnetic field. When accounting for such losses via the GSF method, all track parameters relevant to the bending-plane are expected to improve. Such a parameter is the transverse impact parameter significance: $\trackdO$ divided by its estimated uncertainty $\dOSig$. Since the curvature, in the ATLAS coordinate frame, is positive for negative particles and negative for positive particles, the signed impact parameter significance (i.e.\ multiplied by the sign of the reconstructed electric charge $q$ of the electron) is used. 
Figure~\ref{fig:reco:d0sgnf} shows $q\times \dOSignificance$  for the track associated with the electron, i.e.\ the primary electron track.
A clear improvement in  $q\times \dOSignificance$  for genuine electron tracks fitted with the GSF over tracks with the  ATLAS Global $\chi^2$ Track Fitter is observed; the distribution is narrower and better centred at zero. Figure~\ref{fig:reco:d0sgnf} also shows, for  the ratio of the electron-candidate charge to its momentum $q/p$, the relative difference between the true generator value and the reconstructed value; the GSF method shows a sharper and better-centred distribution near zero with smaller tails.
The reconstruction efficiency for finding both a seed cluster and a GSF track is shown in Figure~\ref{fig:reco:singleElectronRecoEff}~(top). 

\begin{figure}[t]
\centering
\includegraphics[width=0.65\linewidth]{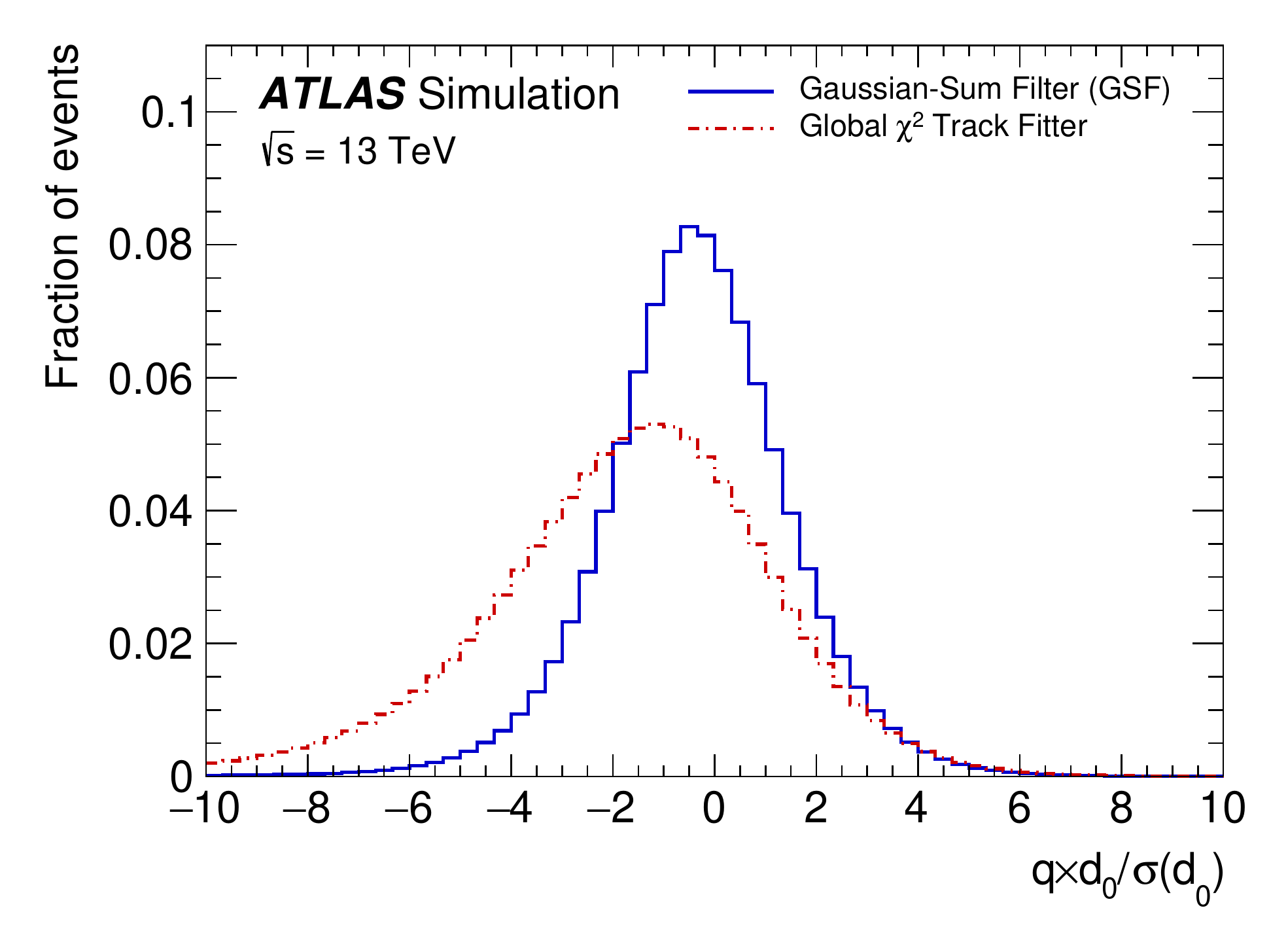}
\includegraphics[width=0.65\linewidth]{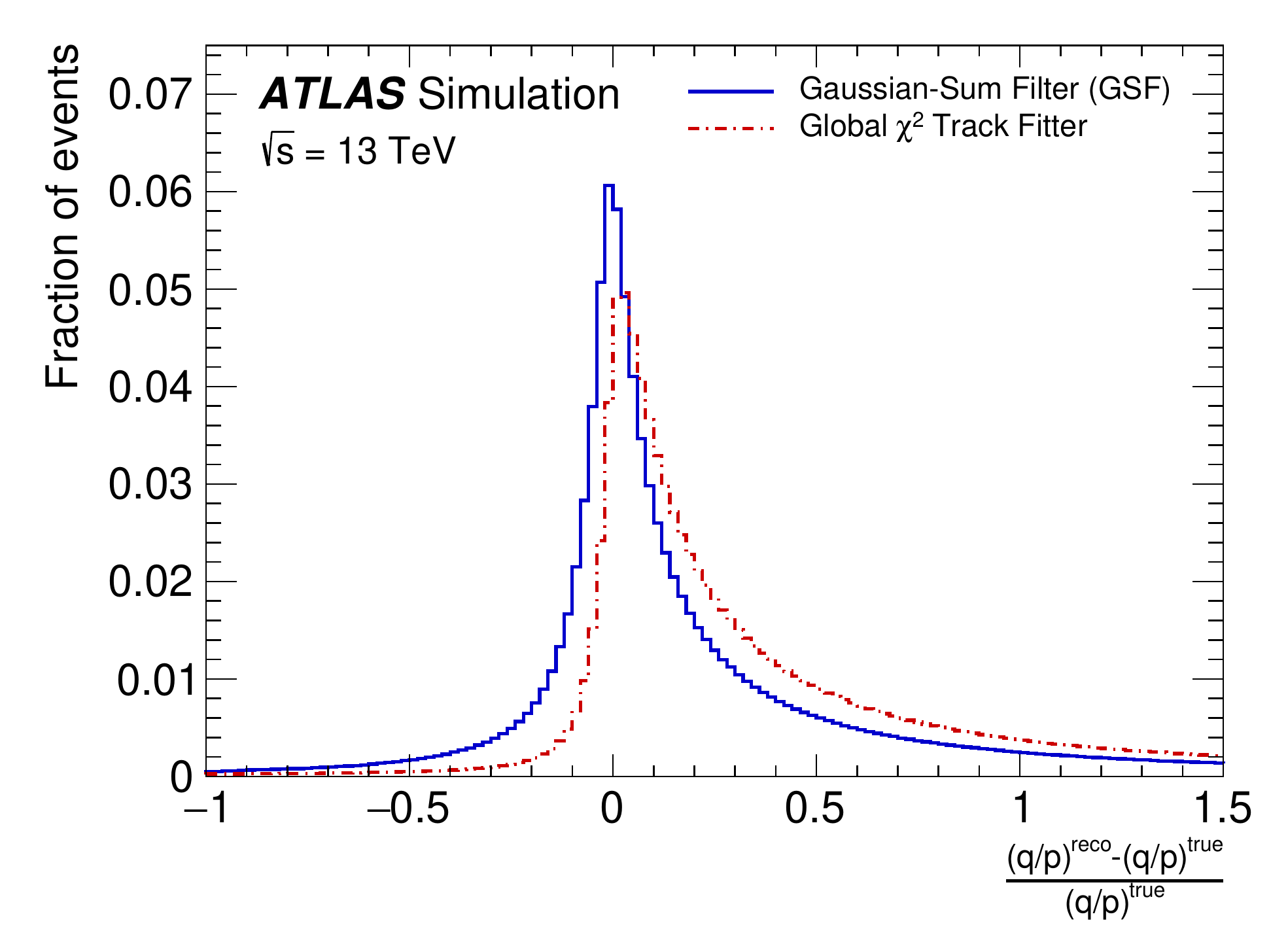}
\caption{Distributions of the reconstructed electric charge of the candidate electron multiplied by the transverse impact parameter significance, $q\times \dOSignificance$ (top) and the relative difference between the reconstructed value of the candidate-electron charge divided by its momentum, $q/p$, and the true generator value (bottom). 
The distributions are shown for tracks fitted with the Global $\chi^2$ Track Fitter (dashed red lines) and for tracks fitted with the GSF (solid blue line). 
The distributions were obtained from a simulated single-electron sample.
\label{fig:reco:d0sgnf}}
\end{figure}

\subsection{Electron-candidate reconstruction} 
\label{reco:candreco}
The matching of the GSF-track candidate to the candidate calorimeter seed  cluster and the determination of the final cluster size complete the electron-reconstruction procedure. This matching procedure is similar to the loose matching discussed above prior to the GSF step, but with stricter requirements; the track-matching in $\phi$ is tightened to $-0.10~<\Delta\phi<~0.05$, keeping the original alternative requirement $-0.10~<\deltaphires<~0.05$ the same. If several tracks fulfil the matching criteria, the track considered to be the primary electron track is selected using an algorithm that takes into account the distance in \eta\ and \phi\ between the extrapolated tracks and the cluster barycentres measured in the second layer of the calorimeter, the number of hits in the silicon detectors, and the number of hits in the innermost silicon layer; a candidate with an associated track with at least four hits in the silicon layers and no association with a vertex from a photon conversion~\cite{PERF-2017-02} is considered as an electron candidate. 
However, if the primary candidate track can be matched to a secondary vertex and has no pixel hits, then this object is classified as a photon candidate (likely a conversion).  A further classification is performed using the candidate electron's $\eoverp$ and \pt, the presence of a pixel hit, and the secondary-vertex information, to determine unambiguously whether the object is only to be considered as an electron candidate or if it should be ambiguously classified as potentially either  a photon candidate or an electron candidate. However, this classification scheme is mainly for the benefit of keeping a high photon-reconstruction efficiency. Since all electron identification operating points described in Section~\ref{sec:identification} require a track with a hit in the innermost silicon layer (or in the next-to-innermost layer if the innermost layer is non-operational), most candidates fall into the `unambiguous' category  after applying an identification criterion.

Finally, reconstructed clusters are formed around the seed clusters using an extended window of size $3\times7$ in the barrel region ($|\eta|<1.37$) or $5\times5$ in the endcap ($1.52<|\eta|<2.47$) by simply expanding the cluster size in $\phi$ or $\eta$, respectively, on either side of the original seed cluster. A method using both elements of the extended-window size is used in the transition region of $1.37<|\eta|<1.52$. The energy of the clusters must ultimately be calibrated to correspond to the original electron energy. This detailed calibration is performed using multivariate techniques~\cite{PERF-2013-05,PERF-2017-03} based on data and simulated samples, and only after the step of selecting electron candidates rather than during the reconstruction step, which relies on approximate EM-scale energy clusters. The energy of the final electron candidate is computed from the calibrated energy of the extended-window cluster while the $\phi$ and $\eta$ directions are taken from the corresponding track parameters, measured relative to the beam spot, of the track best matched to the original seed cluster.

\begin{figure}[t]
\centering
\includegraphics[width=0.48\textwidth]{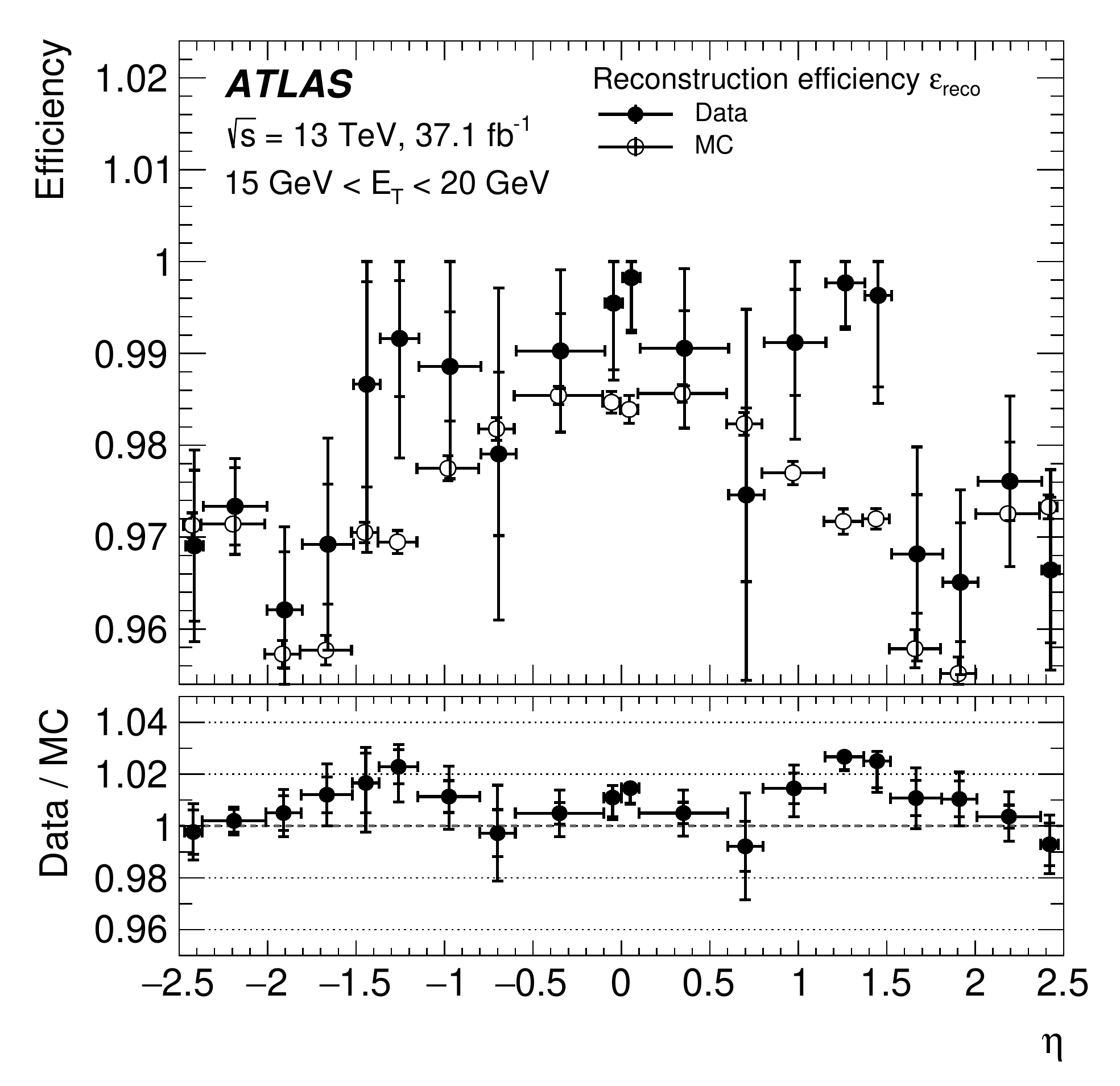}
\includegraphics[width=0.48\textwidth]{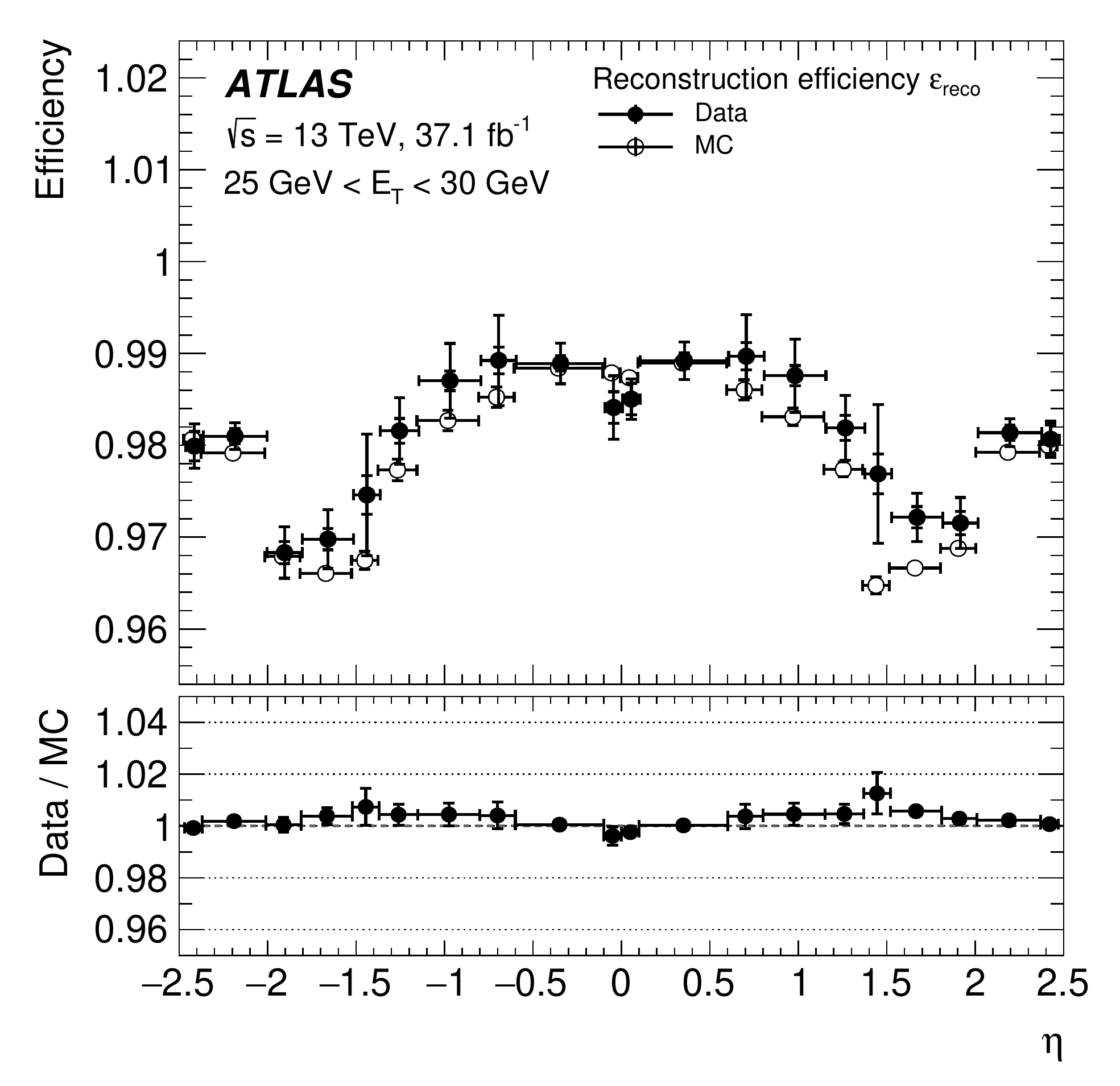}
\includegraphics[width=0.48\textwidth]{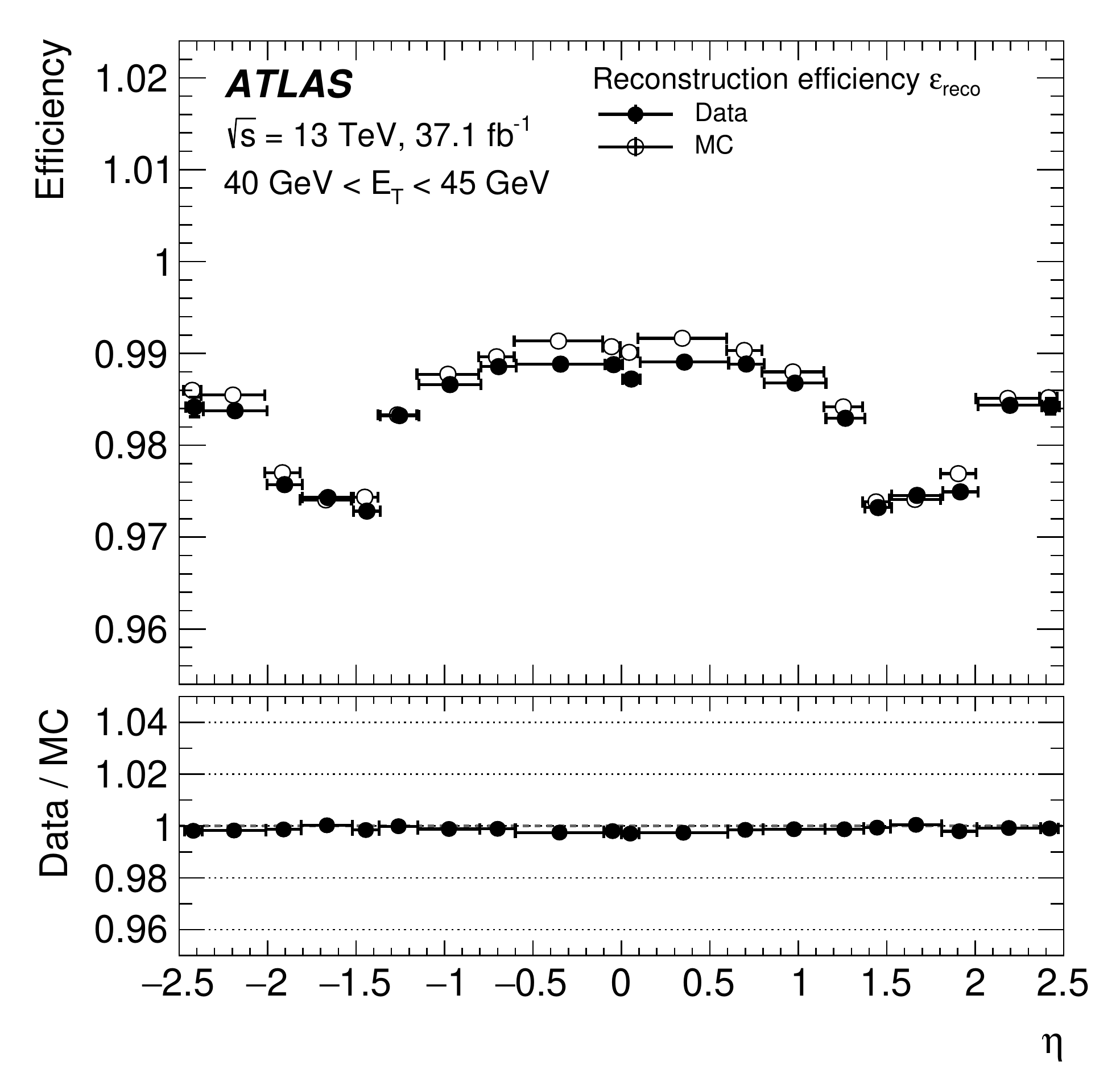}
\includegraphics[width=0.48\textwidth]{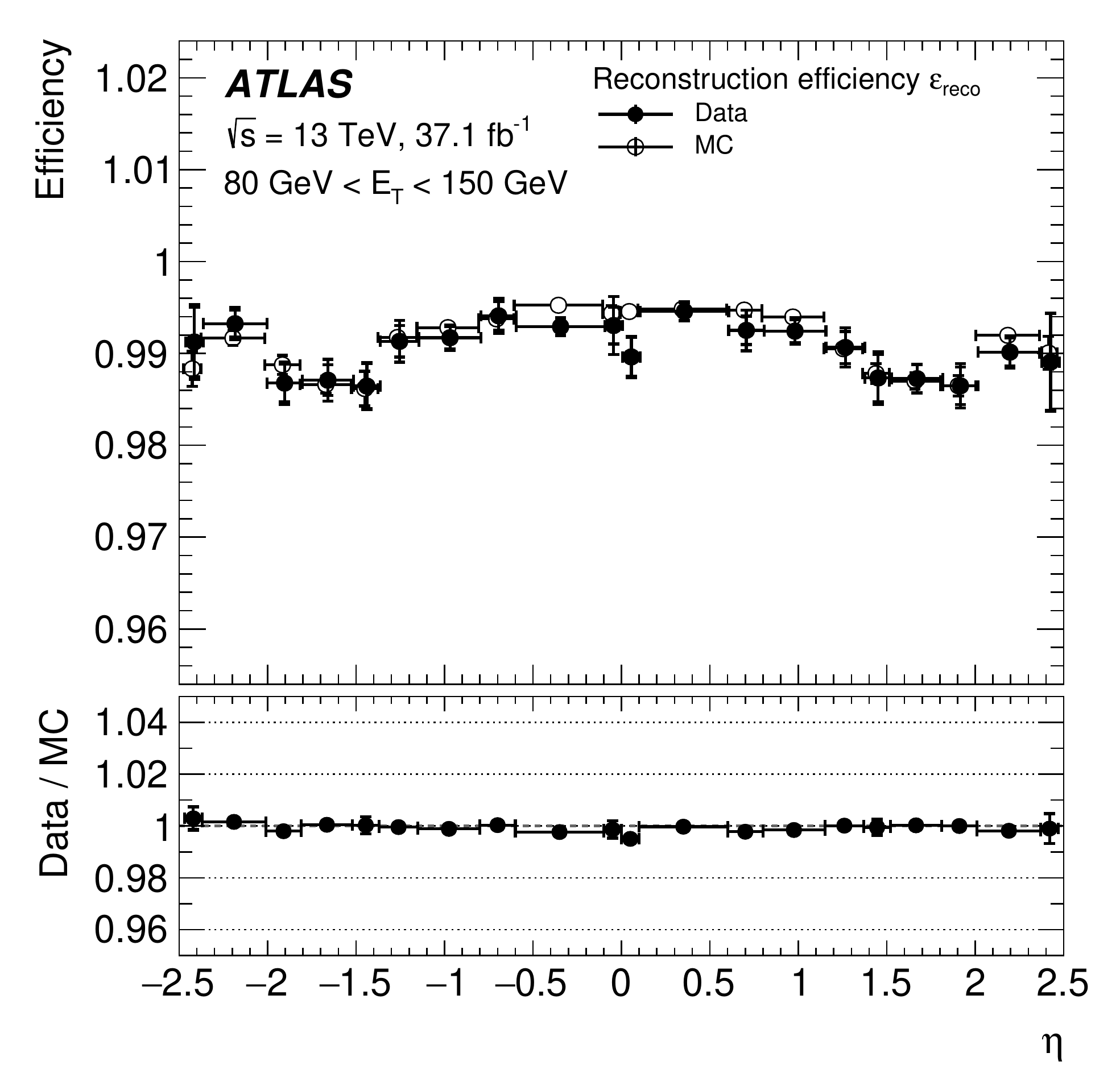}
\caption{ Reconstruction efficiencies relative to reconstructed clusters, \epsreco, evaluated in the 2015--2016 dataset (closed points) and in simulation (open points), and their ratio, using the $Z\rightarrow ee$ process,
as a function of $\eta$ in four illustrative \et\ bins: 15--20~\GeV\ (top left), 25--30~\GeV\ (top right),
40--45~\GeV\ (bottom left), and 80--150~\GeV\ (bottom right). 
The inner uncertainties are statistical while the total uncertainties include both the statistical and systematic components.
}
\label{fig:main_ZeeReco_Reco_vs_eta_ptbins_2016}
\end{figure}

Above $\et=15$~\GeV, the efficiency to reconstruct an electron having a track of good quality (at least one pixel hit and at least seven silicon hits)  varies from approximately 97\% to 99\%. The simulation has lower efficiency than data in the low \et region (\et < 30~\GeV) while the opposite is true for the higher \et region (\et > 30~\GeV), as demonstrated in Figures~\ref{fig:reco:singleElectronRecoEff} and~\ref{fig:main_ZeeReco_Reco_vs_eta_ptbins_2016}, which show the reconstruction efficiency as a function of \ET\ and as a function of $\eta$ in bins of \ET, respectively, from \Zee\ events. All measurements are binned in two dimensions. The uncertainty in the efficiency in data is typically 1\% in the $\et = 15$--20~\GeV\ bin and reaches the per-mille level at higher \et\ and the uncertainty in simulation is almost an order of magnitude smaller than for data. The systematic uncertainty dominates at low \et\ for data, with the estimation of background from clusters with no associated track giving the largest contribution. Below $\et=15$~\GeV, the reconstruction efficiency is determined solely from the simulation; a 2\% (5\%) uncertainty is assigned in the barrel (endcap) region.

\clearpage

\section{Electron identification}
\label{sec:identification}
Prompt electrons entering the central region of the detector ($|\eta|<2.47$) are selected using a
likelihood-based (LH) identification.
The inputs to the LH include measurements from the tracking system,
the calorimeter system, and quantities that combine both tracking and calorimeter information.
The various inputs are described in Table~\ref{tab:IDcuts}
and the components of the quantities described in this table are illustrated
schematically in Figure~\ref{fig:schematic_id_inputs}.
The LH identification is very similar in method to the electron
LH identification used in Run\,1 (2010--2012)~\cite{PERF-2016-01}, but
there are some important differences. To prepare for the start of data-taking with a higher center-of-mass energy and different detector conditions it was necessary to construct probability density functions (pdfs) based on simulated events rather than data events, and correct the resulting distributions for any mismodelling. Furthermore, the efficiency was smoothed as a function of \Et\ and the likelihood was adjusted to allow its use for electrons with $\Et > 300$~\GeV.

\begin{table*}
\caption{Type and description of the quantities used in the electron identification.
The columns labelled ``Rejects'' indicate whether a quantity has significant discrimination power
between prompt electrons and light-flavour (LF) jets, photon conversions ($\gamma$), or
non-prompt electrons from the semileptonic decay of hadrons containing heavy-flavour (HF) quarks ($b$- or $c$-quarks).
In the column labelled ``Usage,'' an ``LH'' indicates that the pdf of this quantity
is used in forming $L_{S}$ and $L_{B}$ (defined in Eq.~(\ref{eq:likelihoods})) and a
``C'' indicates that this quantity is used directly as a selection criterion. 
In the description of the quantities formed using the second layer of the calorimeter, 3$\times$3, 3$\times$5, 3$\times$7, and 7$\times$7 refer to areas of $\Delta\eta \times \Delta\phi$ space in units of $0.025 \times 0.025$.
}
\label{tab:IDcuts}
\scriptsize
\renewcommand{\arraystretch}{1.13}
\begin{center}
\begin{tabular}{|l|l|l|c|c|c|l|}
\hline
Type & Description & Name &  \multicolumn{3}{c|}{Rejects} & Usage  \\
 & & & LF & $\gamma$ & HF &\\
\hline
 Hadronic & Ratio of \et in the first layer of the hadronic calorimeter  & \rhadone & x & x &  & LH \\ 
 leakage & to \et of the EM cluster & & & & & \\
& (used over the range $|\eta| < 0.8$ or $|\eta| > 1.37$)  & & & & & \\
\cline{2-7}
  & Ratio of \et in the hadronic calorimeter &  & & & & \\
  &  to \et of the EM cluster & \rhad & x & x &  & LH  \\
 & (used over the range $0.8 <|\eta| < 1.37$) & & & & & \\
\hline
Third layer of  & Ratio of the energy in the third layer to the total energy in the & & & & &\\
EM calorimeter  & EM calorimeter. This variable is only used for & & & & & \\
                & $\et < 80~\GeV$, due to inefficiencies at high \et, and is& \fIII & x & & & LH \\
                & also removed from the LH for $|\eta| > 2.37$, where it is& & & & & \\
                & poorly modelled by the simulation. & & & & & \\
\hline
Second layer of  & Lateral shower width, $\sqrt{(\Sigma E_i \eta_i^2)/(\Sigma E_i) -((\Sigma E_i\eta_i)/(\Sigma E_i))^2}$, & & & & & \\
EM calorimeter  & where $E_i$ is the energy and $\eta_i$ is the pseudorapidity  & \weta & x & x & & LH \\
 & of cell $i$ and the sum is calculated within a window of 3$\times$5 cells & & & & & \\
\cline{2-7}
& Ratio of the energy in 3$\times$3 cells over the energy in 3$\times$7 cells & \rphi & x & x &  & LH  \\
& centred at the electron cluster position & & & & & \\
\cline{2-7}
& Ratio of the energy in 3$\times$7 cells over the energy in 7$\times$7 cells  & \reta & x & x & x & LH  \\
& centred at the electron cluster position & & & & & \\
\hline
First layer of       & Shower width, $\sqrt{(\Sigma E_i (i-i_\mathrm{max})^2)/(\Sigma E_i)}$, where $i$ runs over  &   & & & & \\  
EM calorimeter       & all strips in a window of $\Delta\eta \times \Delta\phi \approx 0.0625 \times 0.2$,   & \wstot & x & x & x & C \\
		     & corresponding typically to 20 strips in $\eta$, and $i_\mathrm{max}$ is the & & & & &                  \\
		     & index of the highest-energy strip, used for \et\ $>$ 150~\gev\ only        &  & & & &   \\
\cline{2-7}
                     & Ratio of the energy difference between the maximum &    & & & &  \\
                     & energy deposit and the energy deposit in a secondary & \deltaEmax & x & x & & LH  \\
                     & maximum in the cluster to the sum of these energies & & & & &   \\
\cline{2-7}     
& Ratio of the energy in the first layer to the total energy  & \fI & x & & & LH  \\
& in the EM calorimeter &  & & & & \\
\hline
Track  & Number of hits in the innermost pixel layer &   $n_\mathrm{Blayer}$ & & x & & C \\
\cline{2-7}
conditions                     & Number of hits in the pixel detector        &    $n_\mathrm{Pixel}$ & & x & & C \\
\cline{2-7}
                     & Total number of hits in the pixel and SCT detectors  &   $n_{\mathrm{Si}}$  & & x & & C \\
\cline{2-7}
                     & Transverse impact parameter relative to the beam-line
		                                                  &       \trackdO  & & x & x & LH \\
\cline{2-7}
                     & Significance of transverse impact parameter &       |\dOSignificance|  & & x & x & LH  \\
                     & defined as the ratio of \trackdO to its uncertainty                     &  & & & &              \\
\cline{2-7}
                     &  Momentum lost by the track between the perigee and the last &   \deltapoverp & x & & & LH \\
                     & measurement point divided by the  momentum at perigee & & & & & \\
\hline
TRT                       & Likelihood probability based on transition radiation in the TRT &   \TRTPID & x & & & LH  \\
\hline
Track--cluster     & $\Delta\eta$ between the cluster position in the first layer &   \deltaeta & x & x & & LH  \\
matching          &  and the extrapolated track & & & & &   \\
\cline{2-7}
&   $\Delta\phi$ between the cluster position in the second layer &   & & & &  \\
&   of the EM calorimeter and the momentum-rescaled  & \deltaphires & x & x & & LH  \\
&   track, extrapolated from the perigee, times the charge $q$  & & & & &  \\
\cline{2-7}
                    & Ratio of the cluster energy to the track momentum, used for            &       $E/p$   & x & x & & C\\
                    & \et $>$ 150~\gev\ only & & & & &  \\
\hline
\end{tabular}
\end{center}
\end{table*}

\subsection{The likelihood identification}

The electron LH is based on the products for signal, $L_{S}$, and for background,
$L_{B}$, of $n$ pdfs, $P$:
\begin{equation}
\label{eq:likelihoods}
L_{S(B)}(\mathbf{x}) = {\displaystyle \prod_{i=1}^{n}P_{S(B),i}(x_{i})},
\end{equation}
where $\mathbf{x}$ is the vector of the various quantities specified in Table~\ref{tab:IDcuts}.
$P_{S,i}(x_{i})$ is the value of the signal pdf for quantity $i$ at value $x_i$ and
$P_{B,i}(x_{i})$ is the corresponding value of the background pdf.
The signal is prompt electrons, while the background is the combination of
jets that mimic the signature of prompt electrons,
electrons from photon conversions in the detector material, and
non-prompt electrons from the decay of hadrons containing heavy flavours.
Correlations in the quantities selected for the LH are neglected.

For each electron candidate, a discriminant $d_{L}$ is formed: 
\begin{equation}
  \label {eq:discriminant}
d_{L} = \frac{L_{S}}{L_{S} + L_{B}};
\end{equation}
the electron LH identification is based on this discriminant.
The discriminant $d_L$ has a sharp peak at unity (zero) for signal (background);
this sharp peak makes it inconvenient to select operating points as it would require extremely fine binning.
An inverse sigmoid function is used to transform the distribution of the
discriminant of Eq.~(\ref{eq:discriminant}):
\begin{equation*}
  \label{eq:inverse_sigmoid_transformation}
  d^\prime_{L} = -\tau^{-1}\ln(d_{L}^{-1} - 1),
\end{equation*}
where the parameter $\tau$ is fixed to 15~\cite{TMVA}.
As a consequence, the range of values of the transformed discriminant no longer varies between
zero and unity.
For each operating point, a value of the transformed discriminant is chosen:
electron candidates with values of $d^\prime_{L}$ larger than this value
are considered signal.
An example of the distribution of a transformed discriminant is shown in
Figure~\ref{fig:discriminant} for prompt
electrons from $Z$-boson decays and for background. This distribution illustrates the effective
separation between signal and background encapsulated in this single quantity.

\begin{figure}[t]
\begin{center}
\includegraphics[width=.65\textwidth]{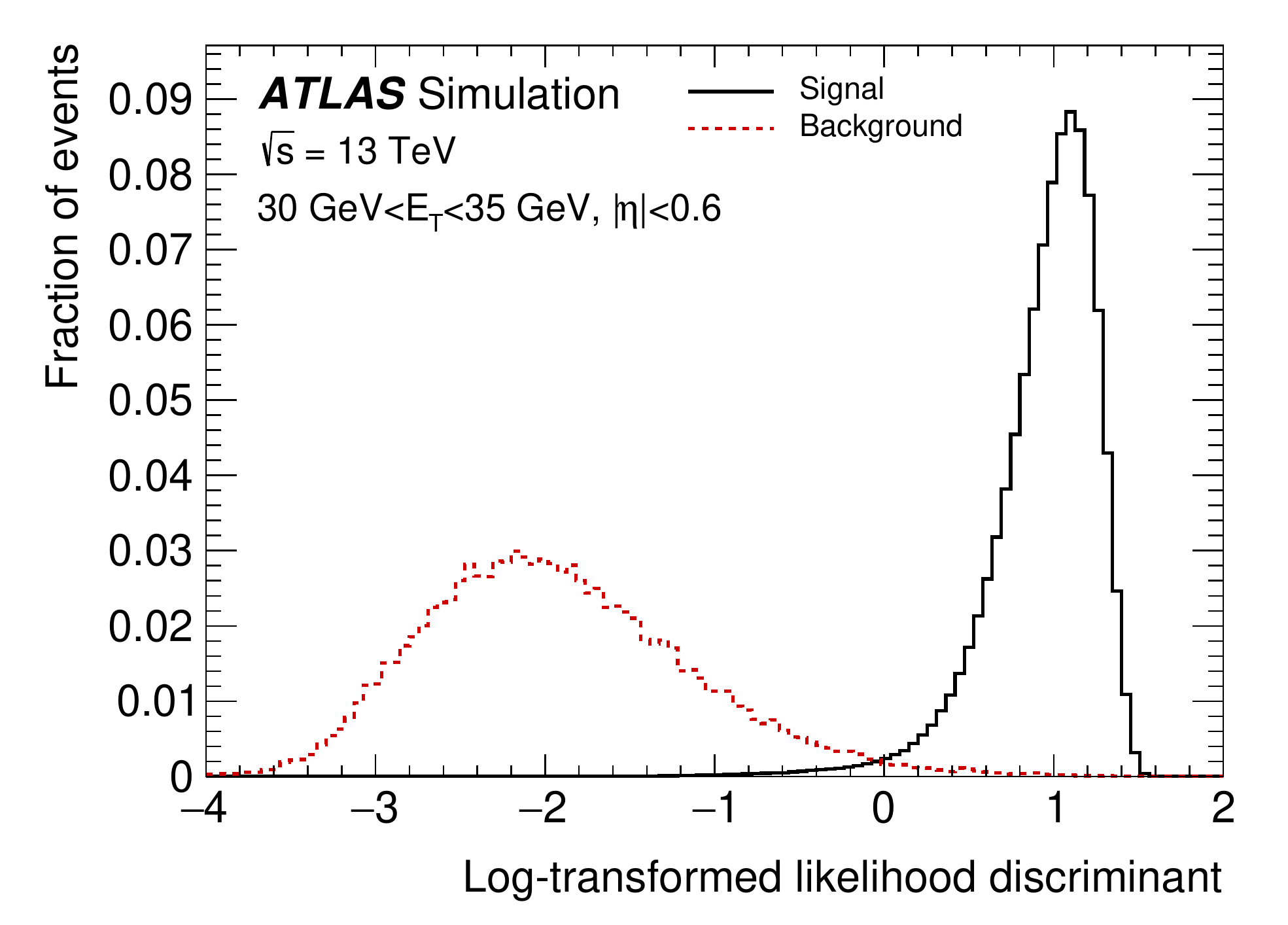}
\caption{
  The transformed LH-based identification discriminant $d^\prime_L$ for reconstructed electron candidates with good quality tracks with
  $30~\GeV < \Et < 35~\GeV$ and \hbox{$|\eta|<0.6$}.
  The black histogram is for prompt electrons in a \Zee simulation sample,
  and the red (dashed-line) histogram is for backgrounds in a generic two-to-two process simulation sample
 (both simulation samples are described in Section~\ref{sec:samples}).
  The  histograms are normalised to unit area.
  }
\label{fig:discriminant}
\end{center}
\end{figure}

There are two advantages to using a LH-based electron identification over a selection-criteria-based (so-called ``cut-based'') identification.
First, a prompt electron may fail the cut-based identification
because it does not satisfy the selection criterion for a single quantity.
In the LH-based selection, this electron can still satisfy the identification
criteria, because the LH combines the information of all of the discriminating
quantities.
Second, discriminating quantities that have distributions too similar
to be used in a cut-based identification without suffering
large losses in efficiency may be added to the LH-based identification without
penalty.
Two examples of quantities that are used in the LH-based identification,
but not in cut-based identifications, are
\rphi and \fI, which are defined in Table~\ref{tab:IDcuts}.
Figure~\ref{fig:rphi_f1_pdfs} compares the distributions of these two
quantities for prompt electrons and background.

\begin{figure}[t]
\begin{center}
  \includegraphics[width=0.65\textwidth]{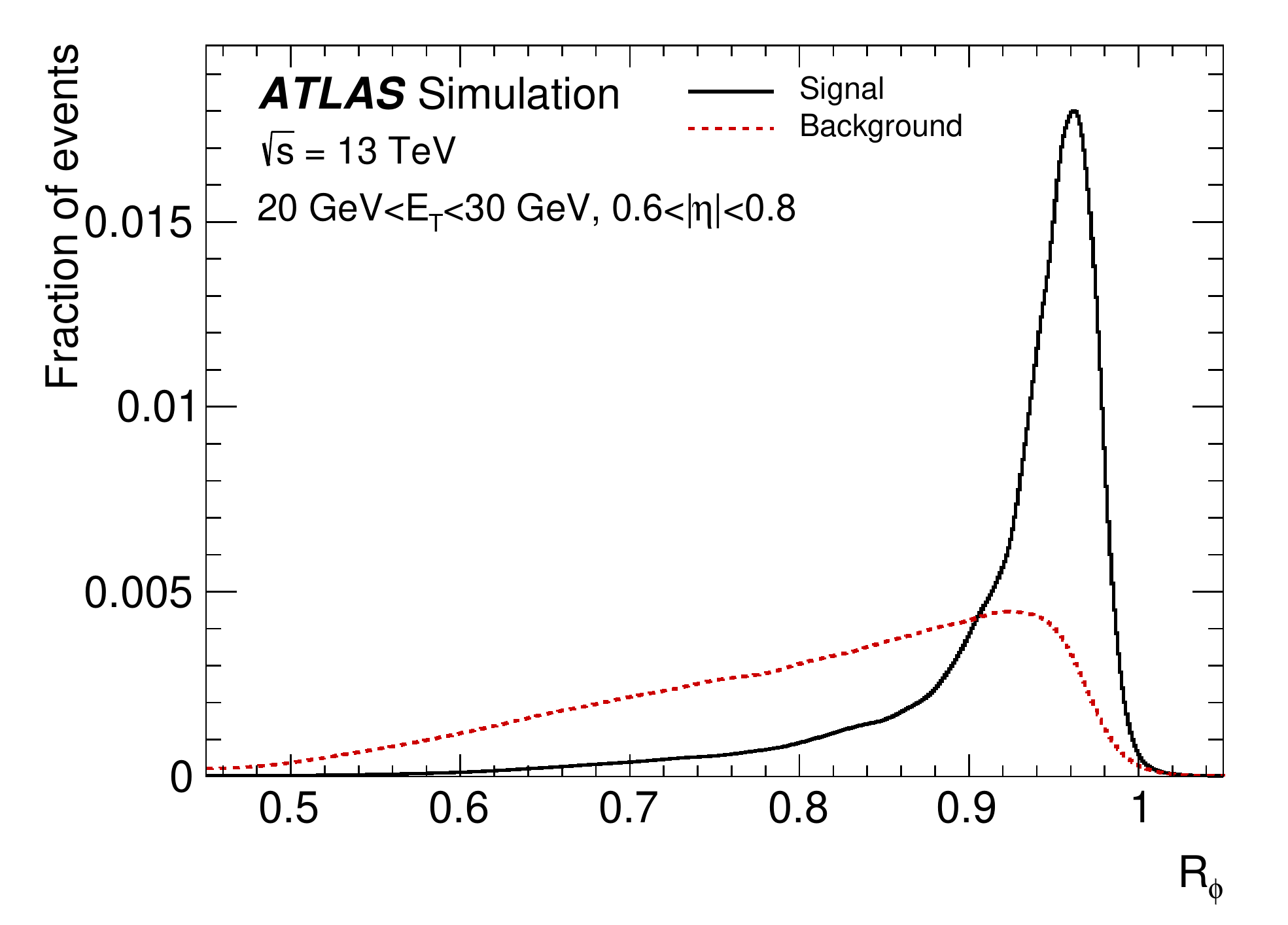}
  \includegraphics[width=0.65\textwidth]{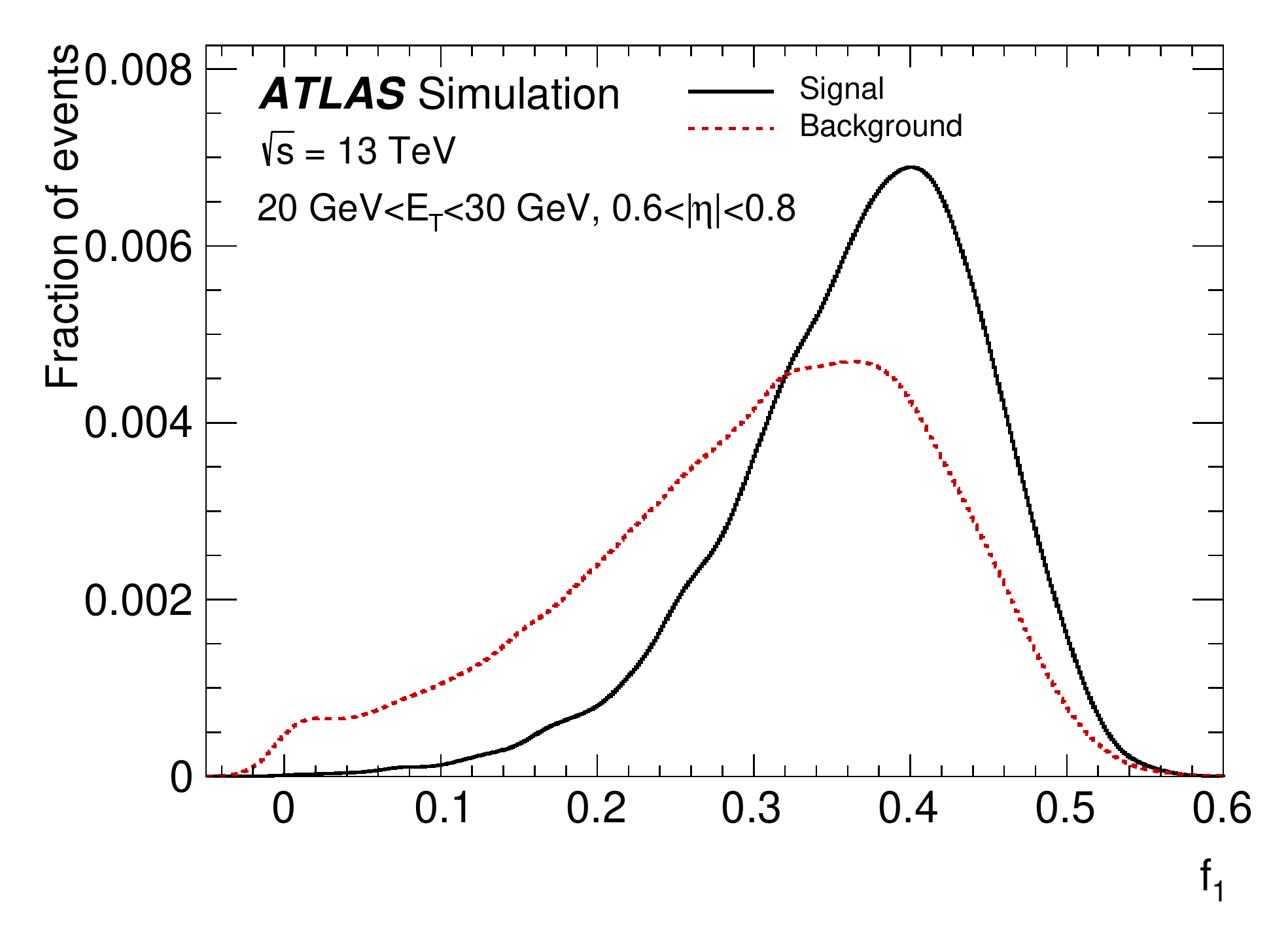}
  \caption{Examples of distributions of two quantities \rphi (top) and \fI (bottom),
    both defined in Table~\ref{tab:IDcuts} and shown for $20~\GeV < \et < 30~\GeV$ and $0.6<|\eta|<0.8$, that would be inefficient if used in a
    cut-based identification, but which, nonetheless, have
    significant discriminating power against background and, therefore,
    can be used to improve a LH-based identification.
    In each figure, the red-dashed distribution is determined from a background simulation sample
    and the black-line distribution is determined from a \Zee simulation sample.
    These distributions are for reconstructed electron candidates before applying any identification. They are smoothed using an adaptive KDE and have been corrected for offsets or differences in widths between the distributions in data and simulation as described in Section~\ref{sec:LHpdfs}.
}
\label{fig:rphi_f1_pdfs}
\end{center}
\end{figure}

\subsection{The pdfs for the LH-identification}
\label{sec:LHpdfs}

The pdfs for the electron LH are derived from the simulation samples
described in Section~\ref{sec:samples}.
As described below, distinct pdfs are determined for each identification
quantity in separate bins of electron-candidate \Et and \eta.
The pdfs are created from finely binned histograms of the individual 
identification quantities.
To avoid non-physical fluctuations in the pdfs arising from
the limited size of the simulation samples,
the histograms are smoothed using an adaptive kernel density
estimation (KDE) implemented in the TMVA toolkit~\cite{TMVA}.

Imperfect detector modelling causes differences between the simulation quantities
used to form the LH-identification and the corresponding quantities in data.
Some simulation quantities are corrected to account for these differences
so that the simulation models the data more accurately and hence the
determination of the LH-identification operating points is made using
a simulation that reproduces the data as closely as possible.
These corrections are determined using simulation and data obtained with the
\Zee\ \tnp method.

The differences between the data and the simulation
typically appear as either a constant offset between the
quantities (i.e., a shift of the distributions) or a difference in the width, quantified here as
the full-width at half-maximum (FWHM) of the distribution of the quantity. In some cases, both shift and width corrections are applied. 
The quantities \fI, \fIII, \reta, \weta and \rphi have \eta-dependent offsets,
and the quantities \fI, \fIII, \rhad, \deltaeta and \deltaphires have differences in
FWHM.

In the case that the difference is a shift, the value in the simulation is
shifted by a fixed (\eta-dependent) amount to make the distribution in
the simulation agree better with the distribution in the data.
In the case of a difference in FWHM, the value in the simulation is scaled by a multiplicative factor.
The optimal values of the shifts and width-scaling factors are determined by
minimising a $\chi^2$ that compares the distributions in the data and the
simulation.
An example of applying an offset is shown in the top panel of
Figure~\ref{fig:data_MC_comparison_shifts}, while
an example of applying a width-scaling factor is shown in the bottom panel
of Figure~\ref{fig:data_MC_comparison_shifts}.

The pdfs for the \Et range of 4.5~\GeV\ to 15~\GeV\ are determined using \Jee Monte Carlo
simulation and the pdfs for $\Et > 15$~\GeV\ are determined using \Zee Monte Carlo simulation.

\begin{figure}[t]
\centering
\includegraphics[width=.65\textwidth]{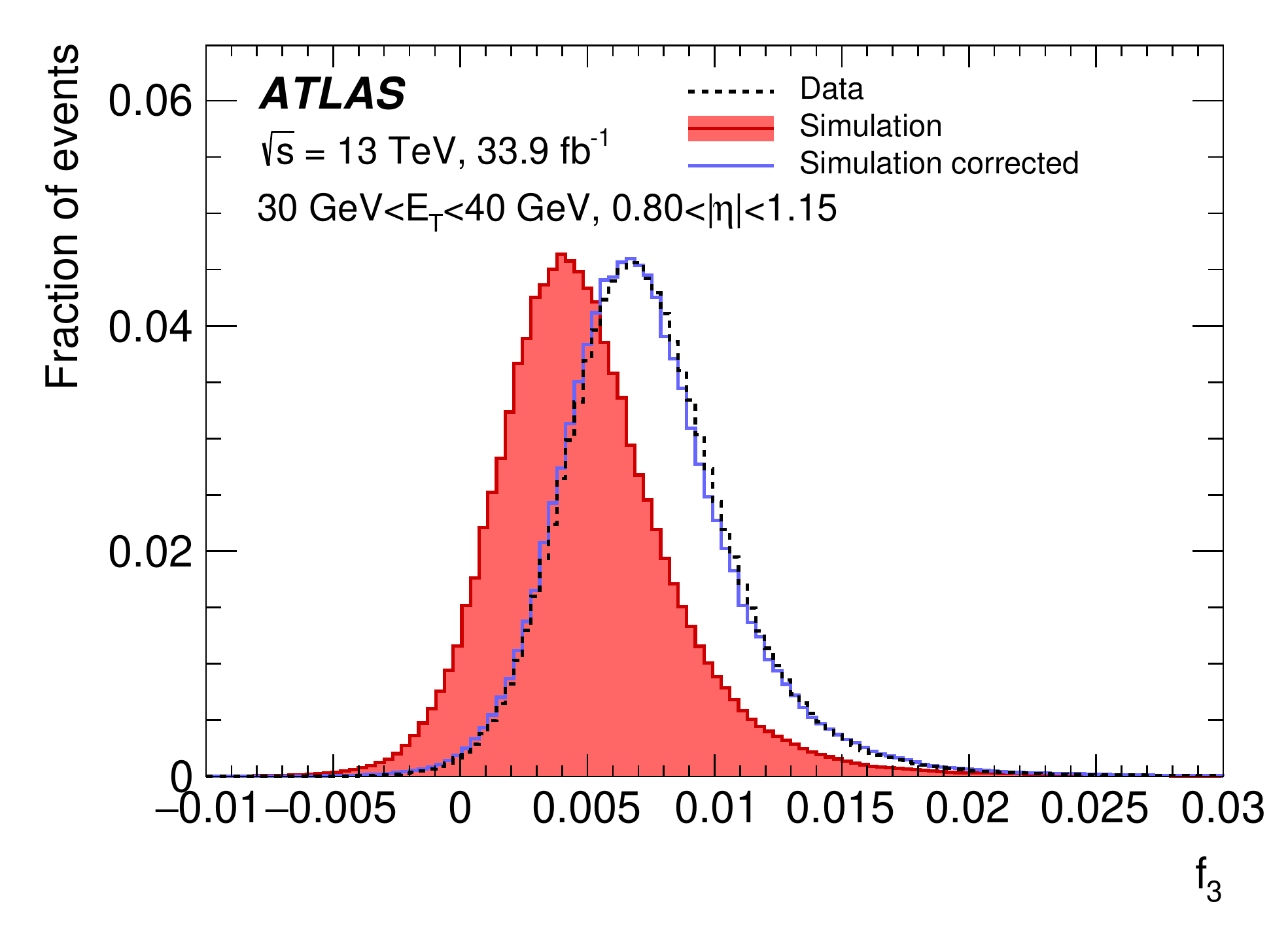}
\includegraphics[width=.65\textwidth]{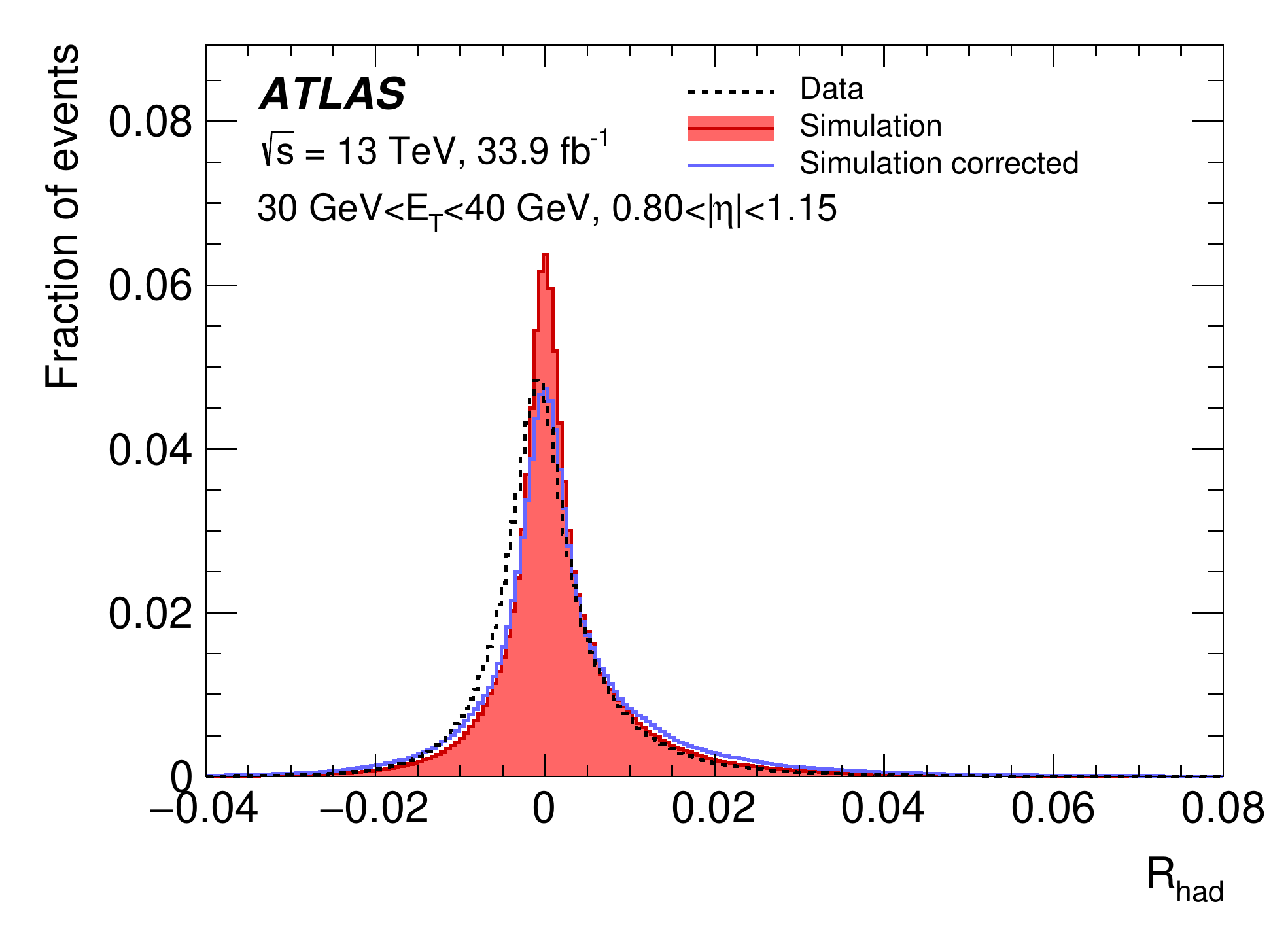}
\caption{The \fIII (top) and \rhad (bottom)
  pdf distributions in data and simulation
  for prompt electrons that satisfy $30~\GeV < \et < 40~\GeV$ and
  $0.80 < \left|\eta\right| < 1.15$.
  The distributions for both simulation and data
  are obtained using the \Zee\ \tnp method.
  KDE smoothing has been applied to all distributions.
  The simulation is shown before (shaded histogram) and after (open histogram)
  applying a constant shift (\fIII, top)
  and a width-scaling factor (\rhad, bottom). 
  Although some $\left|\eta\right|$ bins of $\fIII$ additionally
  have a width-scaling factor, this particular $\left|\eta\right|$ bin only has
  a constant shift applied.
}
\label{fig:data_MC_comparison_shifts}
\end{figure}
\clearpage

\subsection{LH-identification operating points and their corresponding efficiencies}

To cover the various required prompt-electron signal efficiencies and
corresponding background rejection factors needed by the physics analyses
carried out within the ATLAS Collaboration, four fixed values of the
LH discriminant are used to define four operating points.
These operating points are referred to as \VeryLoose, \Loose, \Medium, and \Tight
in the text below, and correspond to increasing thresholds for the
LH discriminant. The numerical values of the discriminant are determined using the simulation.
As shown in more detail later in this section, the efficiencies for identifying
a prompt electron with $\Et = 40$~\GeV\ are 93\%, 88\%, and 80\% for the
\Loose, \Medium, and \Tight operating points, respectively. 

The identification is optimised in bins of cluster~\eta\
(specified in Table~\ref{tab:etabins})
and bins of \Et\ (specified in Table~\ref{tab:etbins}).
The selected bins in cluster~\eta\ are based on calorimeter geometry,
detector acceptances and the variation of the material in the inner detector.
The pdfs of the various electron-identification quantities vary
with particle energy, which motivates the bins in \Et.
The rate and composition of the background also varies with \eta\ and \Et.

\begin{table*}[tb]
\caption{Boundaries in absolute cluster pseudorapidity used to define the nine bins
  for the LH pdfs and LH discriminant requirements.}
\begin{center}
  \begin{tabular}{cccccccccc}
\hline
\multicolumn{10}{c}{Bin boundaries in $|\eta|$}\\
\hline
0.0& 0.6& 0.8& 1.15& 1.37& 1.52& 1.81& 2.01& 2.37& 2.47 \\
\hline
\end{tabular}
\end{center}
\label{tab:etabins}
\end{table*}

\begin{table*}[tb]
\caption{Boundaries in electron transverse energy used to define the seven
  bins for the LH pdfs and the twelve bins for LH discriminant requirements.}
\begin{center}
\begin{tabular}{l|ccccccccccccc}
\hline
\multicolumn{14}{c}{Bin boundaries in \Et [\GeV]}\\
\hline
pdfs         & 4.5& 7& 10& 15& 20&   & 30&   & 40&   &   &    & $\infty$ \\
Discriminant & 4.5& 7& 10& 15& 20& 25& 30& 35& 40& 45& 80& 150& $\infty$ \\
\hline
\end{tabular}
\end{center}
\label{tab:etbins}
\end{table*}

To have a relatively smooth variation of electron-identification efficiency with electron \Et,
the discriminant requirements are varied in finer bins (specified in Table~\ref{tab:etbins}) than the pdfs.
To avoid large discontinuities in electron-identification efficiency at the bin boundaries in electron \Et, the pdf values and discriminant requirements are linearly interpolated between the centres of two adjacent bins in \Et.

All of the operating points have fixed requirements on tracking criteria:
the \Loose, \Medium, and \Tight operating points require at least two hits
in the pixel detector and seven hits total in the pixel and silicon-strip detectors combined.
For the \Medium and \Tight operating points,
one of these pixel hits must be in the innermost pixel layer
(or in the next-to-innermost layer if the innermost layer is non-operational).
This requirement helps to reduce the background from photon conversions.
A variation of the \Loose operating point---LooseAndBLayer---uses the same threshold for the  
LH discriminant as the \Loose operating point and also adds
the requirement of a hit in the innermost pixel layer.
The \VeryLoose operating point does not include an explicit requirement on the innermost pixel layer and requires only one hit in the pixel detector; the goal of this operating point is to
provide relaxed identification requirements for background studies.

The pdfs of some of the LH quantities---particularly \rhad and \reta---are
affected by additional activity in the calorimeter due to pile-up,
making them more background-like.
The number of additional inelastic $pp$ collisions in each event is
quantified using the number of reconstructed primary vertices \nvtx.
In each \eta\ bin and \Et bin, the LH discriminant $d^\prime_{L}$
is adjusted to include a linear variation with \nvtx.
Imposing a constraint of constant prompt-electron efficiency with
\nvtx leads to an unacceptable increase in backgrounds.
Instead, the background efficiency is constrained to remain approximately constant
as a function of \nvtx, and this constraint results in a small ($\le 5$~\%) decrease in signal efficiency with \nvtx.

The minimum \Et of the electron identification was reduced from 7~\GeV\ in Run~1
to 4.5~\GeV\ in Run~2.  
The use of \Jee to determine LH pdfs at low \Et is also new in Run~2.
The push towards lower \Et was motivated in part by searches for supersymmetric particles
in compressed scenarios.
In these scenarios, small differences between the masses of supersymmetric particles can lead
to leptons with low transverse momentum.

Special treatment is required for electrons with $\Et > 80$~\GeV.
The \fIII quantity (defined in Table~\ref{tab:IDcuts}) degrades the capability to distinguish
signal from background because high-\Et electrons deposit a larger fraction of their energy
in the third layer of the EM calorimeter (making them more hadron-like) than low-\Et electrons.
For this reason and since it is not modelled well in the simulation, the pdf for \fIII is removed from the LH for $\Et > 80$~\GeV.
Furthermore, changes with increasing prompt-electron \Et in the \rhad and \fI quantities
cause a large decrease in identification efficiency for $\Et > 300$~\GeV.
Studies during development of the identification algorithm showed that this loss in efficiency was very large for the \Tight operating 
point (the identification efficiency fell from 95\% at $\Et = 300$~\GeV\ to 73\% for $\Et = 2000$~\GeV).
To mitigate this loss, for electron candidates with $\Et > 150$~\GeV,
the LH discriminant threshold for the \Tight operating point is set to be the same as
for the \Medium operating point, and two additional selection criteria are added to the
\Tight selection: \eoverp and \wstot.
The requirement on \wstot depends on the electron candidate \eta, while the requirement on \eoverp is $\eoverp < 10$.
The high value of the latter requirement takes into account the decreased momentum resolution in track fits of a few 100~\GeV\ and above.
With these modifications, good signal efficiency and background rejection
are maintained for very high \et\ electrons in searches for physics beyond the Standard Model, such as
$W^\prime\rightarrow e\nu$.

In Run~1, electron candidates satisfying tighter operating points did not necessarily satisfy the more
efficient looser operating points.
This situation was a result of using different quantities in the electron LH for
the different operating points.
In Run~2, electron candidates satisfying tighter operating points also satisfy less restrictive
operating points, i.e.\ an electron candidate that satisfies the \Tight criteria will
also pass the \Medium, \Loose, and \VeryLoose criteria.

Another important difference in the electron identification between Run~1 and Run~2 is that
the LH identification is used in the online event selection (the high-level trigger, 
HLT) in Run~2, instead of a cut-based identification in Run~1.
This change helps to reduce losses in efficiency incurred by applying the offline identification
criteria in addition to the online criteria. 
The LH identification in the trigger is designed to be as close as possible to the LH used
in offline data analysis; however, there are some important differences.
The \deltapoverp quantity is removed from the LH because it relies on the GSF algorithm
(see Section~\ref{sec:trackreco}), which is too CPU-intensive for use in the HLT.
The average number of interactions per bunch crossing, $\langle \mu \rangle$, is used to quantify
the amount of pile-up, again because the determination of the number of primary vertices,
\nvtx, is too CPU-intensive for the HLT.
Both the \trackdO and \dOSignificance\ quantities are removed from the LH used in the trigger
in order to preserve efficiency for electrons from exotic processes
which  might have non-zero track impact parameters.
Finally, the LH identification in the trigger uses quantities reconstructed in the trigger,
which generally have poorer resolution than the same quantities reconstructed offline. 
The online operating points corresponding to \VeryLoose, \Loose, \Medium, and \Tight
are designed to have efficiencies relative to reconstruction like those of the corresponding offline
operating points.
Due to these differences, the inefficiency of the online selection for electrons fulfilling the same operating point as the offline selection is typically a few percent (absolute), up to 7\% for the \Tight operating point.

The efficiencies of the LH-based electron identification for
the \Loose, \Medium, and \Tight operating points for data and the corresponding data-to-simulation ratios 
are summarised in Figures~\ref{fig:identification_efficiencies} and \ref{fig:identification_efficiencies_nvtx}. They are extracted from \Jee\ and \Zee\ events, as discussed in Section~\ref{sec:efficiencies}.
The variations of the efficiencies with \Et, \eta, and the number of reconstructed primary vertices are shown.
Requirements on the transverse ($\trackdO$) and longitudinal ($\trackzO$) impact parameters measured as the distance of closest approach of the track to the measured primary vertex (taking into account the beam-spot and the tilt of the beam-line) are applied when evaluating the numerator of the identification efficiency.
For the \Tight operating point, the identification efficiency varies from 55\% at $\Et = 4.5$~\GeV\ to 90\% at $\Et = 100$~\GeV, while it ranges from 85\% at $\Et = 20$~\GeV\ to 96\% at $\Et = 100$~\GeV\ for the \Loose operating point. The uncertainties in these measured efficiencies for the \Loose (\Tight) operating point range from 3\% (4\%) at $\Et = 4.5$~\GeV\ to 0.1\% (0.3\%) for $\Et = 40$~\GeV.
As mentioned earlier in this section, simulation was used to determine the discriminant values that define the various operating points, with the intended outcome that the efficiencies would fall smoothly with decreasing electron \Et, while keeping the rapidly increasing background at acceptable levels. The simple offsets and width variations applied to the simulation to account for mismodelling of the EM-calorimeter shower shapes (see Section~\ref{sec:LHpdfs}) work well at higher electron \Et, but are unable to fully correct the simulation at lower electron \Et. This leads to an unintended larger efficiency in data for signal electrons at lower \Et, as can be seen in Figure~\ref{fig:identification_efficiencies}.
The figure also shows the corresponding rise in the data-to-simulation ratios.

\begin{figure}[t]
\centering
\includegraphics[width=.65\textwidth]{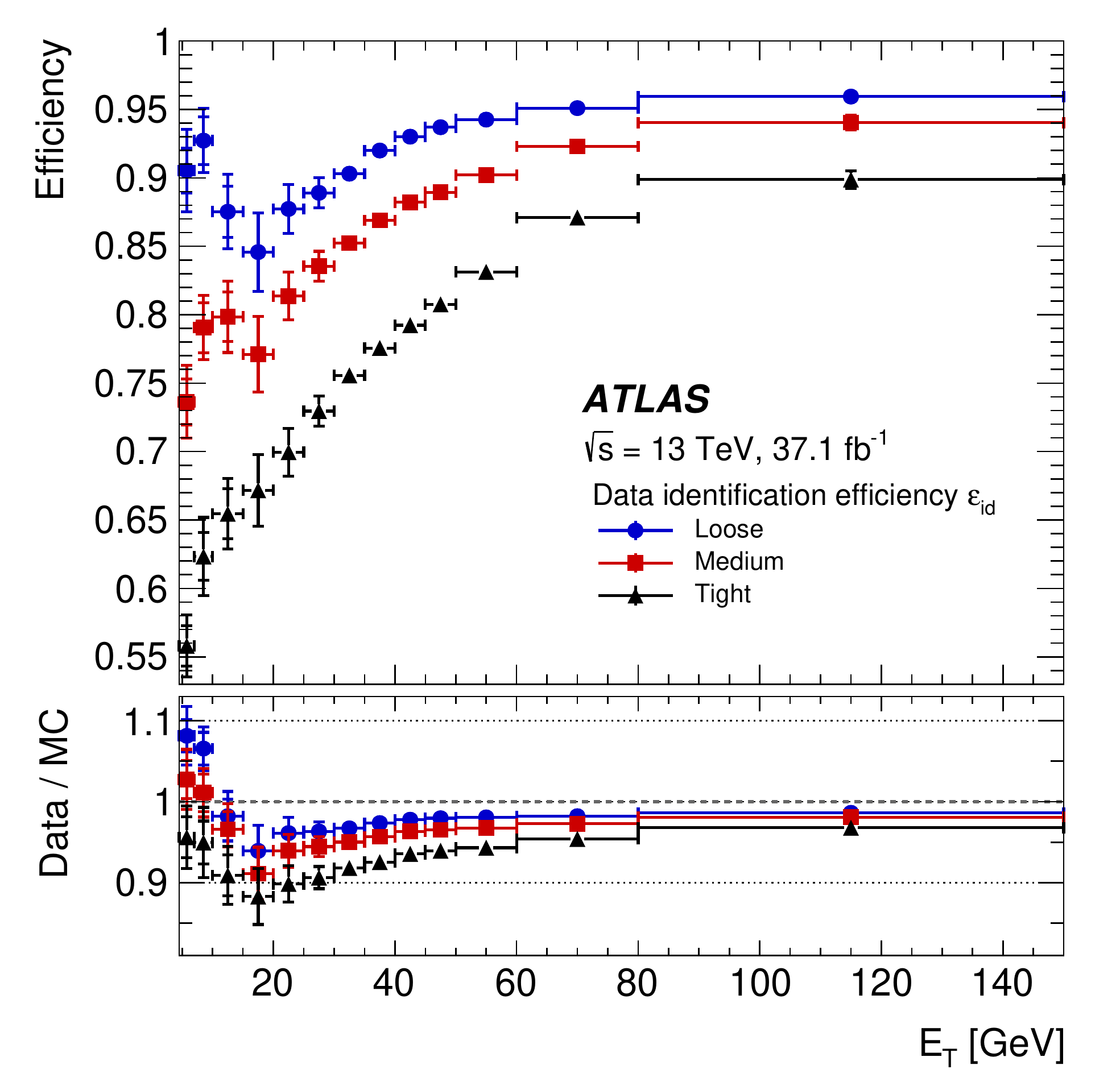}
\includegraphics[width=.65\textwidth]{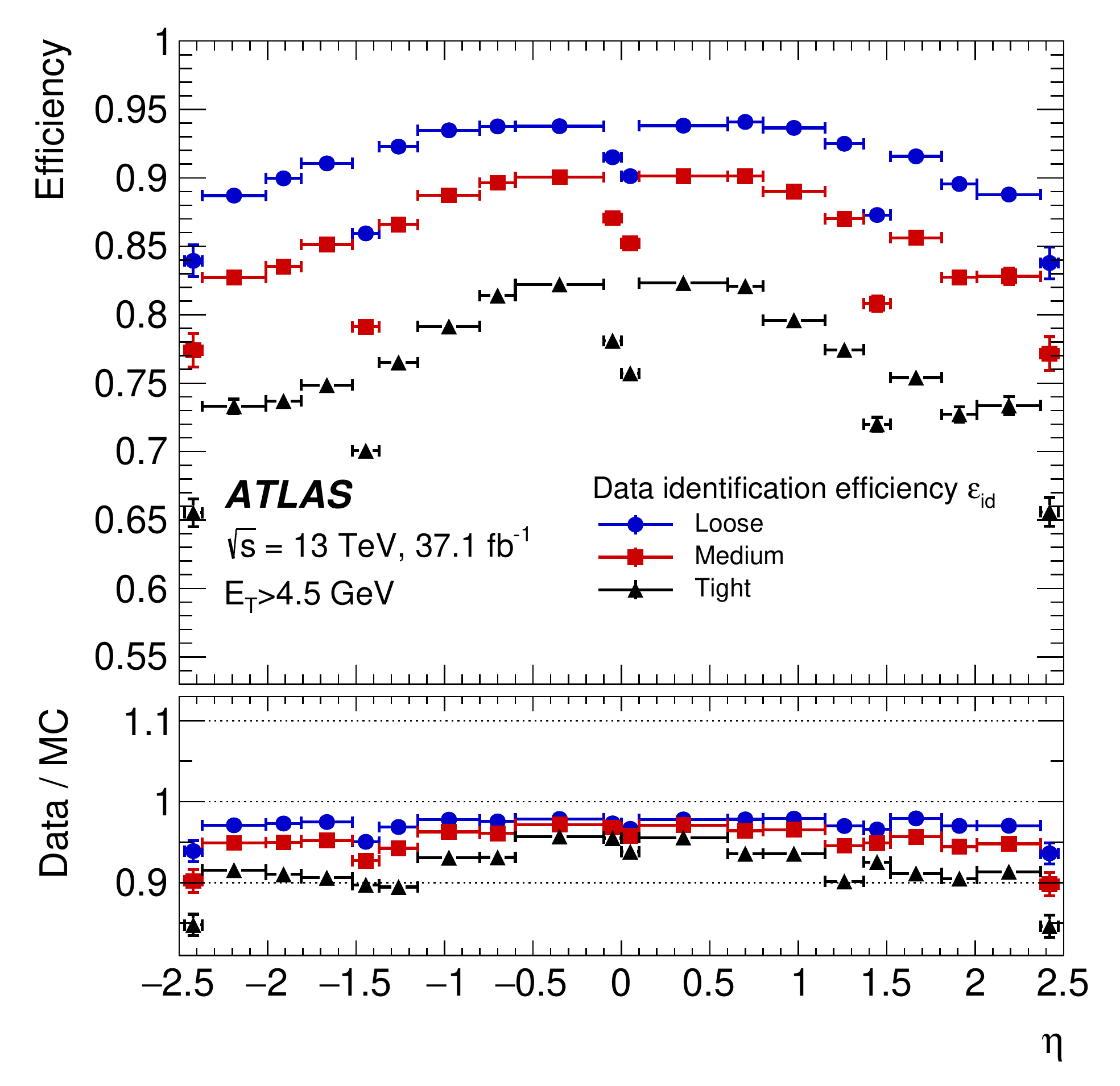}
  \caption{Measured LH electron-identification efficiencies in \Zee events for the
    \Loose (blue circle), \Medium (red square), and \Tight (black triangle) operating points as a function of \Et\ (top) and \eta\ (bottom). The vertical uncertainty bars (barely visible because they are small) represent the statistical (inner bars) and total (outer bars) uncertainties. The data efficiencies are obtained by applying data-to-simulation efficiency ratios that are measured in \Jee\ and \Zee\ events to the \Zee simulation.
For both plots, the bottom panel shows the data-to-simulation ratios.
  }
\label{fig:identification_efficiencies}
\end{figure}

\begin{figure}[t]
  \centering
  \includegraphics[width=.65\textwidth]{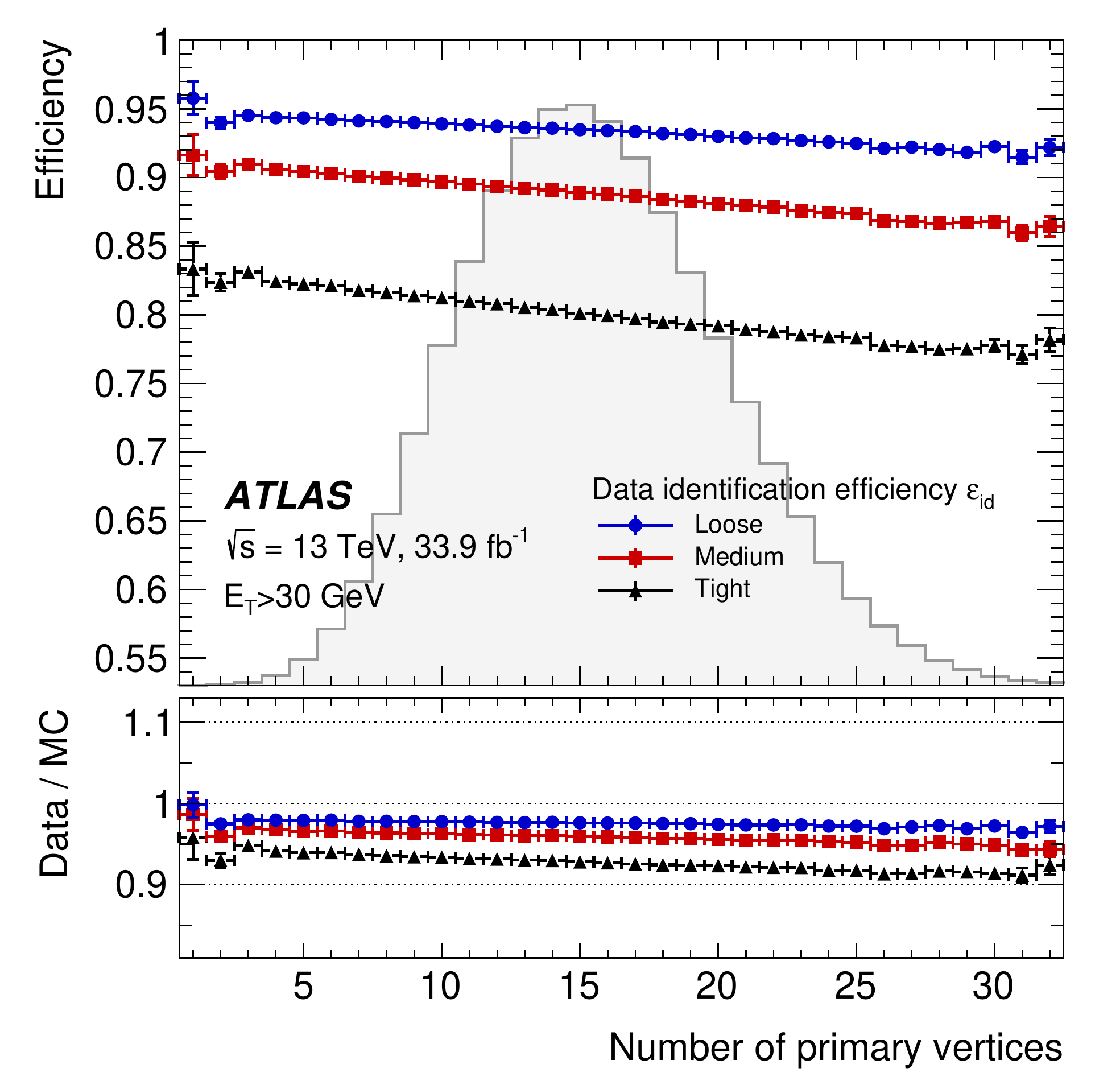}
  \caption{The LH electron-identification efficiencies for electron candidates with $\Et>30$\,\GeV\ for the
    \Loose (blue circle), \Medium (red square), and \Tight (black triangle) operating points as a function of the number of primary vertices
    in the 2016 data using the \Zee\ process.
    The shaded histogram shows the distribution of the number of primary vertices for the 2016 data.
The inner uncertainties are statistical while the total uncertainties include both the statistical and systematic components. 
The bottom panel shows the data-to-simulation ratios.
}
  \label{fig:identification_efficiencies_nvtx}
\end{figure}

The lower efficiencies of the \Medium and \Tight operating points compared to \Loose result in an increased rejection of background; the rejection factors for misidentified electrons from multijet production (evaluated with the two-to-two process simulation sample described in Section~\ref{sec:samples}) increase typically by factors of approximately 2.5 for \Medium and 5 for \Tight compared to \Loose, in the \ET range of $4-50$~\GeV.
Computations and measurements of the rejection, especially absolute rejections, are typically associated with large uncertainties due to ambiguities in the definition of the denominator, and the diversity of the sources of background.
The factors mentioned above are similar to those published in Table~3 of the ATLAS Run-1 publication~\cite{PERF-2016-01} when these considerations are taken into account.

\clearpage
\section{Electron isolation}
\label{sec:isolation}
A considerable challenge at the LHC experiments is to differentiate the prompt production of electrons, muons, and photons in signal processes  (from the hard-scattering vertex, or from the decay of heavy resonances such as Higgs, \Wboson, and \Zboson\ bosons) from background processes such as semileptonic decays of heavy quarks, hadrons misidentified as leptons and photons, and photons converting  into electron--positron pairs in the detector material upstream of the electromagnetic calorimeter. A  characteristic signature of such a  signal is represented by little activity (both in the calorimeter and in the inner detector) in an area of $\Delta \eta \times \Delta \phi$ surrounding the candidate object. However, the production of boosted particles decaying, for example, into collimated electron--positron pairs or the production of prompt electrons, muons, and photons within a busy experimental environment such as in $t\overline{t}$ production can obscure the picture. Variables are constructed that quantify the amount of activity in the vicinity of the candidate object, something usually performed by summing the transverse energies of clusters in the calorimeter or the transverse momenta of tracks in a cone of radius $\Delta R=\sqrt{(\Delta \eta)^2 + (\Delta \phi)^2}$ around the direction of the electron candidate, excluding the candidate itself.

Several components enter into building such isolation variables: identifying the candidate object itself, its direction, and its contribution to the activity within the cone, and summing, in a pile-up and underlying-event robust way, the other activity found within the cone. The two classes of isolation variables considered in this paper are based on calorimeter and tracking measurements, and are respectively discussed in Sections~\ref{sec:caloiso} and ~\ref{sec:trackiso}.

\subsection{Calorimeter-based isolation}
\label{sec:caloiso}

The reconstruction of electron candidates is described in Section~\ref{sec:reconstruction}.
To build an isolation variable, a cone of size $\Delta R$ is then delineated around the candidate electron's cluster position.

The computation of calorimeter-based isolation in the early running period of ATLAS simply summed the transverse energies of the calorimeter cells (from both the electromagnetic and hadronic calorimeters) within a cone aligned with the electron direction, excluding the candidate's contribution. This type of calorimeter-based variable exhibited a lack of pile-up resilience and demonstrated poor data--simulation agreement. A significant improvement was achieved by using the transverse energies of topological clusters~\cite{PERF-2014-07} instead of cells, thus effectively applying a noise-suppression algorithm to the collection of cells. 

Topological clusters are seeded by cells with a deposited  electromagnetic-scale energy of more than four times the expected noise-level threshold of that cell; this includes both electronic noise and the effects of pileup. The clusters are then expanded, in the three spatial directions across all electromagnetic and hadronic calorimeter layers, by iteratively adding neighbouring cells that contain a deposited energy more than twice the noise level. After the expansion around the cluster stops due to a lack of cells satisfying the energy threshold requirements, a final shell of cells surrounding the agglomeration is added to the cluster. The topological clusters used in the isolation computation are not further calibrated: they remain at the electromagnetic scale, regardless of the origin of the particle. 

\begin{figure}[h]
\centerline{\includegraphics[width=0.6\textwidth]{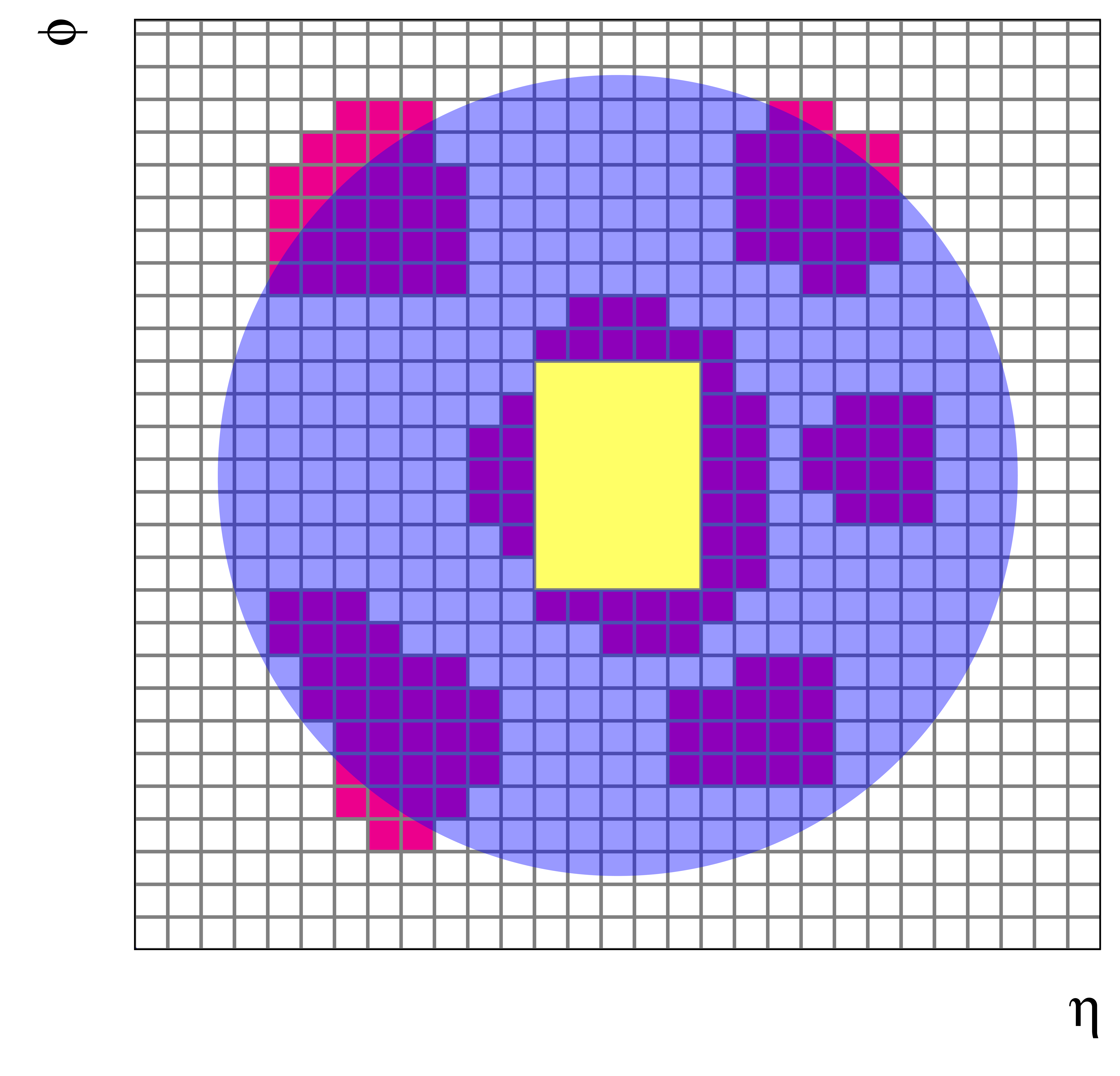}
}
\caption{Schema of the calorimeter isolation method: the grid represents the second-layer calorimeter cells in the $\eta$ and $\phi$ directions. The candidate electron is located in the centre of the purple circle representing the isolation cone. All topological clusters, represented in red, for which the barycentres fall within the isolation cone are included in the computation of the isolation variable. The $5\times 7$ cells (which cover an area of $\Delta \eta \times \Delta \phi = 0.125 \times 0.175$) represented by the yellow rectangle   correspond to the subtracted cells in the core subtraction method.
\label{fig:IsolSchema}
}
\end{figure}

The energies of all positive-energy topological clusters, whose barycentres fall within the cone of radius $\Delta R$, as illustrated in Figure~\ref{fig:IsolSchema}, are summed into the raw isolation energy variable \ETraw. 
This raw isolation energy still includes the energy deposited by the candidate electron, called the core energy \ETcore. 
The core energy is subtracted by removing the cells included in a $\Delta \eta \times \Delta \phi = 0.125 \times 0.175$ rectangle around the candidate's direction, as illustrated in Figure~\ref{fig:IsolSchema}. The advantage of this method is its simplicity and stable subtraction scheme for both the signal and background candidates. A disadvantage of this method is that the candidate object may deposit energy outside of this fixed rectangular area which may be incorrectly assigned as additional activity, requiring an additional leakage correction to the subtracted core energy. 
The core leakage correction is evaluated using samples of simulated single electrons (without additional pile-up activity).
The energy leaking into the cone is then fit to a Crystal Ball function; 
its most probable value $\mu_\mathrm{CB}$  is parameterised as a function of \ET\ and  is used as an estimator of the average leakage, $E_\mathrm{T, leakage}(\ET)$. 
The corrections are derived in ten bins of the associated cluster $\eta$ position.

Figure~\ref{fig:coreptrealEl} shows the isolation energy corrected with a rectangular core, without and with the calculated leakage correction, as a function of the electron \Et for a sample of simulated single electrons which includes the effects of pile-up; a leakage correction is essential when using a rectangular core region.  

\begin{figure}[t]
\centering
\centering
\includegraphics[width=0.70\textwidth]{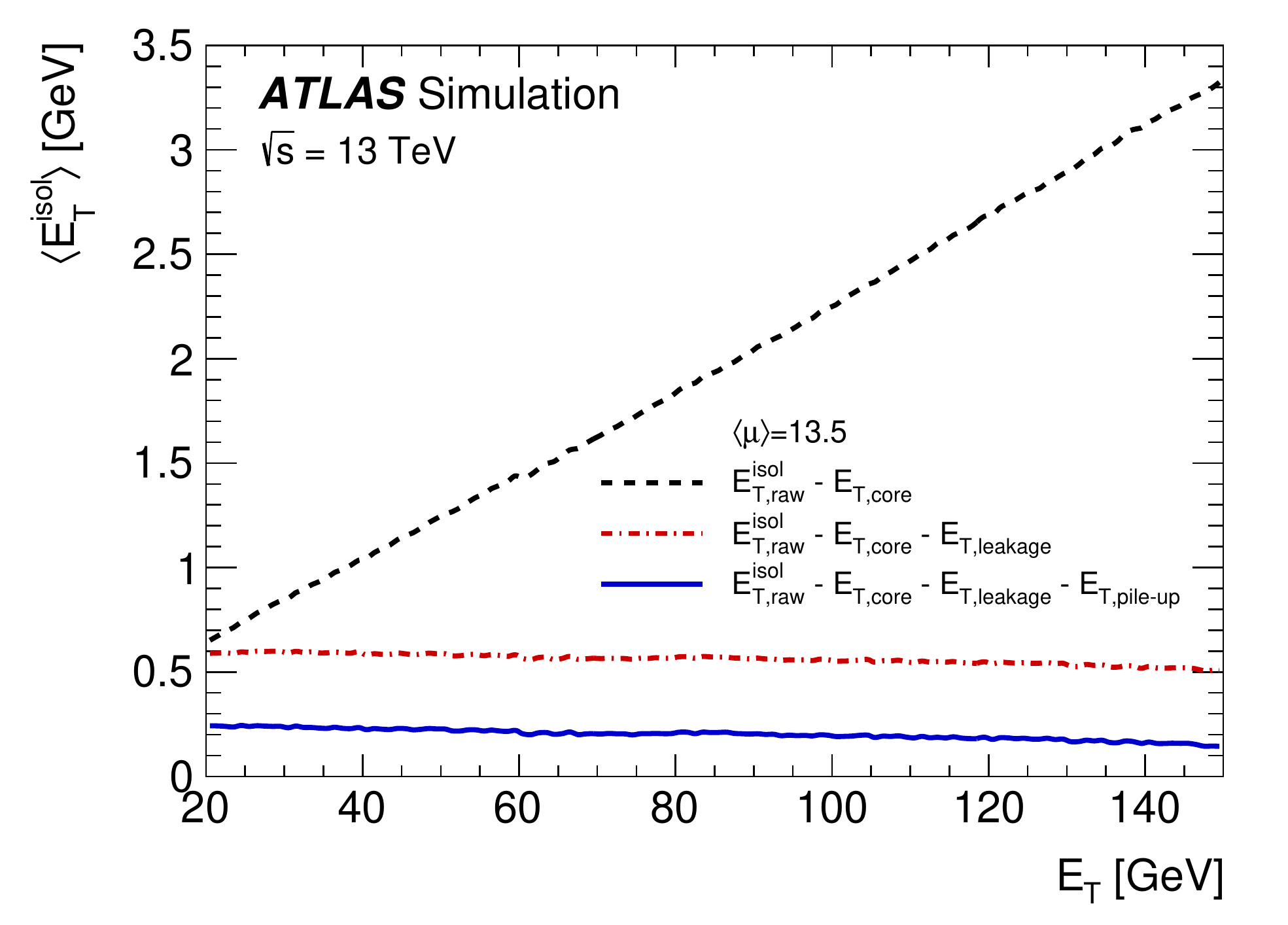}
\caption{ Isolation transverse energy built with a cone of  radius $\Delta R=0.2$, corrected with a rectangular core without (black dashed line) and with (red dot-dashed line) the leakage correction $E_\mathrm{T, leakage}$ as a function of the electron \ET, for a simulated sample with electron candidates in the central region of the detector ($|\eta|<2.47$) that satisfy a \Tight\ electron identification criterion. This figure also shows the size of the pile-up correction $E_\textrm{T,pile-up}$ (the difference between the red dot-dashed line and solid blue line). The curves were obtained from a simulated single-electron sample 
that includes the effects of $\langle \mu \rangle = 13.5$ pile-up.} 
\label{fig:coreptrealEl}
\end{figure}

The pile-up and underlying-event contribution to the isolation cone is estimated from the ambient energy density~\cite{jetarea}. For each event, the entire calorimeter acceptance up to $|\eta|=5$ is used to gather positive-energy topological clusters using the $k_t$ jet-clustering algorithm~\cite{Ellis:1993tq,Catani:1993hr} with radius parameter $R=0.5$, with no jet \pt\ threshold. The area $A$ of each jet is estimated 
and the transverse energy density $\rho$ of each jet is computed as $\rho=\pT/A$.
The median energy density $\rho_\mathrm{median}$ of the distribution of jet densities in the event is used as an estimator of the transverse energy density of the event. For a simulated $Z \to ee$ sample at $\sqrt s = 13$~\TeV\ with average pile-up $\langle \mu \rangle=22$, $\rho_\mathrm{median}$ is approximately 4~\GeV\ per unit of $\eta-\phi$ space in the central $\eta$ region of the calorimeter, decreasing to 2~\GeV\ at $|\eta|=2.5$.  
The pile-up/underlying-event correction is then evaluated as:
\begin{equation*}
E_\textrm{T,pile-up}(\eta) = \rho_\mathrm{median}(\eta) \times \left( \pi \Delta R^2  - A_\mathrm{core}\right),
\end{equation*}
where $\Delta R$ is the radius of the isolation cone and $A_\mathrm{core}$ is the area of the signal core that was subtracted. The $\eta$ dependence of $\rho$ is estimated in two bins: $|\eta|<1.5$ and $1.5<|\eta|<3.0$. Figure~\ref{fig:coreptrealEl} shows the size of this pile-up correction for a simulated single-electron sample with $\langle \mu \rangle = 13.5$ pile-up. 

The fully corrected calorimeter-based isolation variable \ETcorr\ calculated within the cone of radius $\Delta R$ is then obtained after subtracting the components described above, namely
\begin{equation*}
\ETcorr=\ETraw-\ETcore-\ETleak-\ETpile.
\end{equation*}
Figure~\ref{fig:coreptrealEl} shows the resulting distribution as a function of the transverse energy of the electron for a simulated single-electron sample. The distribution is slightly positive since only positive-energy clusters are summed, allowing for only positive fluctuations from noise.

\subsection{Track-based isolation}
\label{sec:trackiso}

The computation of track-based isolation variables uses tracks with $\pT>1$~\GeV, reconstructed within a fiducial region of the inner detector, $|\eta|<2.5$, and that satisfy basic track-quality requirements. This track selection, optimised using candidate muons from simulated $t\overline{t}$ samples, includes a minimum number of hits identified in the silicon detectors and a maximum number of inoperable detector regions crossed by the track. In order to minimise the impact of  pile-up, a requirement is placed on the longitudinal impact parameter, $z_0$, corrected for the reconstructed position of the primary vertex and multiplied by the sine of the track polar angle: $|z_{0}\sin\theta|<3$~mm. 
This requirement on $|z_{0}\sin\theta|$ aims to select tracks that originate from the vertex that is chosen to be the relevant vertex of the process. In most cases, the relevant vertex corresponds to the ``hardest'' vertex of the event, i.e.\ the vertex for which the sum of the squares of the transverse momenta of the associated tracks is the largest; this is the vertex used by default in the track-isolation computation. 
Track-based isolation variables are constructed by summing the transverse momenta of the tracks found within a cone of radius $\Delta R$ aligned with the electron track, excluding the candidate's own contribution.

The track-\pT\ contribution of the candidate electron to the track-isolation variable must be subtracted from the cone. Electrons  can undergo bremsstrahlung radiation with the radiated photons converting into secondary electrons; such additional particles should be counted as part of the initial particle's energy. For this reason, tracks are extrapolated to the second layer of the EM calorimeter. All extrapolated tracks that fall within a $\Delta \eta \times \Delta\phi=0.05\times 0.1$ window around the cluster position are considered to be part of the candidate and are removed from the track-isolation-variable computation. The resulting track-isolation variable is called \pTiso.

Unlike calorimeter isolation, where a cone with a radius much less than $\Delta R = 0.2$ is difficult to build  due to the finite granularity of the calorimeter, the much smaller tracker granularity allows the use of narrower cone sizes. For example, in boosted decay signatures or very busy environments, other objects can be close to the signal lepton direction. For such cases, a variable-cone-size track isolation, \pTvar, can be used, one that progressively decreases in size as a function of the \pT\ of the candidate:
\begin{equation*}
\Delta R = \mathrm{min} \left(\frac{10~\mathrm{GeV}}{\pt[\mathrm{GeV}]},R_\mathrm{max}\right),
\end{equation*}
where $R_\mathrm{max}$ is the maximum cone size (typically $0.2$ to $0.4$). The value of 10~\GeV\ in the argument is derived with a simulated $t\overline{t}$ sample, and designed to maximise the rejection of background.

\subsection{Optimisation of isolation criteria and resulting efficiency measurements}

The implementation of isolation criteria is specific to the physics analysis needs, be it to identify isolated prompt electrons or electrons produced in a busy environment, or to reject light hadrons misidentified as electrons.  Precision measurements with copious signal at lower \pT\ may favour tighter isolation requirements, and be willing to sacrifice some signal in order to ensure high background rejection, whereas searches at high \pT\ may instead favour looser requirements in order to maintain high signal efficiency. Therefore, several isolation operating points were established that use calorimeter-based isolation in a cone of radius $\Delta R=0.2$ (Section~\ref{sec:caloiso}) or track-based isolation using a variable-size cone with $R_\mathrm{max}=0.2$ or  $0.4$ (Section~\ref{sec:trackiso}), or both types of isolation simultaneously. The requirements for each efficiency-targeted operating point are established in bins of electron \et\ and $\eta$ with edges:
\begin{itemize}
\item \et\ [\GeV]: 7, 10, 15, 20, 25, 27.5,     30, 32.5,     35--45 (1~\GeV\ bins),  47.5, 50, 60, 80, $\infty$,
\item $|\eta|$: 0, 0.1, 0.6, 0.8, 1.15, 1.37, 1.52, 1.81, 2.01, 2.37, 2.47.
\end{itemize}

 The operating points are defined in three categories:
\begin{itemize}
\item targeting a fixed value of the isolation efficiency $\epsilon_\mathrm{iso}$, uniform in the $\et, \eta$ of the electron (`Loose' isolation),
\item targeting a fixed value of the isolation efficiency $\epsilon_\mathrm{iso}$, dependent on the \et\ of the electron but uniform in $\eta$ (`Gradient' isolation),
\item imposing fixed requirements on the value of the isolation variable (`Fix'). 
\end{itemize}

Figure~\ref{fig:isoresults} shows the isolation efficiencies measured in data and the corresponding data-to-simulation ratios as a function of the electron \ET\  and $\eta$ for the operating points given in Table~\ref{tab:leptonWP} and for candidate electrons satisfying  \Tight identification requirements. 
The efficiencies that determine the values of the requirements given in Table~\ref{tab:leptonWP}  are evaluated in simulation from a \Jee\ sample for $\et<15$~\GeV\ and from a \Zee\ sample for $\et>15$~\GeV. Since the value of the efficiency is process dependent, this can result in a jump in efficiency at the transition point of $\et=15$~\GeV, as can be seen in the top-left plot of Figure~\ref{fig:isoresults}, which was produced with a \Zee\ sample over the full \et\ range. 
The overall differences between data and simulation are less than approximately 2--4\%; the operating point `Fix (Tight)' has the most significant difference in shape in $\eta$.

\begin{table}[t]
\caption{Definition of the electron-isolation operating points and isolation efficiency $\epsilon_\mathrm{iso}$. For the `Gradient' operating points, the units of \pT\ are \GeV. All operating points use a cone size of $\Delta R=0.2$ for calorimeter isolation and $R_\mathrm{max}=0.2$ for track isolation except for the final entry `Fix (Track)' which uses $R_\mathrm{max}=0.4$. The values are obtained from a simulated $Z\rightarrow ee$ sample where electrons satisfy \Tight identification requirements.}
\centering
\footnotesize
\begin{tabular}{|c||c|c|c|}
\hline
Operating point           & \ETcorr & \pTvar & Total $\epsilon_\mathrm{iso}$ \\
                        & ($\Delta R=0.2$) & ($R_\mathrm{max}=0.2$) & \\ 
\hline \hline
Loose (Track Only)          & -                                              & $\epsilon_\mathrm{iso}=99\%$                   & $99\%$ \\
Loose                   & $\epsilon_\mathrm{iso}=99\%$                   & $\epsilon_\mathrm{iso}=99\%$                   & $98\%$ \\
\hline
Gradient                & $\epsilon_\mathrm{iso}=0.1143\times \pt +92.14\%$ & $\epsilon_\mathrm{iso}=0.1143\times \pt +92.14\%$ & $90 (99)\%$ at $25 (60)~\gev$ \\
Gradient (Loose)           & $\epsilon_\mathrm{iso}=0.057\times \pt +95.57\%$  & $\epsilon_\mathrm{iso}=0.057\times \pt +95.57\%$  & $95 (99)\%$ at $25 (60)~\gev$ \\ \hline
Fix (Loose)             & $\ETcorr/\pt<0.20$  & $\pTvar/\pt<0.15$   & - \\
Fix (Tight)             & $\ETcorr/\pt<0.06$  & $\pTvar/\pt<0.06$  & - \\
Fix (Tight, Track Only)    & -                & $\pTvar/\pt<0.06$  & - \\
Fix (Calo Only)  & $\ETcorr<3.5$~\GeV  & - & -\\ 
Fix (Track $R_\mathrm{max}=0.4$)     & $\ETcorr/\pt<0.11$  & $\pTvar/\pt<0.06$  & -\\ 
\hline
\end{tabular}
\label{tab:leptonWP}
\end{table}

\begin{figure}[t]
\centering
\includegraphics[width=0.48\textwidth]{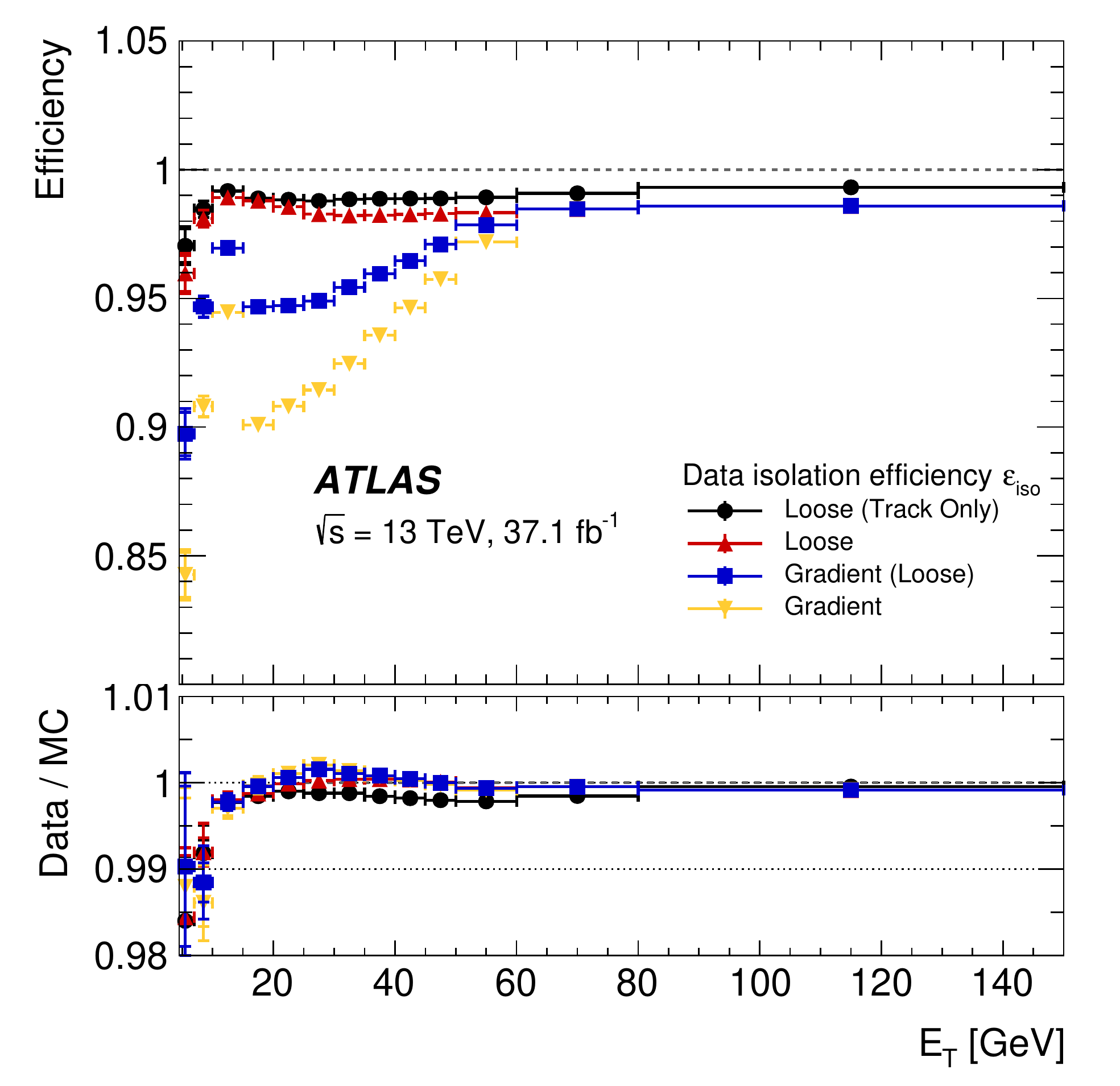}
\includegraphics[width=0.48\textwidth]{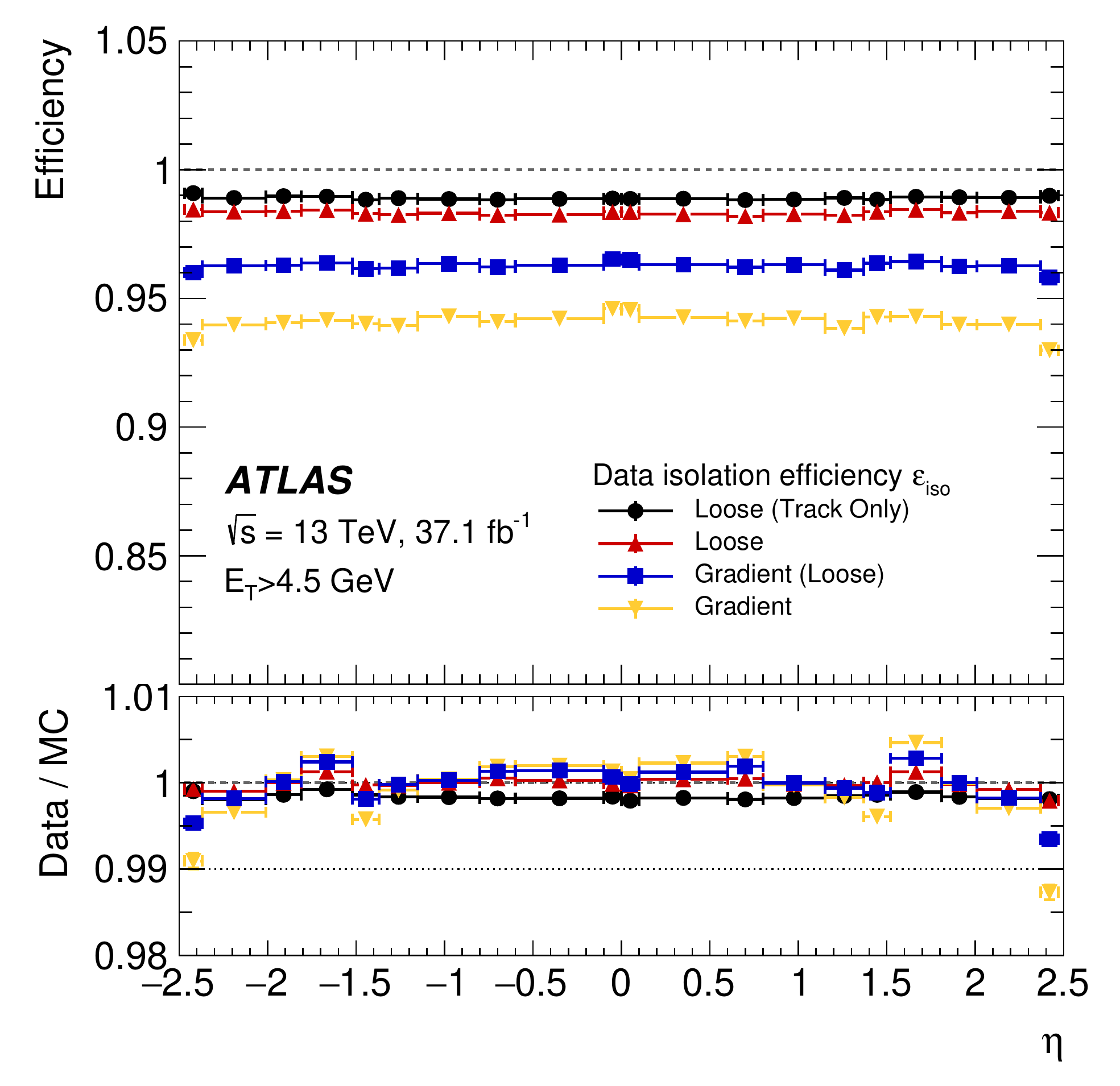}
\includegraphics[width=0.48\textwidth]{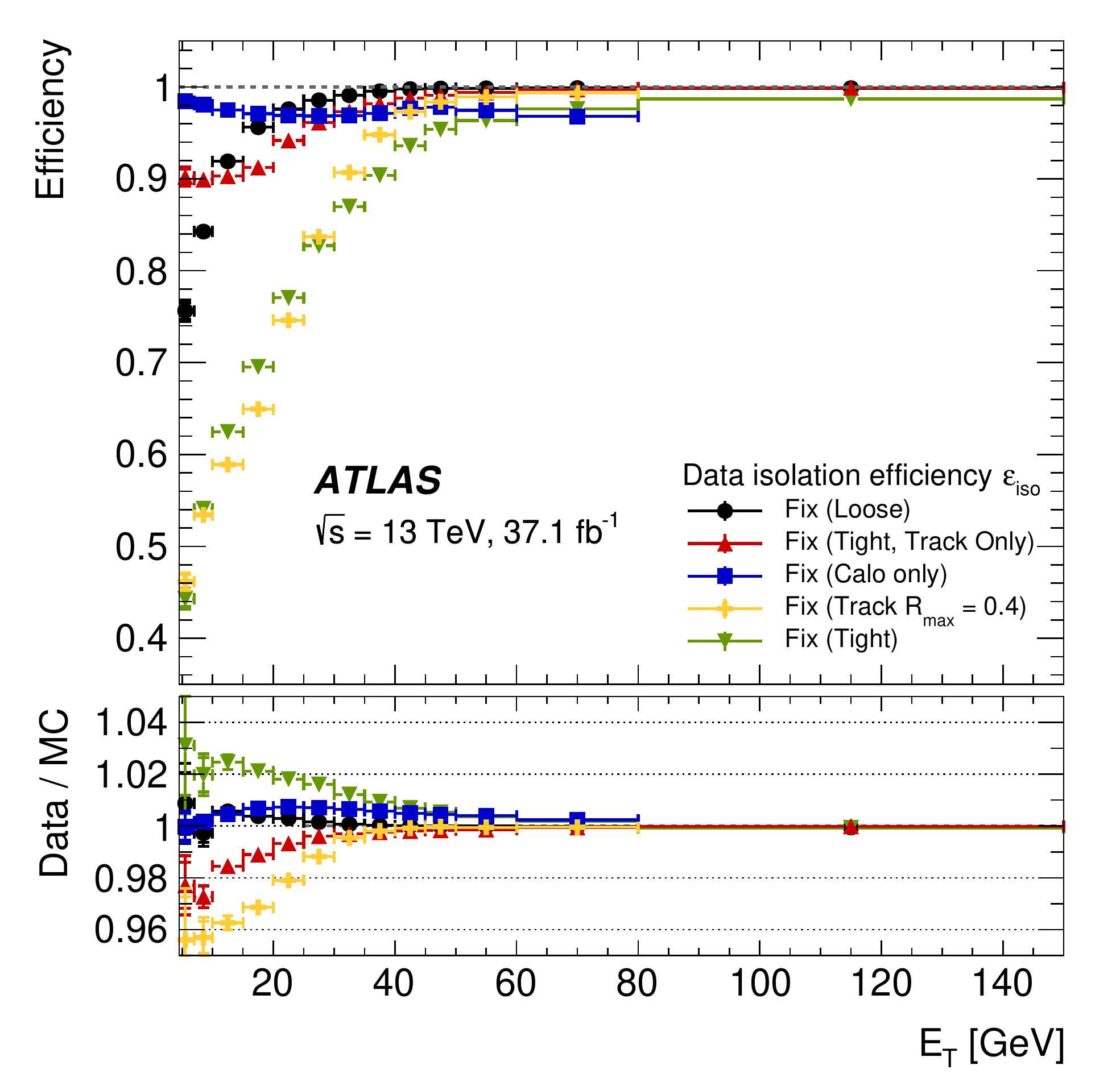}
\includegraphics[width=0.48\textwidth]{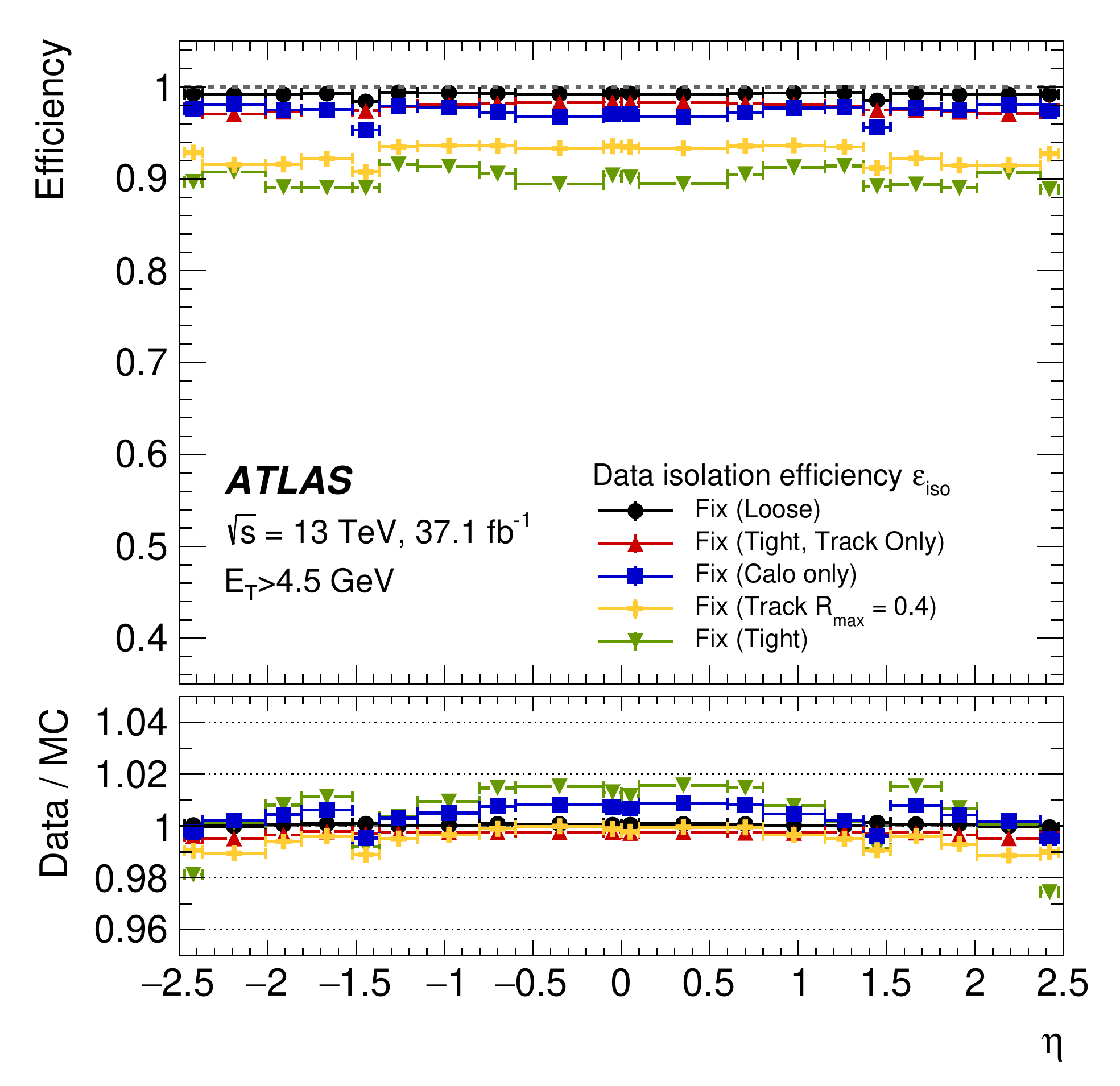}
\caption{Isolation efficiencies for data (in the upper panels) and the ratio to simulation (lower panels) for the operating points given in Table~\ref{tab:leptonWP} as a function of candidate-electron \ET\ (left) and $\eta$ (right) in \Zee\ events. Top plots: efficiencies for optimised operating points targeting specific efficiencies. Bottom plots: efficiencies for operating points where fixed requirements are applied to the isolation variables.  Cone sizes of $\Delta R=0.2$ for calorimeter isolation and $R_\mathrm{max}=0.2$ for track isolation are used (except for ``Fix (Track $R_\mathrm{max}=0.4$)'' where a twice larger radius cone is used). 
The inner uncertainties are statistical while the total uncertainties include both the statistical and systematic components.
} 
\label{fig:isoresults}
\end{figure}
\clearpage

\section{Electron-charge identification}
\label{sec:chargemisid}
\subsection{Reconstruction of electric charge}
\label{chargemisID_reco}

The electric charge of an electron is determined from the curvature of the associated track reconstructed in the inner detector. The misidentification of electron charge can result from the matching of an incorrect track to the electron candidate or from a mismeasurement of the curvature of the primary electron track. 

The probability of bremsstrahlung emission and subsequent photon conversion to an electron--positron pair (see Section~\ref{sec:reconstruction}) depends significantly on the  amount of detector material traversed. As shown in Figure~\ref{fig:chargemisID}~(top, left) for a simulated \Zee\  sample, most electrons for which the wrong charge was assigned are found in the pseudorapidity region $1.5<|\eta|<2.2$, corresponding to a region in the detector with a relatively large amount of inactive material. 

\begin{figure}[t]
  \centering
  {\includegraphics[width=0.49\textwidth]{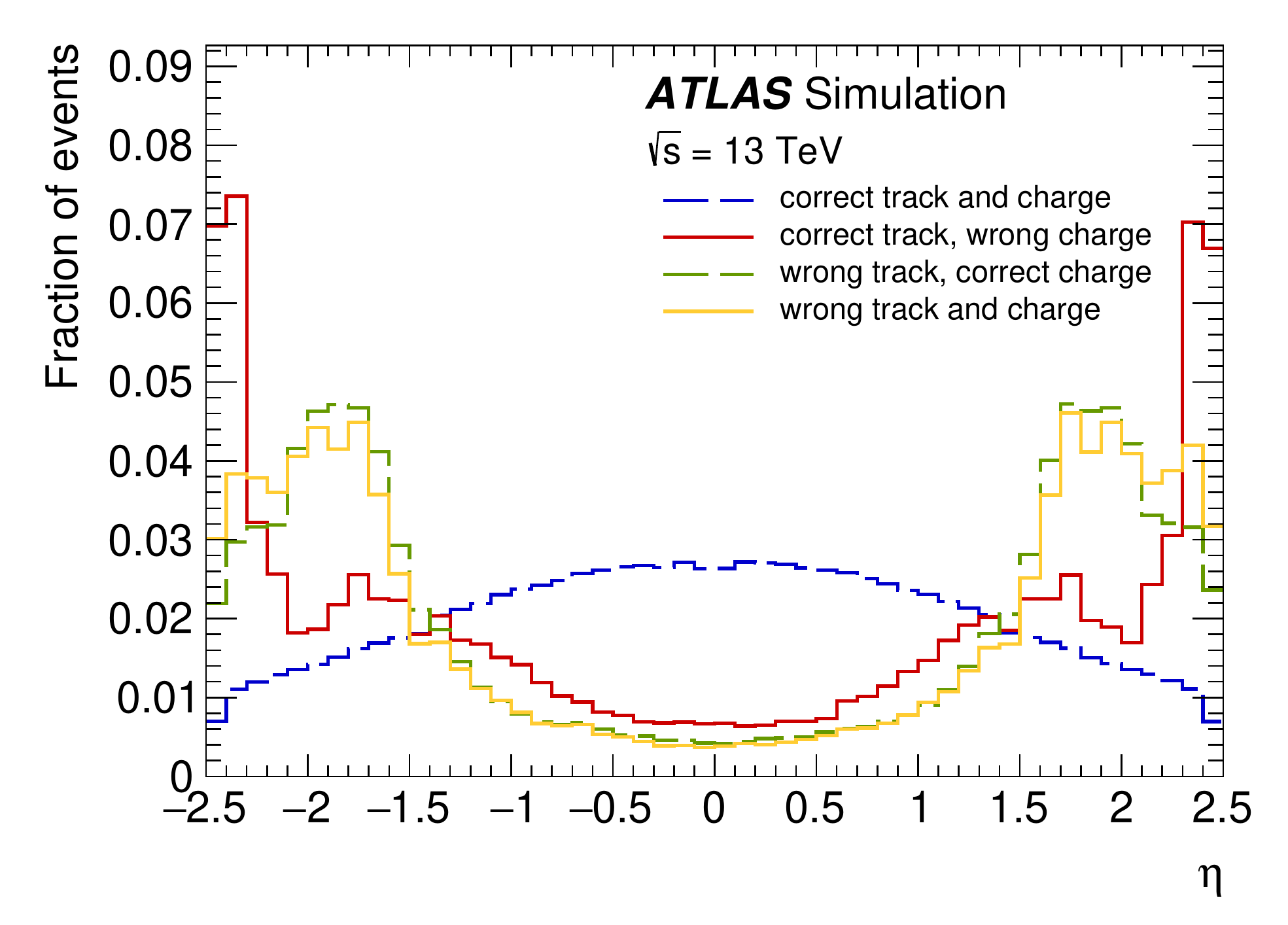}}
  {\includegraphics[width=0.49\textwidth]{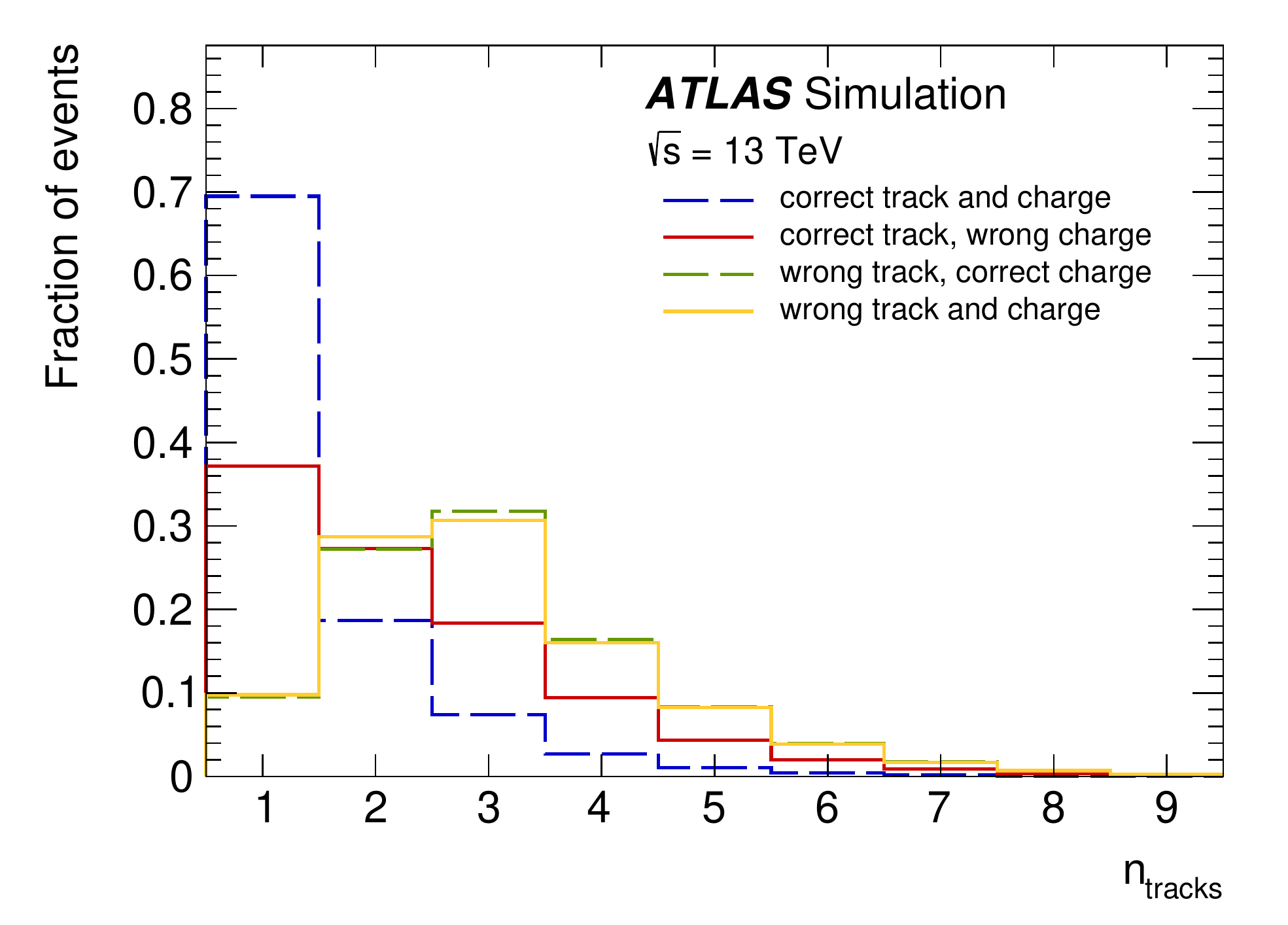}}
  {\includegraphics[width=0.49\textwidth]{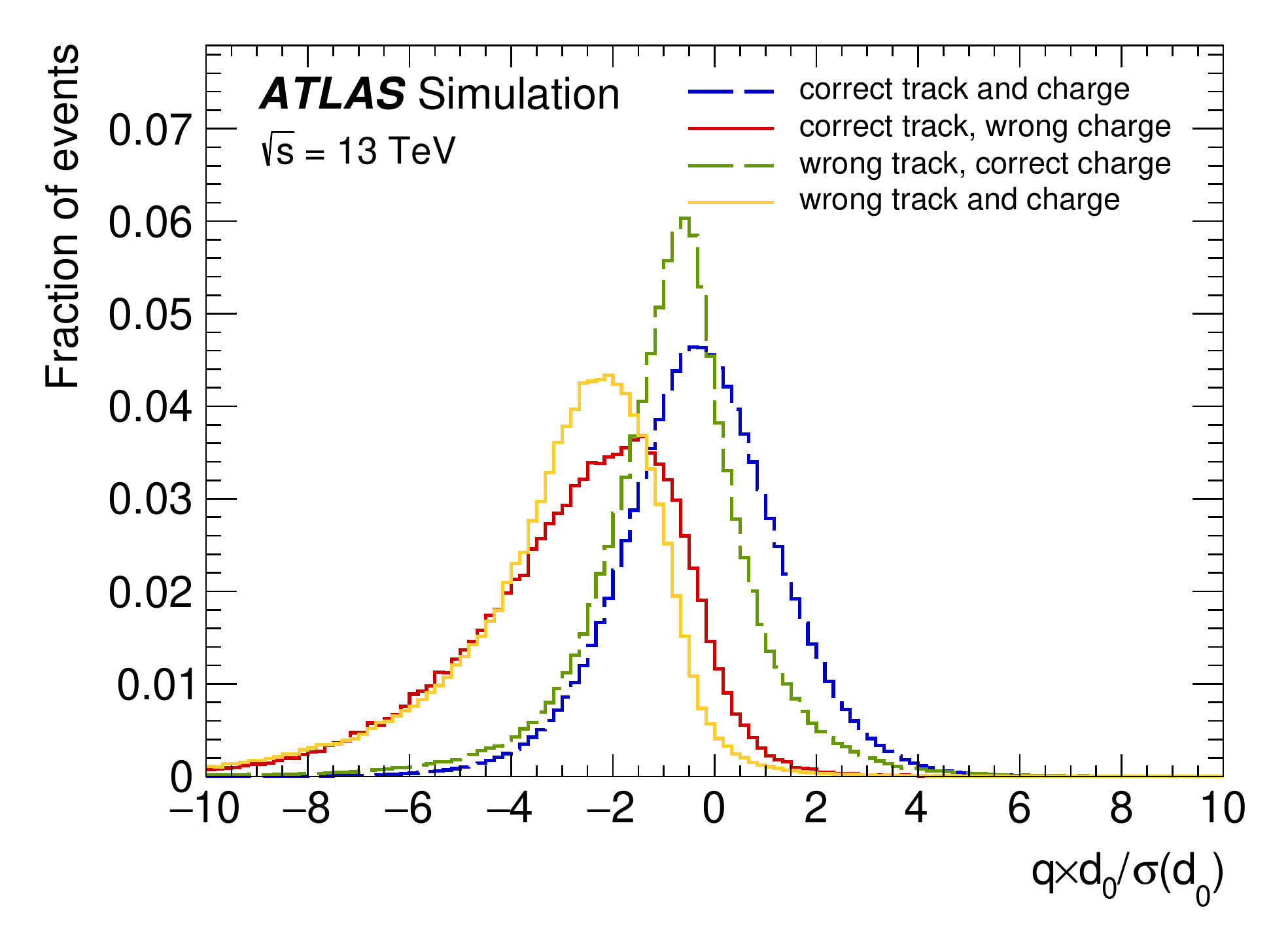}}
  {\includegraphics[width=0.49\textwidth]{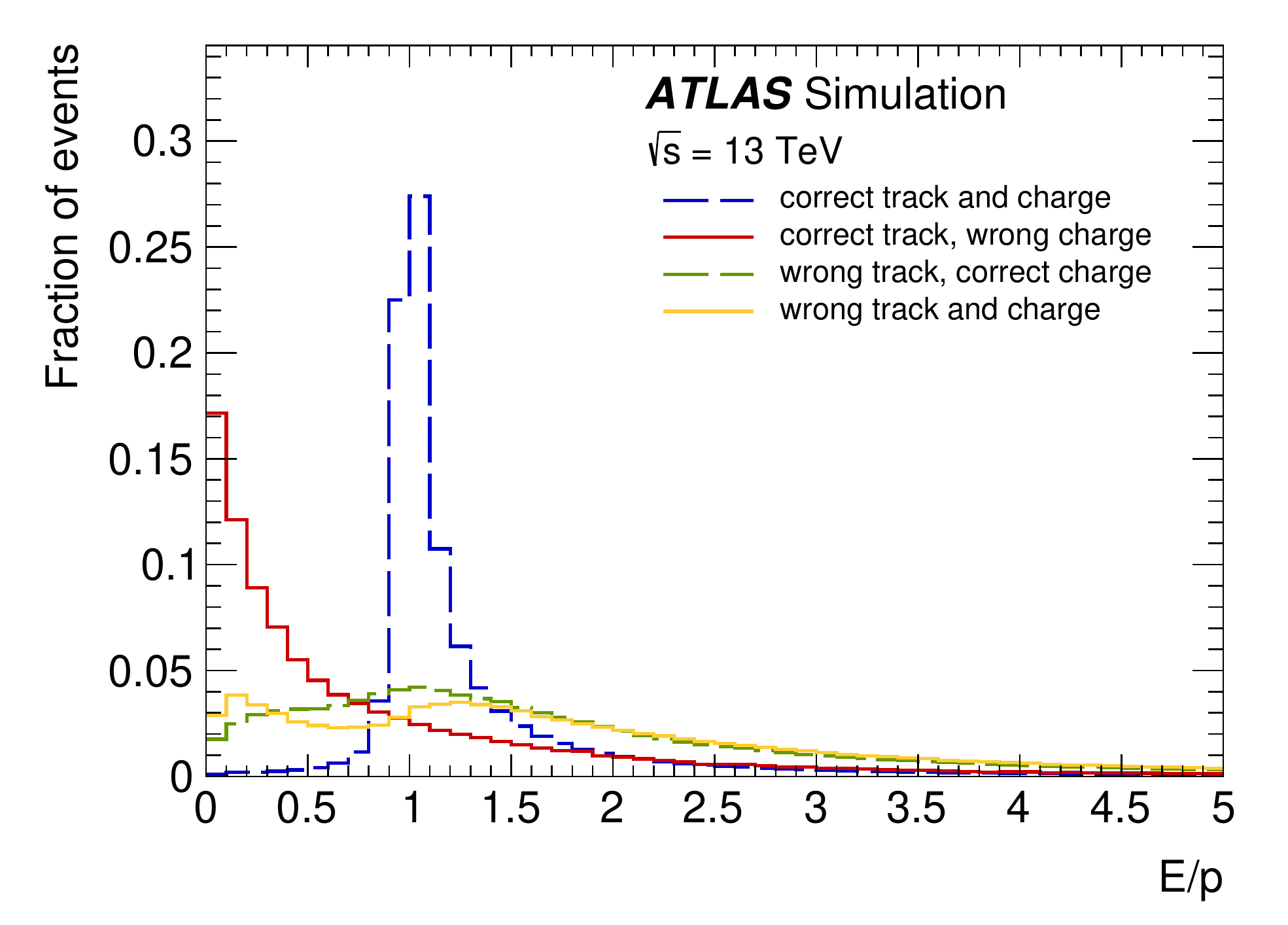}}

\caption{Kinematic and tracking properties of electrons from a simulated \Zee sample. Top, left: track pseudorapidity. Top, right: multiplicity of associated tracks. Bottom, left: electric charge multiplied by the transverse impact parameter significance. Bottom, right:  energy-to-momentum ratio, in each case for electrons with the correctly and incorrectly reconstructed charge with respect to the particle that caused the track and the correctly and incorrectly  assigned track with respect to the primary electron. The distributions are normalised to the same area.\label{fig:chargemisID}}
\end{figure}
\clearpage

Pair production resulting from bremsstrahlung is the main reason why a wrong track can be matched to the electron candidate; 
three tracks in close proximity are present, two of which have the correct charge assignment, causing an ambiguity in the selection of the primary electron track. 
In this context, ``correct track'' means the primary electron track while ``correct charge'' means the track has the same reconstructed charge as the primary electron.
The number of reconstructed tracks passing the requirements described in Section~\ref{reco:candreco} is shown in Figure~\ref{fig:chargemisID}~(top, right); events with electrons that have a wrongly assigned track tend to have higher track multiplicities. Electrons produced in conversions of photons originating from the primary electron have a most probable value of three tracks regardless of whether the charge was correctly reconstructed or not, illustrating the probabilistic nature of the process.

The presence of additional electrons and positrons in close proximity can cause the track to be reconstructed from hits of primary and secondary electrons and positrons.
Significant energy loss can also distort the shape of the track. At very high energies or large pseudorapidities, the tracks can also become  straighter and hence it is difficult to determine their curvature. 
All of these effects impact the ability to correctly fit the track, as can be seen from the comparison of the correct charge and wrong charge distributions for the correctly assigned track.
These effects are more pronounced as a function of increased passive material, located at large values of $\eta$, as shown by comparing the equivalent two distributions in Figure~\ref{fig:chargemisID}~(top, left).

Figures~\ref{fig:chargemisID}~(bottom, left) and \ref{fig:chargemisID}~(bottom, right) show the distributions of the electric charge multiplied by the transverse impact parameter significance $q\times\dOSignificance$ and the $E/p$, respectively,  for electron candidates with correctly and incorrectly reconstructed charge and assigned track. 
As expected, electrons with incorrectly reconstructed charge tend to have a larger magnitude of \dOSignificance. Their $q\times\dOSignificance$ tends to be more negative due to the incorrect measurement of the electric charge. The distributions have a significant fraction of events with $E/p \neq 1$, due to either the wrong track being matched to the candidate electron or the wrong track curvature being  measured, resulting in an erroneous measurement of the electron's momentum.

\subsection{Suppression of charge misidentification}
\label{sec:ECIDS}

The charge-misidentification rate for reconstructed electron candidates can be reduced with an additional selection criterion based on the output discriminant of a boosted decision tree (BDT) trained using  the TMVA toolkit. The training of the BDT is performed on a simulated single-electron sample.
A set of eight variables related to the quality of the chosen primary electron track, the lateral development of the electromagnetic shower (which is expected to be narrower for electrons that don't undergo bremsstrahlung), and combined tracking and calorimeter quantities is used to separate electrons with the correctly reconstructed electric charge from those reconstructed with the wrong charge. The list of these quantities, ranked by their contribution to the discriminating power of the selection (highest to lowest, as evaluated with a simulated \Zee\ sample), is:  $E/p$,  $\Delta \phi_\mathrm{res}$, $q \times d_0$,  the pseudorapidity $\eta$ of the electron, $R_\phi$, $(q/p)/\sigma(q/p)$, $\Delta \phi_1$, and the transverse energy \ET\ of the electron. Table~\ref{tab:IDcuts} provides the definitions of these variables except for the charge-to-momentum ratio $q/p$ divided by its uncertainty $\sigma(q/p)$ and the distance in $\phi$ between the cluster position in the first layer of the calorimeter and the extrapolated track, $\Delta \phi_1$.

The BDT parameters are optimised to achieve the best possible rejection of electrons reconstructed with the wrong charge given an efficiency loss of 3\% for electrons with correctly reconstructed charge. The resulting efficiencies from applying the additional BDT requirement for electrons already identified with a \Medium requirement (see Section~\ref{sec:identification}) are shown in Figure~\ref{fig:chargemisID_efficiency} for candidate electrons from \Zee\ events. The inclusion of the kinematic properties \ET\ and $\eta$ in the BDT introduces a dependence of the efficiency on these quantities (in particular for $\eta$), even for electron candidates reconstructed with the correct charge. The differences observed between data and simulation arise not only from mismodelling of some of the electromagnetic shower shapes in the simulation but also from the modelling of complicated objects such as the beam-spot. For this reason, data-to-simulation scale factors are calculated; they are also shown in Figure~\ref{fig:chargemisID_efficiency}.
At very high $|\eta|$, the efficiency in data is approximately 76\%. The corresponding efficiency in simulation is about 8\% higher, due to mismodelling.
Analyses using same-charge dilepton pairs such as searches for supersymmetry in same-charge signatures~\cite{SUSY-2016-14} or the search for  Higgs boson production in association with a top-quark pair~\cite{HIGG-2017-02} benefit from a suppression of electrons from charge misidentification with the BDT requirement.

\begin{figure}[t]
  \centering
  \includegraphics[width=0.64\textwidth]{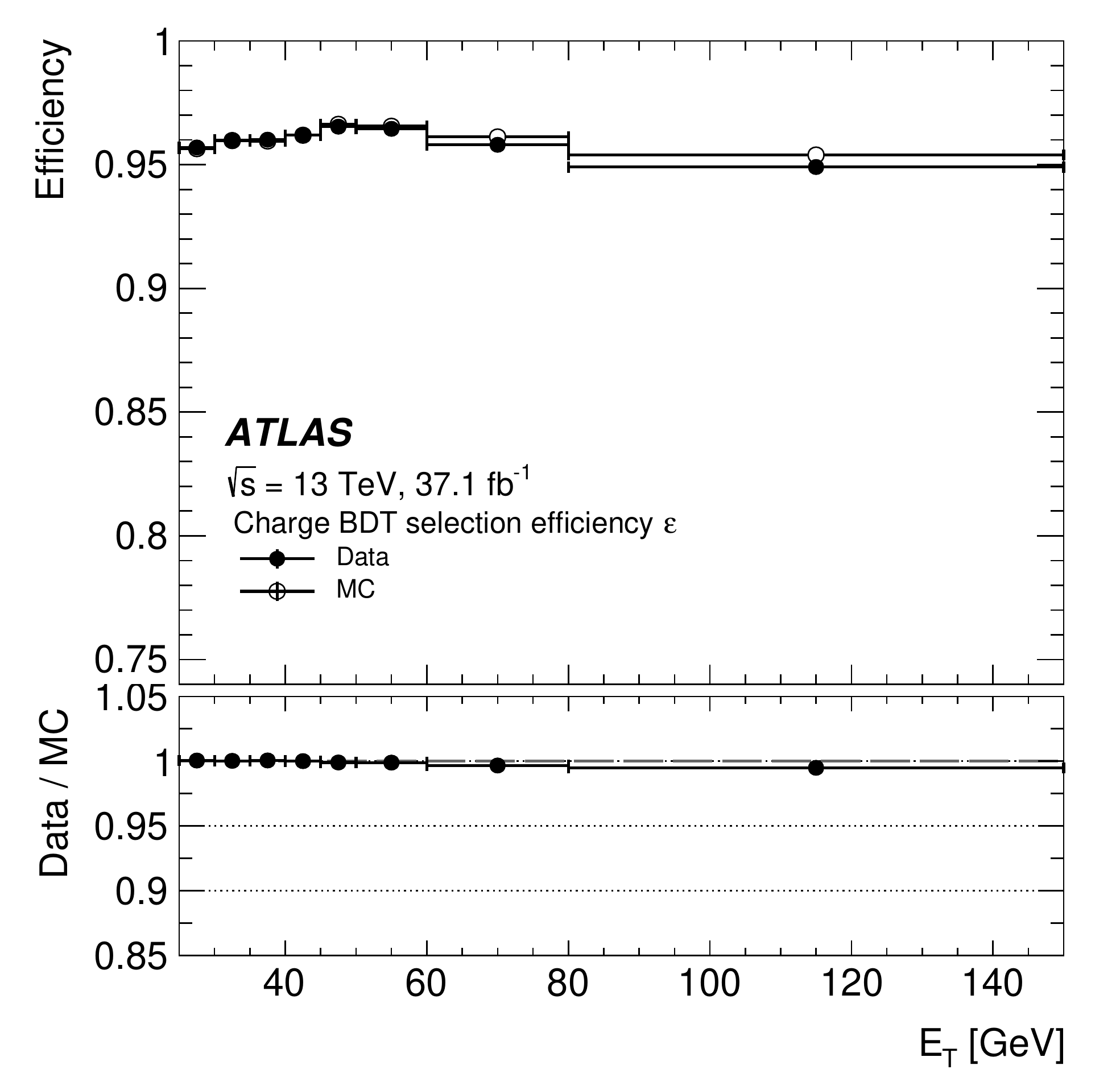}
  \includegraphics[width=0.64\textwidth]{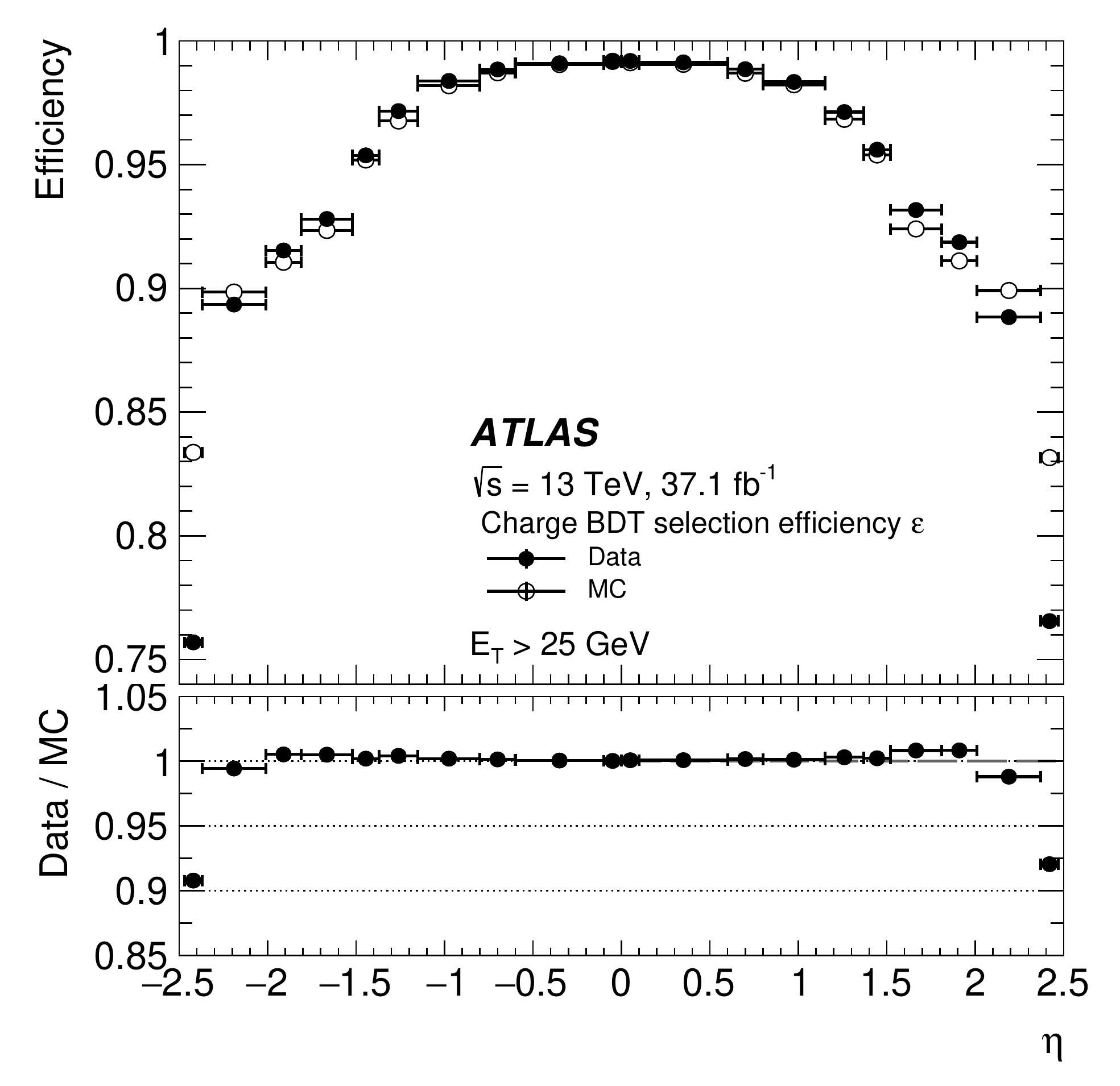}
  \caption{Efficiency of applying the additional BDT selection criterion to choose \Medium-identified electrons reconstructed with the correct charge, as evaluated in \Zee\ events. The criterion is based on a BDT discriminant optimised to select electrons with a correct charge assignment with an efficiency of 97\%.
    Top: vs. \ET; bottom: vs. $\eta$. Open points: data; closed points: simulation; and lower panels: data-to-simulation ratios. The uncertainties are smaller than the markers and hence not visible.}  \label{fig:chargemisID_efficiency}
\end{figure}
\clearpage

\subsection{Measurement of the probability of charge misidentification}
\label{sec:chargemisID_rates}

A measurement of the probability of charge misidentification is performed using a sample of \Zee\ events where both electrons are required to satisfy the same identification and isolation criteria.
The probability $\epsilon$ for an electron to be reconstructed with the wrong charge in a given $(\eta,\ET)$ bin is obtained from the numbers of events where both electrons have the same (SC) and opposite electric charges (OC).  The average \textit{expected} number of  same-charge events for a pair of electrons falling into their respective $(\eta,\ET)$ bins $i$ and $j$, $N^{\mathrm{SC}}_{ij}$, follows from the total number of \textit{observed} events $n^{\mathrm{OC+SC}}_{ij}$ in these same two $(\eta,\ET)$ bins regardless of charge, corrected for the charge-misidentification probability in each bin $\epsilon_{i,j}$:
\begin{equation*}
N^{\mathrm{SC}}_{ij} = n^{\mathrm{OC+SC}}_{ij}[(1-\epsilon_i)\epsilon_j + (1-\epsilon_j)\epsilon_i].\label{eq:nevents}
\end{equation*}
The probabilities $\epsilon_i$ and $\epsilon_j$ are obtained by maximising the Poisson likelihood to observe the number of same-charge events $n^{\mathrm{SC}}_{ij}$ in these bins.
The non-prompt-background contributions to  $n^{\mathrm{OC}}_{ij}$ and $n^{\mathrm{SC}}_{ij}$ are estimated by taking equal-size sideband invariant-mass regions about the \Zboson-boson peak and linearly interpolating to estimate the number of events under the mass peak.  

Systematic uncertainties associated with the \Zboson-boson mass window are evaluated by increasing and decreasing its range.
Systematic uncertainties from other sources are evaluated by not performing the background subtraction, by taking differences between probabilities in events where both electrons fired the trigger and probabilities in events where only one electron fired the trigger, and by comparing the charge-misidentification rates extracted using the likelihood method with the rates extracted from a comparison with the true charge as determined in simulation. 

Figure~\ref{fig:chargemisID_rates} shows the charge-misidentification rates in data and simulation for \Medium\ electrons in \Zee\ events as a function of \ET and $|\eta|$, demonstrating the impact of applying the BDT requirement to suppress charge misidentification, as explained in Section~\ref{sec:ECIDS}. As expected, the charge-misidentification rates are larger at high \ET and  $|\eta|$. Two identification operating points are also contrasted showing the impact of using a \Tight\ identification requirement without the BDT requirement compared with using \Medium\ identification but also including this BDT requirement.

The knowledge of electron charge-misidentification rates is crucial for the estimation of background from charge misidentification in measurements, e.g.\ in the observation of the electroweak  $W^\pm W^\pm jj$ production~\cite{STDM-2017-06} or the search for doubly charged Higgs boson production~\cite{EXOT-2016-07}.
To account for disagreements between data and simulation, data-to-simulation ratios, extracted from the efficiencies described above, are provided as correction factors for analyses to account for this mismodelling in the simulation.

\begin{figure}[t]
  \centering
  \includegraphics[width=0.64\textwidth]{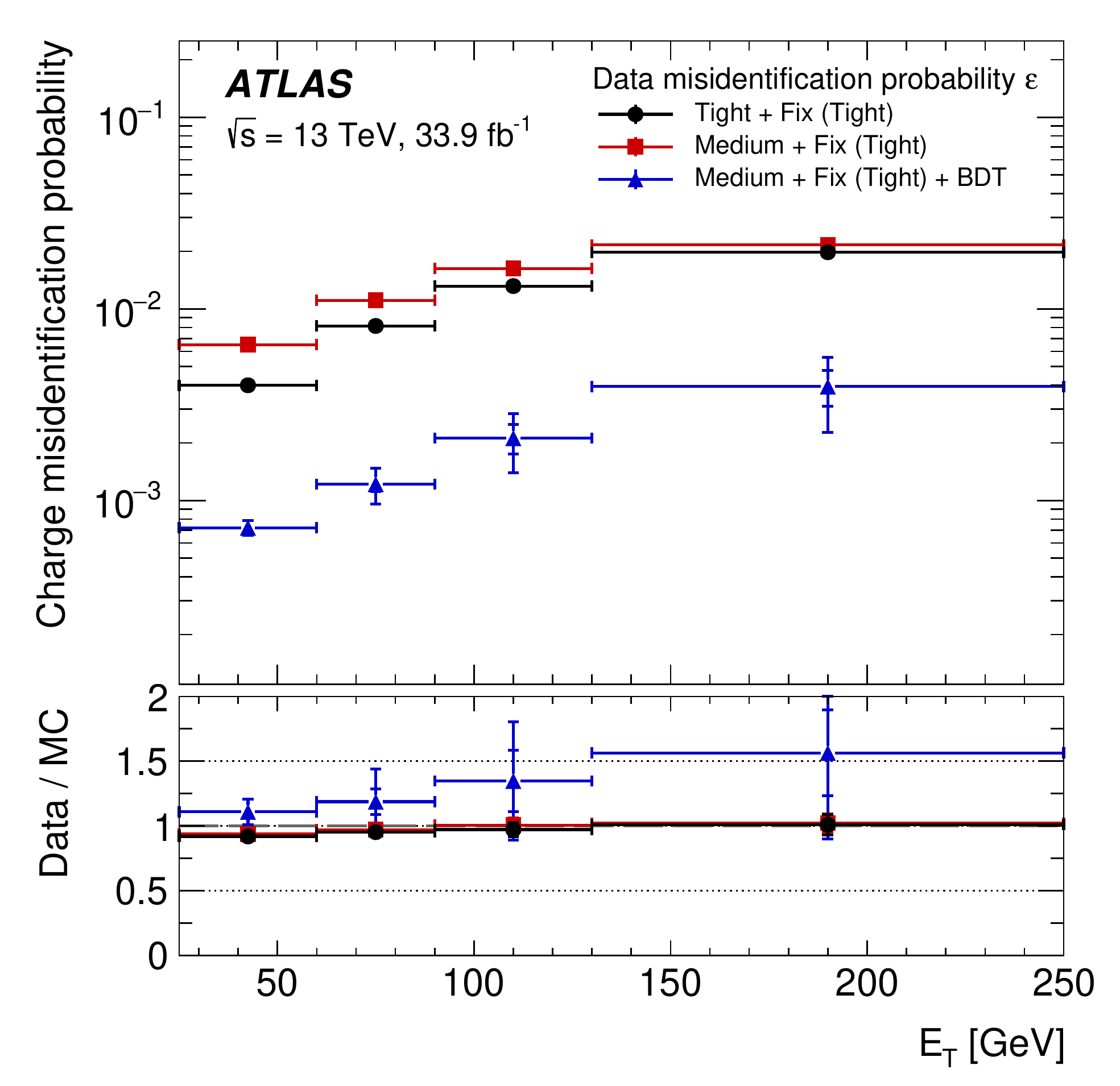}
  \includegraphics[width=0.64\textwidth]{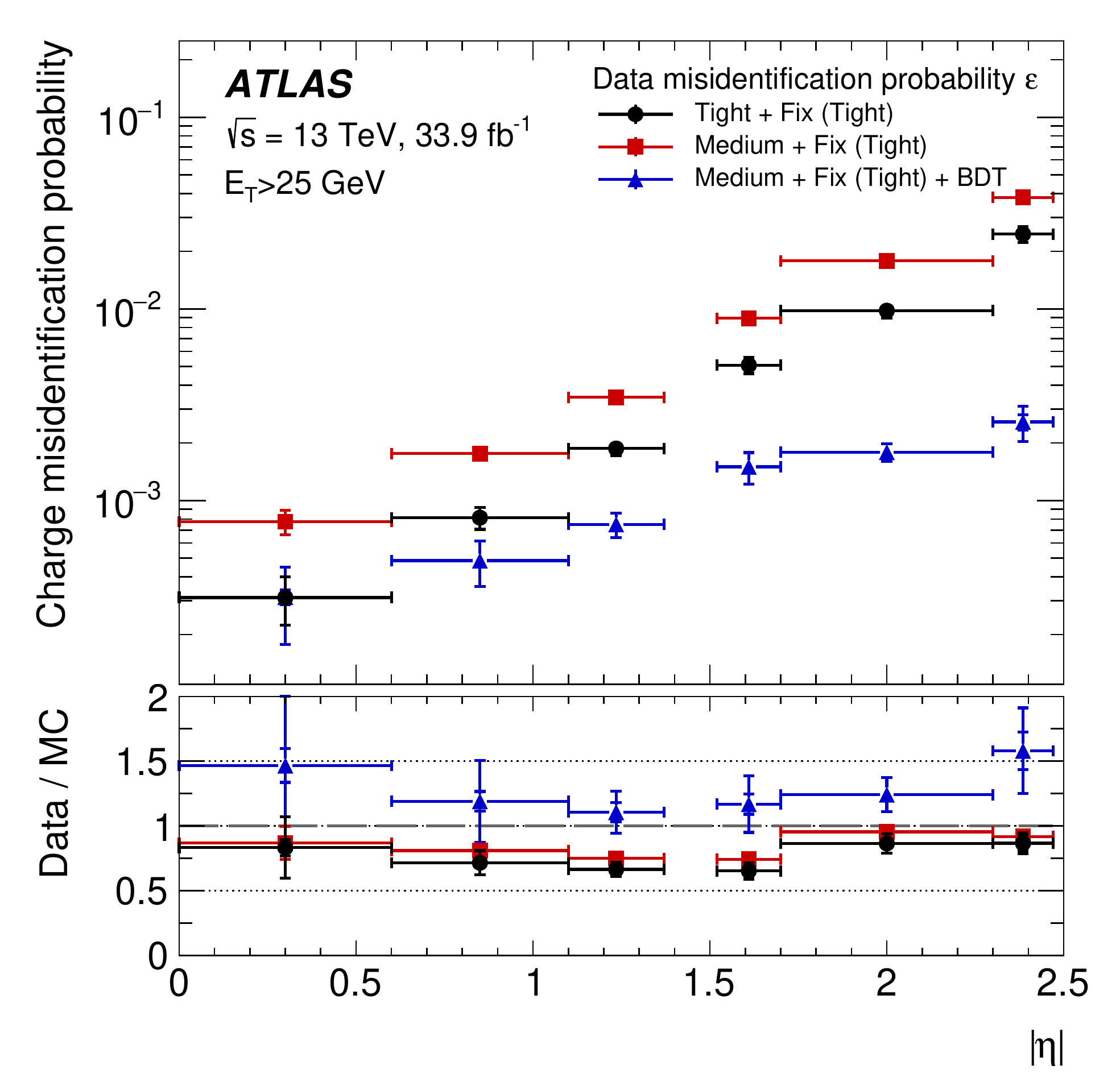}
  \caption{Charge-misidentification rates in 2016 data and simulated \Zee\ events as a function of \ET (top) and $|\eta|$ (bottom) showing the impact of applying the BDT requirement to suppress charge misidentification (red squares vs. blue triangles). \Medium identification with BDT vs.\ \Tight identification without BDT is also explored (black circles vs. blue triangles). The Fix~(Tight) isolation requirement is applied in all cases (see Table~\ref{tab:leptonWP}).
The inner uncertainties are statistical while the total uncertainties include both the statistical and systematic components.
  }
  \label{fig:chargemisID_rates}
\end{figure}
\clearpage

\section{Usage of electron selections in physics measurements}
\label{sec:examples}
 
Efficiencies for electron reconstruction, identification, isolation, and charge identification are shown for several operating points  in Sections~\ref{sec:reconstruction}, \ref{sec:identification}, \ref{sec:isolation}, and \ref{sec:chargemisid}, respectively. The product of these efficiencies  (and their corresponding data-to-simulation correction factors) is relevant in physics measurements. In this section, electron efficiencies for data from a  \Zee\ sample are presented for the combination $\epsreco \times \epsid \times \epsiso$ (see Eq.~(\ref{eq:EffDef})) for some typical operating points used in ATLAS physics analyses. The results are shown in Figure~\ref{fig:alleff} for the following operating points  (listed as Identification~+~Isolation):
\begin{itemize}
\item \Loose~+~Fix (Loose): used in cross-section measurements of $H\rightarrow 4\ell$~\cite{HIGG-2016-25} and Standard Model $ZZ$ production~\cite{STDM-2016-15},
\item \Medium~+~Gradient (Loose): used in searches for direct top-squark-pair production in final states with two leptons~\cite{SUSY-2016-17} and to identify the \Zboson-boson candidate in measurements of Standard Model $WZ$ production~\cite{STDM-2018-03},
\item \Tight~+~Gradient: used in measurements of single-top-quark production in association with a \Zboson or \Wboson boson~\cite{TOPQ-2016-12,TOPQ-2016-14}  and Standard Model $WW$ production~\cite{Aaboud:2019nkz},
\item \Tight~+~Fix (Track $R_\mathrm{max}=0.4$) used in cross-section measurements of gluon--gluon fusion and vector-boson-fusion Higgs-boson production in the $H \rightarrow WW^* \rightarrow e\nu\mu\nu$ decay channel~\cite{HIGG-2016-07}.
\end{itemize}

\begin{figure}[t]
\centering
\includegraphics[width=0.51\textwidth]{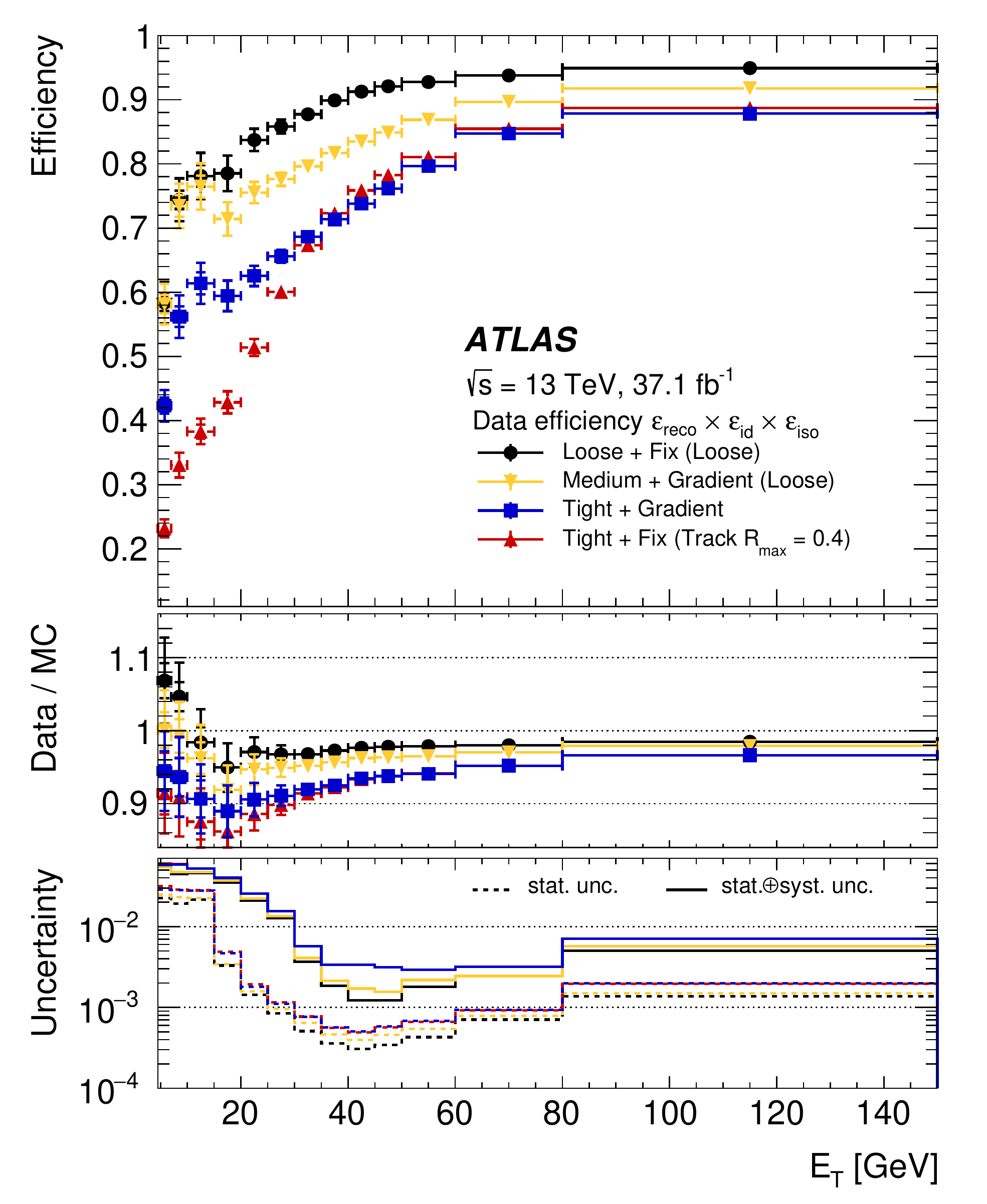}
\includegraphics[width=0.51\textwidth]{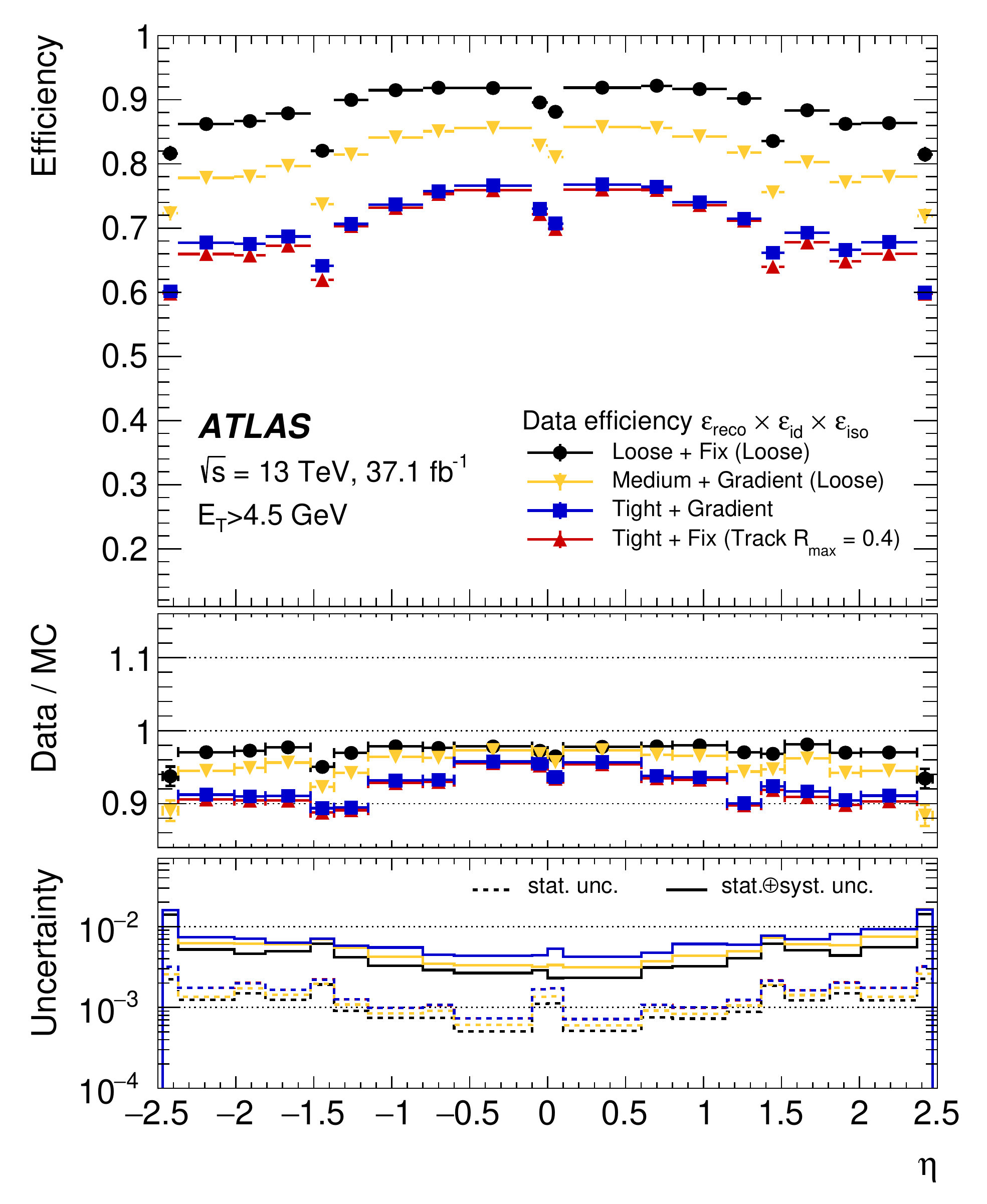}
\caption{The product of reconstruction, identification, and isolation efficiencies $\epsreco \times \epsid \times \epsiso$ for data from a  \Zee\ sample as a function of electron \ET (top) and $\eta$ for $\ET>4.5$~\GeV\ (bottom), for the following operating points:  ``\Loose~+~Fix~(Loose)''~(black circles), ``\Medium~+~Gradient~(Loose)''~(yellow upside-down triangles), ``\Tight~+~Gradient''~(blue squares) and ``\Tight~+~Fix''~(Track $R_\mathrm{max}=0.4$)~(red triangles). The inner uncertainties are statistical while the total uncertainties include both the statistical and systematic components.
The lower panels show data-to-simulation ratios as well as the relative statistical and total uncertainties (statistical and systematic added in quadrature) applicable to both the data efficiencies and correction factors.
\label{fig:alleff}
}
\end{figure}

\FloatBarrier 

\section{Conclusions}
\label{sec:conclusion}
The focus of this paper is to document the methods used by the ATLAS experiment at the LHC in Run~2 to reconstruct, identify, and isolate prompt-electron candidates  with high efficiency, as well as to suppress electron-charge misidentification. 
Efficiency measurements as evaluated with \lumi~\ifb\ of 13~\TeV\ $pp$ collision data  recorded in 2015--2016 are then compared with those determined from simulation. 

The method used to reconstruct electron candidates
is explained for candidates in the central region of the detector ($\abseta < 2.47$). Illustrative measurements of the reconstruction efficiency are presented for candidates with transverse energy $\ET > 4.5$~\GeV.  For $\ET > 15$~\GeV, the reconstruction efficiencies range from 96\% to 99\%, with  the data uncertainties typically 1\% for $\ET = 15-20$~\GeV\ and reaching the per-mille level at higher \et.

The likelihood method used to identify electrons given the existence of a reconstructed electron candidate is also explained for candidates in the central region of the detector ($\abseta<2.47$). The minimum \Et of the identified electron is reduced from 7~\GeV\ in Run~1 to 4.5~\GeV\ in Run~2. Benchmark efficiency measurements for three identification operating points are provided: 
for the \Tight operating point, the identification efficiency varies from 55\% at $\Et = 4.5$~\GeV\ to 90\% at $\Et = 100$~\GeV, while it ranges from 85\% at $\Et = 20$~\GeV\ to 96\% at $\Et = 100$~\GeV\ for the \Loose operating point. 
The uncertainties in these measured efficiencies for the \Loose (\Tight) operating point range from 3\% (4\%) at $\Et = 4.5$~\GeV\ to  0.1\% (0.3\%) for $\Et = 40$~\GeV.

The ability to identify electron candidates isolated from any other local activity in the detector is documented. Two varieties of isolation variables, calorimeter- and tracking-based, are considered. 
Since the implementation of isolation criteria is specific to the physics analysis needs, several operating points are illustrated; their typical isolation efficiencies are measured in data and determined from simulation, ranging from approximately 90\% for the tightest operating points to nearly 99\% for the loosest, for electrons with $\Et = 40$~\gev. 

Possible sources of electron-charge misidentification are explored and an algorithm to suppress such effects is outlined. 
The efficiency for correct charge reconstruction is illustrated for candidates passing the \Medium identification requirements. The efficiency is approximately 96\% in both data and simulation and is mostly independent of \ET, while it varies from 90\% to 99\% over most of the electron $\eta$ region except at very high $|\eta|$ where the efficiency in data is approximately 76\%. The corresponding efficiency in simulation is about 8\% higher, due to mismodelling.

Finally, since total efficiencies are used in ATLAS physics analyses, the product of the electron efficiencies $\epsreco \times \epsid \times \epsiso$ measured in data and presented in this paper for some typical operating points demonstrate the impact of these efficiencies on ATLAS measurements and searches for new physics.

\section*{Acknowledgements}

We thank CERN for the very successful operation of the LHC, as well as the
support staff from our institutions without whom ATLAS could not be
operated efficiently.

We acknowledge the support of ANPCyT, Argentina; YerPhI, Armenia; ARC, Australia; BMWFW and FWF, Austria; ANAS, Azerbaijan; SSTC, Belarus; CNPq and FAPESP, Brazil; NSERC, NRC and CFI, Canada; CERN; CONICYT, Chile; CAS, MOST and NSFC, China; COLCIENCIAS, Colombia; MSMT CR, MPO CR and VSC CR, Czech Republic; DNRF and DNSRC, Denmark; IN2P3-CNRS, CEA-DRF/IRFU, France; SRNSFG, Georgia; BMBF, HGF, and MPG, Germany; GSRT, Greece; RGC, Hong Kong SAR, China; ISF and Benoziyo Center, Israel; INFN, Italy; MEXT and JSPS, Japan; CNRST, Morocco; NWO, Netherlands; RCN, Norway; MNiSW and NCN, Poland; FCT, Portugal; MNE/IFA, Romania; MES of Russia and NRC KI, Russian Federation; JINR; MESTD, Serbia; MSSR, Slovakia; ARRS and MIZ\v{S}, Slovenia; DST/NRF, South Africa; MINECO, Spain; SRC and Wallenberg Foundation, Sweden; SERI, SNSF and Cantons of Bern and Geneva, Switzerland; MOST, Taiwan; TAEK, Turkey; STFC, United Kingdom; DOE and NSF, United States of America. In addition, individual groups and members have received support from BCKDF, CANARIE, CRC and Compute Canada, Canada; COST, ERC, ERDF, Horizon 2020, and Marie Sk{\l}odowska-Curie Actions, European Union; Investissements d' Avenir Labex and Idex, ANR, France; DFG and AvH Foundation, Germany; Herakleitos, Thales and Aristeia programmes co-financed by EU-ESF and the Greek NSRF, Greece; BSF-NSF and GIF, Israel; CERCA Programme Generalitat de Catalunya, Spain; The Royal Society and Leverhulme Trust, United Kingdom. 

The crucial computing support from all WLCG partners is acknowledged gratefully, in particular from CERN, the ATLAS Tier-1 facilities at TRIUMF (Canada), NDGF (Denmark, Norway, Sweden), CC-IN2P3 (France), KIT/GridKA (Germany), INFN-CNAF (Italy), NL-T1 (Netherlands), PIC (Spain), ASGC (Taiwan), RAL (UK) and BNL (USA), the Tier-2 facilities worldwide and large non-WLCG resource providers. Major contributors of computing resources are listed in Ref.~\cite{ATL-GEN-PUB-2016-002}.

\printbibliography
\clearpage
 
\begin{flushleft}
{\Large The ATLAS Collaboration}

\bigskip

M.~Aaboud$^\textrm{\scriptsize 34d}$,    
G.~Aad$^\textrm{\scriptsize 99}$,    
B.~Abbott$^\textrm{\scriptsize 125}$,    
D.C.~Abbott$^\textrm{\scriptsize 100}$,    
O.~Abdinov$^\textrm{\scriptsize 13,*}$,    
B.~Abeloos$^\textrm{\scriptsize 129}$,    
D.K.~Abhayasinghe$^\textrm{\scriptsize 91}$,    
S.H.~Abidi$^\textrm{\scriptsize 164}$,    
O.S.~AbouZeid$^\textrm{\scriptsize 39}$,    
N.L.~Abraham$^\textrm{\scriptsize 153}$
% ...
\bigskip
\\

% ...
$^{13}$Institute of Physics, Azerbaijan Academy of Sciences, Baku; Azerbaijan.\\
% ...
$^{34}$$^{(a)}$Facult\'e des Sciences Ain Chock, R\'eseau Universitaire de Physique des Hautes Energies - Universit\'e Hassan II, Casablanca;$^{(b)}$Centre National de l'Energie des Sciences Techniques Nucleaires (CNESTEN), Rabat;$^{(c)}$Facult\'e des Sciences Semlalia, Universit\'e Cadi Ayyad, LPHEA-Marrakech;$^{(d)}$Facult\'e des Sciences, Universit\'e Mohamed Premier and LPTPM, Oujda;$^{(e)}$Facult\'e des sciences, Universit\'e Mohammed V, Rabat; Morocco.\\
% ...
$^{39}$Niels Bohr Institute, University of Copenhagen, Copenhagen; Denmark.\\
% ...
$^{91}$Department of Physics, Royal Holloway University of London, Egham; United Kingdom.\\
% ...
$^{99}$CPPM, Aix-Marseille Universit\'e, CNRS/IN2P3, Marseille; France.\\
$^{100}$Department of Physics, University of Massachusetts, Amherst MA; United States of America.\\
% ...
$^{125}$Homer L. Dodge Department of Physics and Astronomy, University of Oklahoma, Norman OK; United States of America.\\
% ...
$^{129}$LAL, Universit\'e Paris-Sud, CNRS/IN2P3, Universit\'e Paris-Saclay, Orsay; France.\\
% ...
$^{153}$Department of Physics and Astronomy, University of Sussex, Brighton; United Kingdom.\\
% ...
$^{164}$Department of Physics, University of Toronto, Toronto ON; Canada.\\
% ...
$^{a}$ Also at Borough of Manhattan Community College, City University of New York, NY; United States of America.\\
$^{b}$ Also at California State University, East Bay; United States of America.\\
$^{c}$ Also at Centre for High Performance Computing, CSIR Campus, Rosebank, Cape Town; South Africa.\\
$^{d}$ Also at CERN, Geneva; Switzerland.\\
$^{e}$ Also at CPPM, Aix-Marseille Universit\'e, CNRS/IN2P3, Marseille; France.\\
% ...
$^{*}$ Deceased

\end{flushleft}



\end{document}